%% file: main_r1.tex
\documentclass[fleqn,usenatbib,useAMS]{mnras}
\usepackage{newtxtext,newtxmath}
\usepackage[T1]{fontenc}
\usepackage{amsmath}
\usepackage{xcolor}
\usepackage{xspace}
\usepackage{xurl}
\usepackage{array}
\usepackage{graphicx}
\usepackage{orcidlink}
\usepackage{rotating}
\usepackage{listings}
\newcommand{\kms}{\ensuremath{\mathrm{km\,s}^{-1}}\xspace}
\newcommand{\hcop}{HCO$^{+}$\xspace}
\newcommand{\app}{$\sim$\xspace}

\newcommand{\cms}{cm$^{-2}$\xspace}

\newcommand{\updates}[1]{\textcolor{black}{#1}}

\defcitealias{Longmore2025}{Paper~I}
\defcitealias{Ginsburg2025}{Paper~II}
\defcitealias{Walker2025}{Paper~III}
\defcitealias{Lu2025}{Paper~IV}
\defcitealias{Hsieh2025}{Paper~V}

\title[ACES HNCO and \hcop]{ALMA Central molecular zone Exploration Survey (ACES) III:\\ Molecular line data reduction and HNCO and \hcop data}

\input{affiliation_macros.tex}
\input{affiliation_labels}
\input{authors}

\date{Accepted 2026 February 02. Received 2026 February 02; in original form 2025 July 18}

\pubyear{2025}

\begin{document}
\label{firstpage}
\pagerange{\pageref{firstpage}--\pageref{lastpage}}
\maketitle

\begin{abstract}
The ALMA Central molecular zone Exploration Survey (ACES) large program has observed the inner \app 200~pc of the Milky Way at 3~mm (Band 3) using ALMA's 12m, 7m, and Total Power arrays. With an angular resolution of \app 2\arcsec\, ACES provides a contiguous, multi-scale view of the Central Molecular Zone (CMZ) via the dust continuum and a suite of molecular lines. We present an overview of the molecular line data processing for ACES and describe the first data release. We showcase the HNCO (4-3) and \hcop (1-0) data, which were targeted at high spectral resolution (0.2~\kms) to trace the kinematics of the molecular gas in the CMZ. The HNCO and \hcop maps are compared with previous single-dish CMZ surveys and discrete ALMA observations of CMZ clouds to demonstrate the quality of the data. \updates{We highlight the ubiquity of parsec-scale, linear absorption features traced by \hcop. Their origin is unknown, and ACES provides the first opportunity to study these enigmatic features throughout the CMZ. We release the HNCO and \hcop cubes for all 45 ACES fields, along with the full cube mosaics which combine all fields into a contiguous mosaic of the CMZ. We additionally provide advanced products of these full mosaics, including integrated and peak intensity, noise, and position-velocity maps. These products provide substantial legacy value for the community, offering an unparalleled view of the physical and kinematic structure of the dense gas in the CMZ.}
\end{abstract}


\section{Introduction}
\label{sec:intro}

The Central Molecular Zone (CMZ; inner few hundred parsecs of the Milky Way) is extreme compared to typical star forming environments in the disc of the Galaxy, exhibiting higher gas temperatures \citep[e.g.][]{Ao2013, Ginsburg2016, Immer2016, Krieger2017}, average densities \citep[e.g.][]{Mills18}, pressures \citep[e.g.][]{Myers2022}, cosmic ray ionisation rates \citep[e.g.][]{Indriolo2015, Oka2019}, stronger magnetic field \citep[e.g.][]{Pillai2015, Mangilli2019, Lu2024}, and elevated turbulence as evidenced via broad line-widths on parsec scales \citep[e.g.][]{Henshaw2016b}. The star formation rate (SFR) in the CMZ is observed to be at least an order of magnitude lower than expected based on common dense-gas--SFR relations \citep[e.g.][]{Longmore2013b, Barnes_17, Lu19}. While the extreme environmental conditions are likely intimately linked to the relative dearth of star formation \citep{Kruijssen2014, Federrath16}, a comprehensive understanding has not yet been achieved.

To better understand how the environment shapes the nature and location of star formation throughout the CMZ, it is necessary to capture a multi-scale view of the molecular gas, from global scales (\app 100~pc) down to the scales of individual star/cluster forming regions (\app 0.1~pc). To do so simultaneously in a homogeneous fashion has not been feasible previously, due to the distance of the CMZ (commonly approximated to be at the distance of the Galactic Centre at 8.2~kpc; \citealt{GravityCollaboration2019}) combined with the large angular scale on the sky. 

Several molecular line surveys have observed the CMZ on large scales using single dish facilities, such as the H$_{2}$O southern Galactic Plane Survey (HOPS; e.g., \citealt{Walsh2008, Walsh2011}), the Mopra CMZ survey \citep{Jones2012, Jones2013}, the APEX CMZ survey \citep{Ginsburg2016}, and the Survey of Water and Ammonia in the Galactic Center (SWAG; \citealt{Krieger2017}). Surveys probing smaller physical scales have been limited by the required combination of high sensitivity at high angular resolution. The SMA CMZoom survey \citep{Battersby2020} observed all of the gas above a column density of 10$^{23}$~\cms at an angular resolution of \app 3\arcsec\,\,(0.1~pc), but this did not represent a contiguous mosaic, and was mostly sensitive to compact structures. Many ALMA projects have been undertaken to observe the small scale structure in CMZ clouds, but this has been done in a piecemeal, heterogeneous manner (see \citealt{Henshaw2023} for a comprehensive overview of such observations).

To address this need for a contiguous, uniform, and multi-scale observation of the molecular gas in the CMZ, the ALMA Large Program ‘ALMA Central molecular zone Exploration Survey’ (ACES) was designed to observe the central \app 200~pc of the Galaxy with ALMA's 12m, 7m, and Total Power (TP) arrays in Band 3 (3~mm) at an angular resolution of \app 2\arcsec.

In this work, we present an overview of the spectral line data processing for ACES, which has generated the largest mosaics ALMA has yet produced, covering $>$1000 arcmin$^2$ with 2$\arcsec$ (0.05\,pc) resolution. This paper focuses primarily on the description and release of the HNCO (4-3) and \hcop (1-0) data, which are the main kinematic tracers of ACES and were observed at high spectral resolution (\app 0.2~\kms), with HNCO being the primary kinematic tracer of the dense molecular gas. These data open a new window in kinematic precision, resolution and sensitivity, tracing the complex gas motion (orbital rotation, shear, collapse under self-gravity, outflows due to feedback, infall towards the central supermassive black hole, etc.) across the whole CMZ down to the sub-sonic regime for the first time. This new understanding of gas kinematics is pivotal to ACES' main science goal of building a global understanding of star formation, feedback, and the mass flows and energy cycles across the Central Molecular Zone \citep{Kruijssen2014, Krumholz17, Sormani2022}.

The spectral setup of ACES targeted HNCO (4-3) and \hcop (1-0) in dedicated narrow bandwidth spectral windows. The setup additionally included four more spectral windows, which targeted other spectral lines and 3~mm dust continuum at coarser spectral resolution. \updates{This paper is part III of a series of initial survey papers. We refer the reader to \citet[][hereafter \citetalias{Longmore2025}]{Longmore2025} for an overview of the survey design, \citet[][hereafter \citetalias{Ginsburg2025}]{Ginsburg2025} for a description of the continuum data processing, \citet[][hereafter \citetalias{Lu2025}]{Lu2025} for an overview of the intermediate bandwidth spectral windows, and \citet[][hereafter \citetalias{Hsieh2025}]{Hsieh2025} for an overview of the broad bandwidth spectral windows.} Though this paper focuses on the HNCO and \hcop data, the majority of the data processing steps are common to all molecular line data, unless otherwise stated. 

Section \ref{sec:obs} presents an overview of the observations and spectral setup. Section \ref{sec:processing} provides a detailed description of the data reduction pipeline, along with a discussion of the specific quirks and caveats relating to the HNCO and \hcop data. Section \ref{sec:analysis} presents a brief analysis of the HNCO and \hcop data products and compares them with previous observations. Section \ref{sec:conclusions} gives a summary of this work.

\section{Observations}
\label{sec:obs}

ACES was observed as an ALMA Large Program (ID: 2021.1.00172.L., PI: S. Longmore) across observing cycles 8 and 9. For a complete overview of the survey design, we refer the reader to \citetalias{Longmore2025}. Here, we provide a brief overview of the observations, with details relevant to the data products presented (see also other data release papers: \citetalias{Ginsburg2025, Lu2025, Hsieh2025}).

\begin{figure*}
    \centering
    \includegraphics[width=\textwidth]{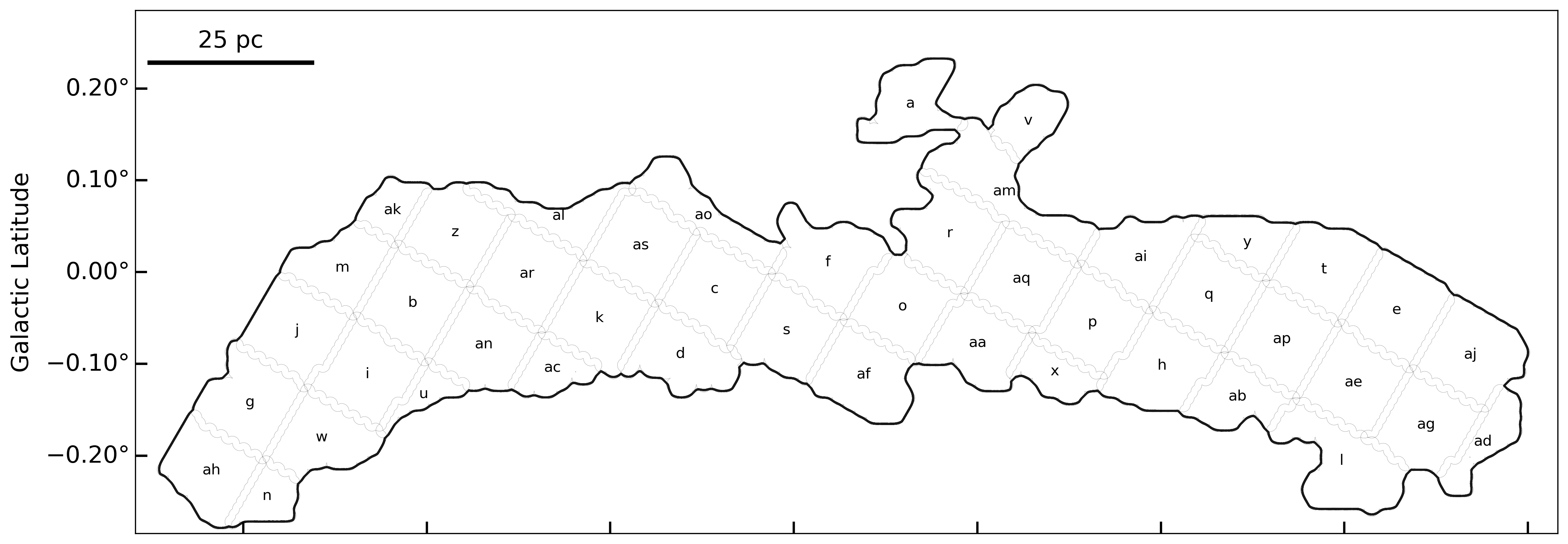}
    \includegraphics[width=\textwidth]{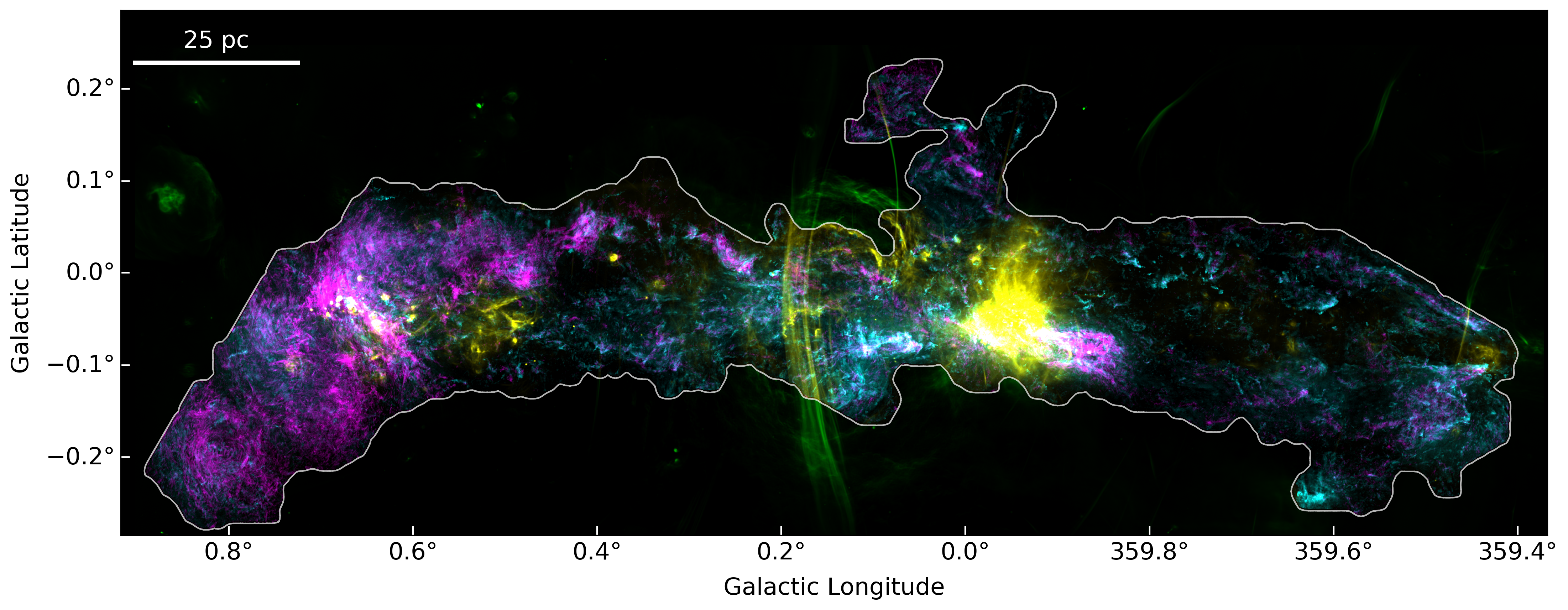}
    \caption{The top panel shows the coverage of ACES. The grey contours show the footprint of the 45 individual ALMA mosaics, where the alphabetical ID in the centre of each corresponds to the field ID (see Section \ref{sec:obs} and Tables \ref{tab:hcopcubestats} and \ref{tab:hncocubestats}). The solid black outline shows the overall footprint of ACES resulting from the combination of all 45 mosaics. The bottom panel shows a three-colour image of the CMZ. Magenta: ACES HNCO peak intensity (Section \ref{sec:moments}), Cyan: ACES \hcop peak intensity (Section \ref{sec:moments}), Yellow: MeerKAT 1.28~GHz continuum \citep{Heywood2022}. The white contour corresponds to the overall footprint of ACES.}
    \label{fig:overview}
\end{figure*}

The spatial setup of the survey was designed to observe a contiguous mosaic of the entire CMZ, targeting all material above a H$_{2}$ column density of 10$^{22}$~\cms. To do this requires $>5000$ pointings with the main ALMA 12m array in Band 3, which far exceeds the mosaic limit of 150 pointings set by ALMA observatory for an individual Scheduling Block (SB). As such, the survey was split into 45 sub-mosaics, named \texttt{a}--\texttt{as}, such that when combined, the result is a contiguous image of the CMZ (see Figure \ref{fig:overview} and \citetalias{Longmore2025, Ginsburg2025}). At the time of writing, ACES represents the largest mosaic ever observed with ALMA. 

Each field (i.e. each sub-mosaic) was observed using a single 12m scheduling block in configuration C-3 or C-4, a 7m SB, and a Total Power (TP) SB. The observations were made between October 2021 and September 2023. 
A discussion of how these SBs from the different arrays are combined is given in Section \ref{sec:comb}.

The spectral coverage of the survey was set up to obtain a mix of spectral resolutions and bandwidths across six spectral windows (SPWs). In the upper sideband there are two broad, lower spectral resolution SPWs that are dedicated to continuum sensitivity, along with specific lines of interest and serendipitous line detection. In the lower sideband there are two medium bandwidth SPWs with intermediate spectral resolution, and two high spectral resolution, narrow bandwidth SPWs. These lower sideband SPWs are dedicated to specific lines of interest, with the two narrowest SPWs targeting \hcop (1-0) and HNCO (4-3) -- the two main kinematic tracers of the survey, and the primary focus of this paper. Full details of the spectral setup are given in Table \ref{tab:spectral_setup}. 

% Suggestion from Alvaro: Include a sketch of the spectral setup with SPWs, key lines, etc.

While the spectral set up of all fields is generally uniform, the central frequency of 21 fields was shifted, with an equivalent velocity offset ranging from -30 to +40~\kms. The reason for these offsets is that the two SPWs targeting HNCO and \hcop are very narrow (58.59 MHz in width, equivalent to $\sim200$~\kms), and the line emission in the CMZ spans a large velocity range (exceeding $\pm 100$\,\kms). As such, it was not possible to use a single central frequency that captured this range across the full survey footprint, and so offsets were introduced for fields at the extremes of this range, to ensure that the bulk of the emission was within the frequency range of the SPWs. These offsets were estimated using prior knowledge of gas velocities based on the many previous CMZ observations \citep[see][and references therein]{Henshaw2023}. Tables \ref{tab:hcopcubestats} and \ref{tab:hncocubestats} give these offsets per field. These two SPWs are necessarily narrow bandwidth as we targeted HNCO and \hcop at high spectral resolution (channel widths \app 0.03~MHz / 0.1~\kms) in order to resolve the kinematics of the gas down to the expected thermal line-width (see \citetalias{Longmore2025} for more details regarding the science goals of ACES).

\input{Tables/spectral_setup}

\section{Data processing}
\label{sec:processing}

\subsection{Calibration and QA}
\label{sec:cal}

All ACES data were calibrated and imaged using CASA \citep{CASATeam2022}, following standard pipeline calibration procedures with the same CASA version as the original ALMA pipeline reduction. Most of the data were processed (i.e. calibration restoration and imaging) with CASA 6.2.1.7, although some of the later datasets were processed with 6.4.1.12, as the project spanned multiple observing cycles and thus multiple pipeline releases. Later versions of CASA, both standard and pipeline, were used for subsequent re-imaging of individual problematic fields. We refer the reader to Table 2 in \citetalias{Ginsburg2025} for the full details. 

Calibration was performed by restoring the ALMA pipeline calibration, generally with no changes to the input parameters. In a small number of cases, calibration issues were identified and addressed if necessary. An example of this is that the phase calibrator J1744-3116 consistently showed absorption in the \hcop window (12m SPW 29). In general this was flagged appropriately during QA2, however, some of the data were delivered without the necessary flagging. In some cases, the data were returned to ALMA for QA3 reprocessing, while other, more minor cases were flagged manually. In contrast to the calibration, pipeline imaging did require a substantial amount of manual parameter changes to address various issues when imaging specific fields (see sections \ref{sec:contsub} and \ref{sec:imaging}).

We adopted a quality assessment (QA) workflow within the data reduction team to perform additional QA beyond that performed by ALMA at the QA2 stage. Our internal QA consisted of checking the weblog and pipeline-imaged products for every \updates{Scheduling Block} (SB) across all three arrays. \updates{We note that all 7m and TP SBs have multiple Execution Blocks (EBs) per SB. The 12m SBs generally have multiple EBs also, though this is not always true, as in some cases only a single execution was required to achieve the requested RMS (see Table 2 in \citetalias{Ginsburg2025} for a full list of all ACES 12m executions).} We used GitHub for tracking \updates{our QA} process, whereby every SB has an associated issue containing information about the observations, links to the weblog, and comments documenting the QA history and relevant fixes to problems that were discovered\footnote{\updates{All ACES GitHub issues can be found at \url{https://github.com/ACES-CMZ/reduction_ACES/issues}.}}.

In a handful of cases, significant problems were uncovered via this internal QA, and the ALMA QA3 process was initiated to address the issues. The most significant case related to the OFF positions used in the TP observations, which are discussed in more detail in Section \ref{sec:tp_baselines}. Another example relevant to this paper is that we discovered an issue in the HNCO cube for the 12m data for field \texttt{af}, in which the emission seemed to strongly diverge, with amplitudes increasing significantly towards the upper edge of the frequency range. \updates{We verified that this signal was not due to divergence during cleaning, but rather was present in the actual visibility data.} This dataset underwent QA3, where it was discovered that a single sub-scan was corrupted but was not appropriately flagged during QA2.  

\subsection{Continuum subtraction}
\label{sec:contsub}

The continuum subtraction was originally performed using the standard ALMA pipeline procedure, using the continuum selection made by the \texttt{findcont} task \updates{(see Section 6 of \citealt{Hunter2023} for a detailed description of this algorithm)}. While this worked well in many cases, we found that the continuum selection was poor in some fields/spectral windows, particularly for fields with strongly varying continuum emission as a function of position (e.g. Sgr B2). The result was a poor continuum subtraction, where residual continuum emission was present throughout the affected fields. This was more problematic in the broader spectral windows, where there are many more spectral lines present due to the larger bandwidth.

Additionally, the pipeline used a default polynomial fit order of 1 when performing the continuum subtraction. This was problematic specifically for the HNCO and \hcop spectral windows, as their narrow bandwidths would often be filled entirely by the broad line emission, leaving little-to-no true continuum channels. This was particularly problematic in the cases where \texttt{findcont} identified a small number of continuum channels at only one edge of the window, which, when combined with a fit order of 1 would introduce an artificial slope into the continuum subtracted spectrum\footnote{This issue was addressed in a later pipeline release (2024.1.0.8, CASA 6.6.1-17).}.

Though this slope issue only affected a subset of the data, for the sake of consistency we reran the continuum subtraction and cube imaging for \textit{all} of the 12m and 7m data, this time using a polynomial fit order of 0 to mitigate the introduction of artificial spectral slopes. We then ran all image cubes through \texttt{statcont} \citep{Sanchez-Monge2018} to remove any residual continuum emission that resulted from the aforementioned poor continuum channel selection in fields with strongly varying continuum emission as a function of position. Although this issue only affected a subset of the fields (e.g., near Sgr B2, Sgr A*, etc.), we \updates{applied this approach to} all fields and SPWs for consistency.

\subsection{Imaging}
\label{sec:imaging}

We re-imaged all 12m and 7m data for all fields and SPWs to account for the continuum subtraction issues discussed in the previous section, along with other field specific issues which are discussed in this section. 

The majority of the imaging was performed in an automated way, using the \texttt{tclean} task in CASA\updates{, which performs deconvolution of interferometric data using a specified algorithm like CLEAN \citep{Hogbom1974, CASATeam2022}.} In the majority of cases, imaging was performed with the same parameters as those selected by the ALMA pipeline \citep{Hunter2023}. In some cases, it was necessary to manually specify some parameters to correct various issues in the original pipeline imaging. In particular, divergence in the cubes was a common occurrence, especially in the broad continuum spectral windows (12m SPWs 33 and 35). In these cases, our typical approach was to iteratively increase the \texttt{cyclefactor} parameter until the offending channels were no longer diverging. The \texttt{cyclefactor} parameter controls the threshold triggering a major cycle in \texttt{tclean}; higher values mean fewer cleaning iterations per minor cycle. Increasing this parameter therefore forces more frequent recalculations of the residual image, which minimises the chance of error accumulation that may result in divergence. We found that using a \texttt{cyclefactor} value of 2.0 typically fixed divergence issues, but in one extreme case (field \texttt{af}, 12m SPW 35) it was necessary to use a value of 5.0, which significantly increased the cleaning time.

The pipeline-delivered products for the 12m data were often impacted by various flavours of size mitigation, whereby the ALMA pipeline restricts the output product sizes by e.g. spectrally-averaging, decreasing the imaged area, increasing pixel sizes, or dropping full SPWs altogether. The most common forms of size mitigation affecting ACES were the spectral averaging, where it was common for 12m datasets to be binned by a factor of 2, and dropping SPWs, typically the 12m SPW 33. In such cases, it was necessary to manually update/specify the relevant cleaning parameters and rerun the imaging for the affected data. The full parameter list can be found using the \texttt{merge\_tclean\_commands.py} script in the ACES GitHub repository\footnote{\updates{The full ACES codebase used for processing the data can be found at }\url{https://github.com/ACES-CMZ/reduction_ACES}}.

\updates{We note that we did not apply any JvM residual scaling correction to the final images \citep{Jorsater1995}. This correction factor, which is calculated as the ratio of the volume of the clean beam to the dirty beam, is a rescaling of the residuals due to the differing units of the cleaned model (Jy / clean beam) and the residual (Jy / dirty beam). While this correction has been justified and implemented for some ALMA Large Programs \citep[e.g.,][]{Czekala2021, Cunningham2023}, it is not standard practice nor is it recommended by ALMA, and other Large Programs do not apply this correction \citep[e.g.,][]{Leroy2021, Sanchez-Monge2025}. \citet{Sanchez-Monge2025} discuss this in the context of the ALMAGAL Large Program and find that the correction factors can be very significant and sensitive to the nature of the observations. They also note that such large correction factors can artificially decrease the noise in the corrected images, to values well below the expected theoretical noise. Suggested workarounds for this latter point are to either avoid the scaling when measuring the noise, or to only apply the scaling factor to channels in the residual that still contain real signal \citep{Czekala2021, Cunningham2023, Sanchez-Monge2025}.}

\updates{Given the uncertainties surrounding the JvM correction and when/how best to apply it, we chose not to apply it to the ACES data for this initial release.}

\subsection{Parallelisation}
\label{sec:parallelisation}

When re-imaging the data, we initially attempted to use CASA's native parallel cleaning capabilities within the \texttt{tclean} task. While this generally performed well on individual computers for team members undertaking manual re-imaging, it proved problematic on the high-performance computing system used for the bulk of our data processing, where MPI CASA instances were frequently violating memory limits and consequently crashing.

We therefore implemented a different parallelisation approach, whereby each dataset was divided into smaller, contiguous channel chunks. Each chunk was then assigned to an independent CASA \texttt{tclean} job, submitted as part of a \texttt{SLURM} \citep{slurm} array job. This allowed multiple channel ranges to be imaged simultaneously. Once all individual channel chunks were processed, the resultant image cubes and associated products were merged to produce the final cubes.

\subsection{Array combination}
\label{sec:comb}

Once the QA workflow, calibration and re-imaging steps were completed, the individual products (measurement sets and cubes) for each SB for all three arrays were ready to be combined to obtain 12m7mTP cubes for all fields. There are various methods used to combine interferometric data with single dish data, and there is much debate about which methods are ‘best’ (see e.g. \citealt{Plunkett2023} for a detailed discussion). Given the size of the ACES data and the complexity of molecular line emission in the CMZ, we tested three different approaches for data combination to explore the trade-off between the resulting image fidelity and the computation time. These methods are discussed in the following subsections.

\subsubsection{Joint-deconvolution + feather}
\label{sec:joint_decon_feather}

We initially combined the data following the procedure that is typically adopted in the literature, i.e. joint-deconvolution of the 12m+7m visibility data via the \texttt{tclean} task, followed by combining the 12m+7m cube with the TP cube in \updates{the Fourier domain} using CASA task \texttt{feather}. 

The main steps for the joint imaging plus feathering for each field were:
\begin{enumerate}
    \item Identify the relevant 12m and 7m measurement sets (MSs) and TP image, then split out the science target and relevant science SPW for the 12m and 7m data via the \texttt{split} task.
    % \item Put the 12m and 7m MSs on a common velocity grid via the \texttt{mstransform} task
    % \item Concatenate the transformed 12m and 7m MSs via the \texttt{concat} task.
    \item Jointly image the 12m+7m data via the \texttt{tclean} task.
    \item Regrid the TP data using the 12m+7m cube as a template via the \texttt{imregrid} task\updates{, which is an algorithm that re-samples an image onto a new coordinate grid \citep{CASATeam2022}.}
    \item (If necessary) reorder the TP axes to match those of the 12m+7m via the \texttt{imtrans} task.
    \item Combine the 12m+7m cube and the TP cube via the \texttt{feather} task.
\end{enumerate}

The initial \texttt{tclean} parameters for the joint-deconvolution for each field were extracted from the pipeline 12m imaging of the same field. If cleaning parameters were manually adjusted for the 12m data (Section \ref{sec:imaging}), these updated parameters were also used in the joint deconvolution. However, this approach resulted in several challenges. The inclusion of the 7m greatly exacerbated the negative bowls surrounding regions of bright emission. This problem was somewhat mitigated by greatly relaxing the automasking parameters associated with the \texttt{auto-multithresh} masking technique in \texttt{tclean} \citep{Automultithresh}. 

Adding the 7m data also frequently led to the cleaning diverging at the field edges, likely due to the widespread, bright molecular line emission that extends well beyond the field edges that is well recovered by the 7m data. The solution to mitigating this divergence was the same as described in Section \ref{sec:imaging} -- to increase the \texttt{cyclefactor} parameter in \texttt{tclean}. In extreme test cases, the \texttt{cyclefactor} value had to be increased up to 7.0.

The combination of relaxed automasking and large cyclefactor values often resolved the issues with the joint 12m+7m imaging. However, these two changes \textit{significantly} increased the imaging time.

\subsubsection{‘Model-Assisted CLEAN plus Feather’ (MACF)}
\label{sec:joint_decon_startmodel}

In addition to the typical joint imaging + feather technique, we also explored the utility of using the TP data as a startmodel for the joint imaging of the 12m+7m data. This technique has been presented in various forms in the literature, and is presented in detail (along with other data combination methods) in \citet{Plunkett2023}, where it is called the ‘Model-Assisted CLEAN plus Feather’ (MACF) method, which uses the single dish image as a startmodel and then also combines the single dish data post-cleaning via feathering.

This method is broadly similar to the standard joint imaging and feather approach, except now the large scale emission is initialised in the clean model by passing the TP data to \texttt{tclean} via the \texttt{startmodel} parameter. In order to do this, a few more steps are required beyond those listed in Section \ref{sec:joint_decon_feather}:

\begin{enumerate}
    \item A dirty image of the 12m+7m data must be created to provide a template and a primary-beam file for the subsequent steps.
    \item The re-gridded and re-ordered TP cube must be converted to units of Jy/pixel.
    \item The converted TP cube must be multiplied by the 12m+7m primary beam in order to apply the primary beam attenuation, such that the fluxes are correct after the primary beam correction post-cleaning.
\end{enumerate}

Although the TP data are used to initialise the clean model in the MACF method, the TP data are still feathered with the jointly imaged cube to recover the zero spacing information and therefore the total flux.

As discussed in the previous subsection, we had to relax the automasking and increase the \texttt{cyclefactor} when imaging the 12m+7m data together, and the same is true with this TP startmodel approach. In fact, we found that the imaging was even more prone to divergence along the field edges when initialising with the large scale information, therefore requiring even larger \texttt{cyclefactor} values. 

We also found that this method seemed to add too much weight to the TP data, resulting in greater total flux than expected. This is potentially due to the fact that the TP data are being used twice in the procedure, first to initialise the cleaning model, and then again in the final feathering step. 

Additionally, we found that \texttt{tclean} breaks when using a startmodel with the parameter \texttt{parallel} set to \texttt{True}. A potential workaround for this would be to follow the parallelisation approach discussed in Section \ref{sec:parallelisation}, though this would require additional steps for chunking up the data across the three different arrays.

\updates{Another potential issue with this method is that by providing the TP image as a startmodel, we are not providing a model of the true sky brightness as it is convolved with the single dish beam. As noted in Appendix A.2 of \citet{Kauffmann17b}, deconvolving the single dish map is recommended to avoid signal from compact structures entering the reduction twice (once correctly via the interferometer data and once blurred via the single dish map); supplying a convolved TP startmodel risks precisely this. This convolved initial model could also potentially bias the deconvolution process towards smoother structure, and potentially degrade the effective resolution. These issues, combined with the fact that the TP data is explicitly used twice (once as a startmodel and again in the final feather step), likely leads to the overestimation of flux that we observed in our tests. Given these fundamental issues and the practical challenges of long computation times, we opted not to explore this method further.}

\subsubsection{Feather-only}
\label{sec:feather_only}

As mentioned earlier, it is fairly standard practice to combine interferometric data in visibility space via joint imaging, followed by combination with single dish data in \updates{the Fourier domain}. However, it is also possible to combine images from multiple interferometric datasets in \updates{Fourier} space using e.g. the \texttt{feather} task in CASA. Given that we already re-imaged all 12m and 7m data and therefore had good quality image products from all 3 arrays, we decided to explore combining the data using only feathering. Though this is not necessarily standard procedure, this approach has been used in the literature \citep[e.g.][]{Barnes2021}. \updates{The order of combination for the feather-only approach was chosen to follow a hierarchical process, combining datasets that cover adjacent spatial scales. The 7m and TP data were combined first to produce a complete map of the large- and intermediate-scale emission. This combined, lower-resolution map was then feathered with the 12m data to incorporate the smaller scale information.}

The feather-only approach is relatively straightforward, consisting of the following steps for a given field:

\begin{enumerate}
    \item Locate the 12m, 7m, and TP image cubes.
    \item (If necessary) reframe the TP cube to match the rest frequency of the 7m data via the \texttt{imreframe} task.
    \item (If necessary) reorder the TP axes to match the 7m data via the \texttt{imtrans} task.
    \item Combine the 7m and TP cubes via the \texttt{feather} task.
    \item Combine the 7m+TP cube with the 12m cube via the \texttt{feather} task.
\end{enumerate}

Given the relative simplicity of the feathering procedure (Fourier transform, combine, transform back to image space), this combination approach is \textit{substantially} faster than joint deconvolution, especially in our case where the necessary cleaning parameters for joint imaging results in much longer imaging timescales. Note that we did not change any of the weighting parameters, always using the default \texttt{feather} parameters.

\subsubsection{Comparison between data combination methods}
\label{sec:comb_compare}

Having discounted the MACF method due to overly long processing times and the potential to overestimate the total flux, we compared the results of the joint clean and feather approach (Sec \ref{sec:joint_decon_feather}) with the feather-only approach (Sec \ref{sec:feather_only}) to decide how to proceed for the final data combination.

\begin{figure*}
    \centering
    \includegraphics[width=\textwidth]{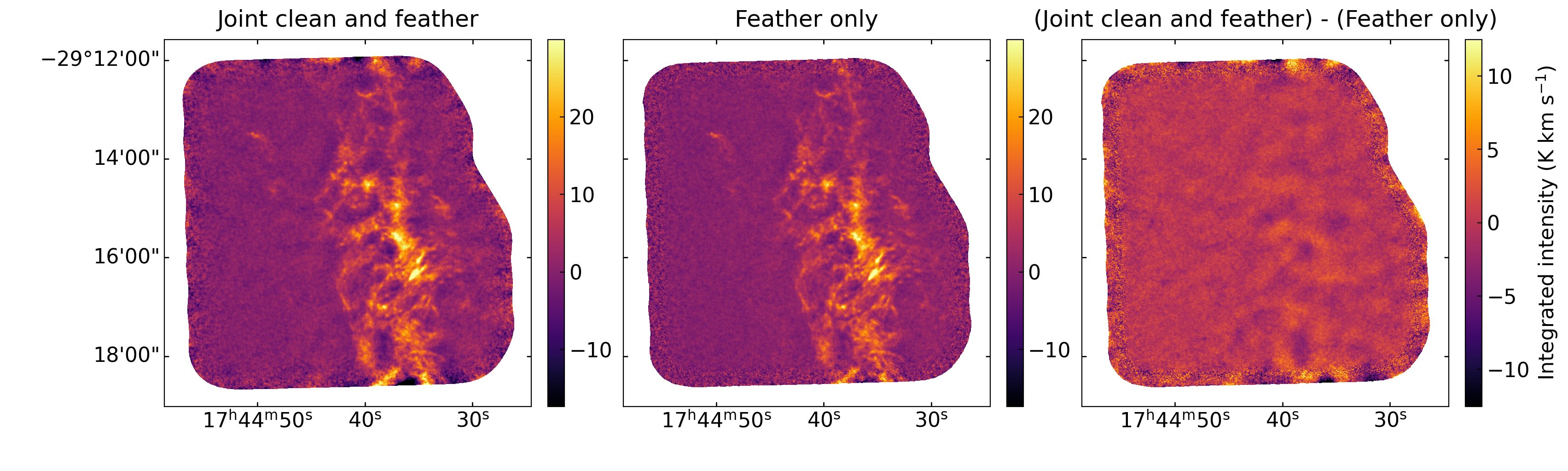}
    \includegraphics[width=\textwidth]{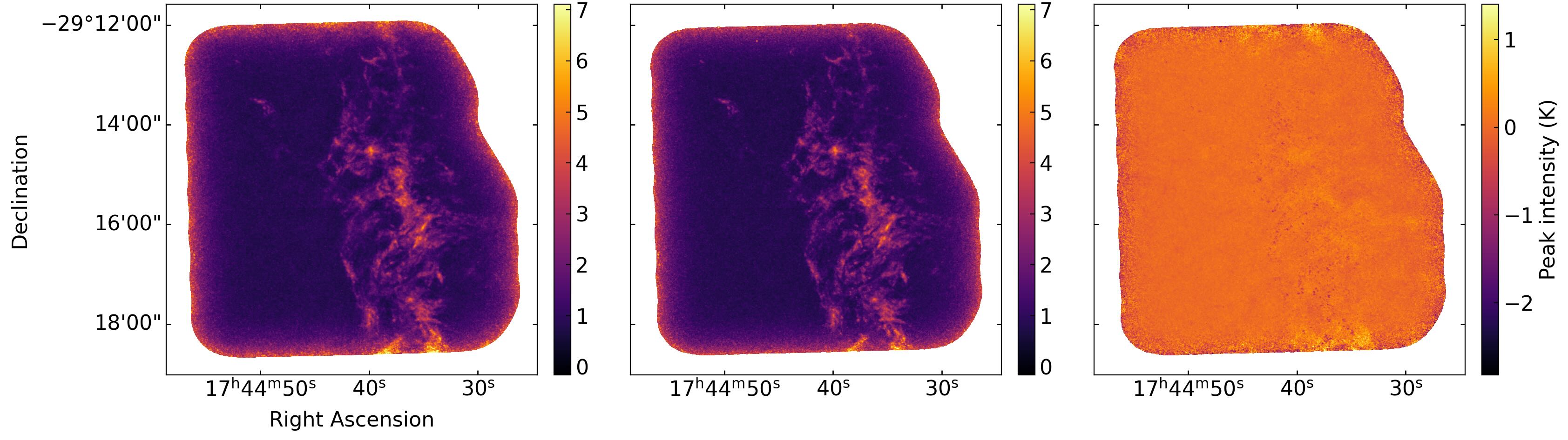}
    \caption{Comparison for field \texttt{t} between the two different array combination methods presented in Section \ref{sec:comb}. The top row shows integrated intensities over the subset of channels used for testing (see Figure \ref{fig:HNCO_imaging_comparison+meanspec}), while the bottom row shows peak intensities. The left column shows the results from the joint imaging of the 12m and 7m data, feathered with the TP data (Sec. \ref{sec:joint_decon_feather}). The central column shows the result from the feather only approach (Sec. \ref{sec:feather_only}). The right column shows the difference (joint clean and feather - feather only). Note that unlike the majority of figures in this paper, these are shown with Right Ascension and Declination to minimise unnecessary white-space.}
    \label{fig:HNCO_imaging_comparison}
\end{figure*}

\begin{figure}
    \centering
    \includegraphics[width=\columnwidth]{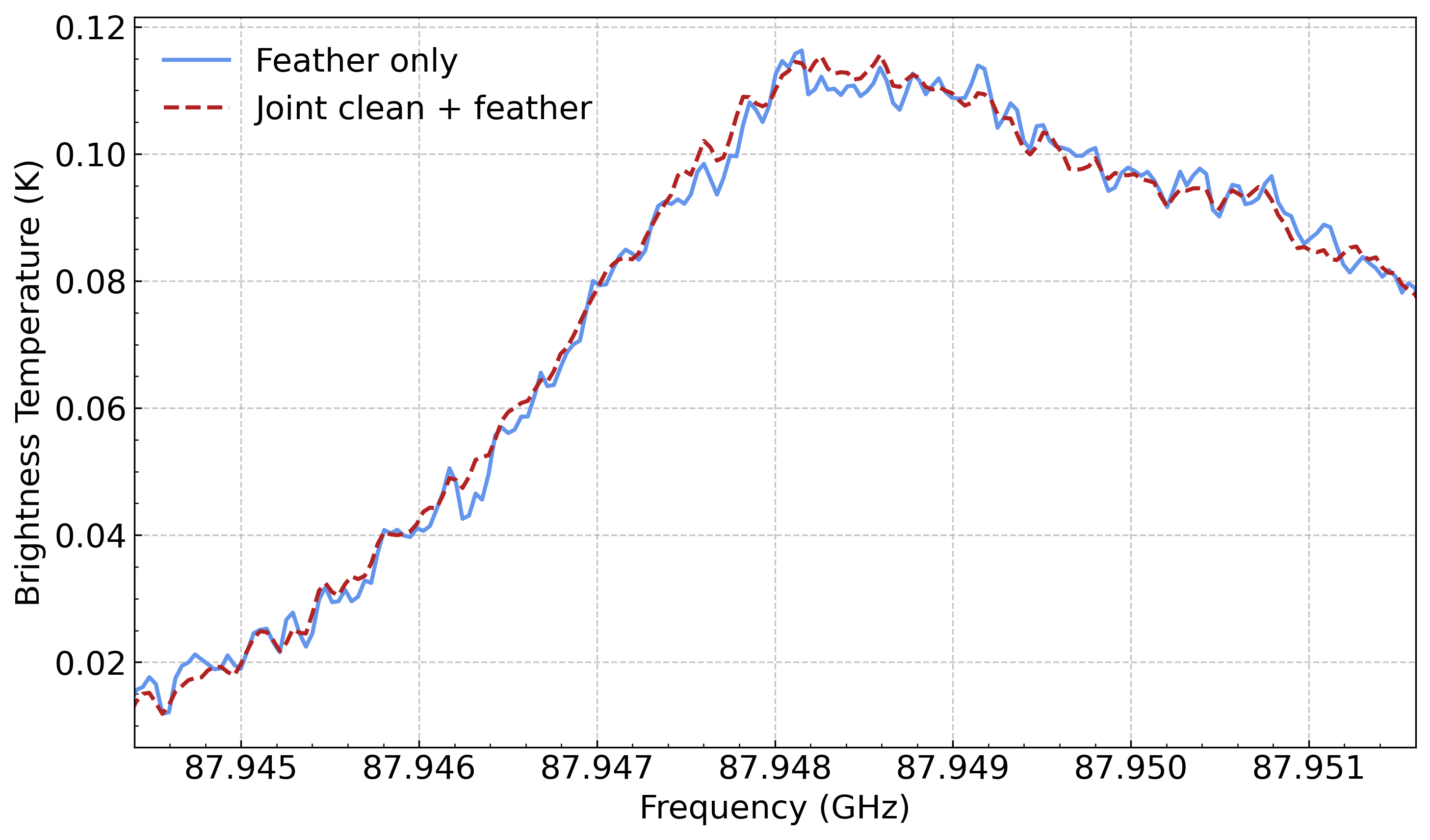}
    \caption{Comparison for field \texttt{t} between the two different array combination methods presented in Section \ref{sec:comb}. Shown are the mean spectra for two HNCO cubes, resulting from the different methods. The dashed red line shows the mean spectrum of the jointly imaging and feathered cube (Sec. \ref{sec:joint_decon_feather}). The solid blue line shows the mean spectrum of the feather-only cube (Sec. \ref{sec:feather_only}). Note that these tests were performed on a small subset of the channels, hence the restricted frequency coverage in the figure.}
    \label{fig:HNCO_imaging_comparison+meanspec}
\end{figure}

Figure \ref{fig:HNCO_imaging_comparison} shows a comparison of the integrated and peak HNCO intensities of these two approaches for region \texttt{t}, while Figure \ref{fig:HNCO_imaging_comparison+meanspec} shows a comparison of the spatially averaged HNCO spectra for the same region. Overall we find that the two approaches yield comparable results, with only minor differences. The joint clean and feather method does do a better job of recovering some large scale emission, though this also enhances the negative bowls surrounding the regions of bright emission. As can be seen in Figure \ref{fig:HNCO_imaging_comparison+meanspec}, we also find that both the intensity and the shape of the line profile are \updates{very similar} between the two approaches. \updates{We confirm that both methods effectively recover the Total Power integrated flux (see Appendix \ref{subsec:flux_recovery}).}

Given that the joint imaging method takes significantly longer than the feather-only approach, and that the differences in the resulting images are relatively minor, we opted to proceed with feather-only so as to not delay the data release. We may explore the potential for improving the efficiency of the joint deconvolution approach, in which case we would deliver the products as a future data release.

\subsection{Mosaicking}
\label{sec:mosaics}

Having created the combined array cubes for each field individually, we then combine all fields for each spectral window to produce a contiguous cube mosaic spanning the full ACES survey area, effectively creating a giant mosaic of sub-mosaics.

This mosaicking approach involves first convolving the cubes to a uniform beam size, which is determined by finding a common beam across all beams for the 45 separate sub-mosaics. Each cube has only a single beam common to all spectral channels, as we set the parameter \texttt{commonbeam = True} at the imaging stage. The resulting beam sizes for the HNCO and \hcop mosaics are (3.20\arcsec $\times$ 2.15\arcsec) and (3.23\arcsec $\times$ 2.77\arcsec), respectively. 

Each channel is then processed individually by reprojecting the data onto a common coordinate grid\footnote{This is done by explicitly defining a template header with a fixed spatial footprint, which is then updated to have the correct number of channels and spectral information for a given molecular line.}. The pixels are also re-binned such that the final mosaics for both HNCO and \hcop have 0.5\arcsec\ pixels, such that we have \app 5-6 pixels per beam. 

During the combination, data from each field are weighted according to the corresponding 12m weight image files that are produced by \texttt{tclean}, where each weight image is the Fourier transform of the gridded visibility weights, and is related to the primary beam response as Primary Beam = $\sqrt{weights}$ (for all pointings, normalised to 1.0). This weighting approach suppresses the noisier edges of individual fields, resulting in a contiguous mosaic with improved uniformity.

We note that for the mosaicking of the HNCO and \hcop cubes, we smooth the data by a factor of 2 along the spectral axis. This is done purely to reduce the file sizes, as the full mosaic cubes for these lines at native channel spacing are \app 1~TB each. By smoothing, we reduce the mosaics to a somewhat more manageable \app 500~GB. Other line mosaics and individual sub-mosaic cubes for all lines (including HNCO and \hcop) use their native channel widths (see Table \ref{tab:spectral_setup}).

\subsection{Total Power baseline issues}
\label{sec:tp_baselines}

Both sensitivity to large angular scales and recovery of total flux on all scales are crucial to the overall goals of ACES. As such, we observed all fields with the Total Power antennas to obtain the zero spacing data. The TP data were processed by the ALMA single dish pipeline during QA2, and we did not do any re-processing of the data once delivered. We did however perform internal QA following the procedure outlined in Section \ref{sec:cal} to verify the quality of the data.

During this internal QA, we discovered multiple issues affecting the baselines in the TP spectra. The problems broadly fall into two categories, which are discussed in the following subsections.

\subsubsection{Poor line identification and baseline subtraction}
\label{sec:tp_line_id}

We found that the ALMA single-dish (SD) pipeline often did not do a satisfactory job of correctly identifying channels that contained line emission, which is crucial for the subsequent baseline subtraction. This was particularly problematic for the narrow spectral windows (TP SPWs 21 and 23), where the broad line-widths of the large scale HNCO and \hcop emission often fill a significant fraction of the bandwidth. \updates{As shown in Table \ref{tab:spectral_setup}, the bandwidth of these SPWs is \app 200~\kms, which is comparable to the range of velocities observed in the CMZ (see e.g., \citealt{Henshaw2023} and references therein).}

Due to the broad, bandwidth filling nature of the spectral line emission in the TP SPWs 21 and 23, the line profile sometimes continues beyond the edge of the spectral window. Although we minimised the chance of this happening by implementing field specific velocity shifts (see Section \ref{sec:obs}), there were still regions in which this occurred.

In cases where the line emission spilled over one of the SPW edges (i.e. the other edge was line-free), this resulted in a subtraction of line emission and sometimes the introduction of an artificial slope in the spectrum, due to the fact that the SD pipeline uses a default polynomial fit order of 1 (this was also a problem with the interferometric pipeline, see Section \ref{sec:contsub}).

\subsubsection{Baseline ripples}
\label{sec:tp_ripples}

We also found that the TP data \updates{for all SPWs} are affected by significant, periodic ripples in the baseline of the spectra. Such ripples are known to occur in single dish telescopes, and in the case of the TP antennas is potentially due to standing wave formation during signal transmission\footnote{See \url{https://help.almascience.org/kb/articles/why-is-there-a-spectral-baseline-ripple-in-total-power-observations-of-bright-targets} (last accessed on 2025-11-27)}. 

Baseline ripples are commonly observed in TP data when the ON and OFF positions are at significantly different elevations. It is therefore ideal to select OFF positions that are clean from line contamination and as close to the target as possible. However, this is particularly challenging in the Galactic plane and more so towards the Galactic centre.

The earliest TP observations for ACES used an OFF position that was \app 5 degrees away, which is the observatory limit in Band 3. Upon inspecting these early TP data, we noticed the ripples and initiated QA3 with ALMA to try to resolve the issue. Due to the difficulty of finding a clean OFF position near the Galactic centre, the best alternative position that the observatory could find was \app 3.9 degrees away. While this new position was found to suppress the amplitude of the ripples by a factor of \app 1.5, it did not remove them entirely.

Analysis of the ripples for a given region showed that they are consistent regardless of position in the image, and that they are typically periodically spaced. Power spectra measured in line-free channels \updates{ of the spatially-averaged spectra} revealed dominant ripple frequencies at \app 33~MHz, with a secondary peak at \app 24~MHz. These characteristic frequencies persisted \updates{to within 10 -- 15\%} between fields, for data of the same field with the old and updated OFF positions, and even in entirely unrelated data from different ALMA projects at different frequencies. These results suggest that the frequency of the ripples is largely independent of observing frequency, and again points to standing waves originating in the antennas. \updates{While the frequency/period of the dominant ripple is consistent between fields, the amplitude is more variable, ranging from \app 0.1 - 5 mK in the subset of cubes that were used for testing.}

Though the baseline ripples affect all the spectral windows in our TP data, they are more prominently visible in the broader spectral windows due to the larger frequency coverage. The typical width of the dominant ripple is equivalent to \app 90~\kms. 

We explored the possibility of subtracting the ripples from the spectrum by iteratively fitting and subtracting sinusoidal components. \updates{For each region and SPW, we first generated a high signal-to-noise spectrum by averaging the cube over its full spatial extent. To avoid contamination from spectral lines, we then automatically identified line-free channels using an iterative sigma-clipping approach, which masks channels lying more than 3~$\sigma$ above the median of the spectrum. The ripple was then modelled by fitting a series of sinusoidal functions to these line-free channels using a non-linear least-squares fitting approach (\texttt{scipy.optimize.curve\_fit}).} The resulting baseline model was then subtracted from the full spectrum to yield the baseline corrected data. 

\begin{figure*}
    \centering
    \includegraphics[width=\textwidth]{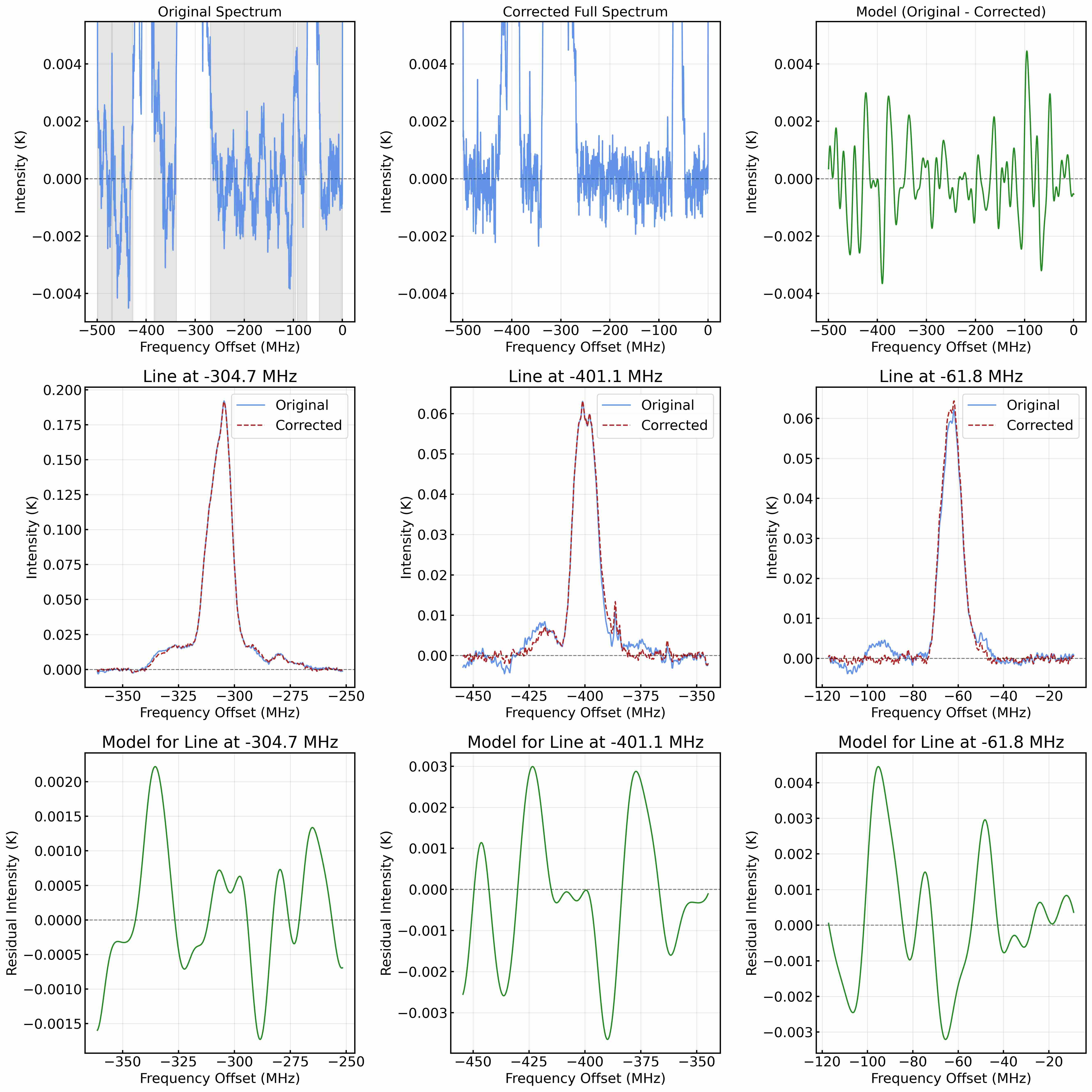}
    \caption{Example showing the Total Power baseline ripple procedure (Section \ref{sec:tp_baselines}) for field \texttt{ak} (TP SPW 19). The spectra are taken from the ALMA single-dish pipeline-produced cube, and are spatially-averaged across the whole field. \textit{Top left}: Mean spectrum of the original cube. Grey shaded regions show the identified line-free channels. \textit{Top centre}: Corrected mean spectrum after subtracting the model of the baseline ripples. \textit{Top right}: Model (original - corrected). Note that the y-axes of the panels in the top row have been cropped to highlight the baseline structure. \textit{Middle row:} the three central panels show zoom-ins of the before (blue solid line) and after (red dashed line) baseline correction spectra for the three most prominent lines. \textit{Bottom row:} the three bottom panels show the same frequency (x-axis) zoom-ins as the middle row, this time showing the model (original - corrected). In all panels, the black dashed horizontal line denotes the zero level for reference.}
    \label{fig:tp_ripples}
\end{figure*}

An example of this analysis for field \texttt{ak}, TP SPW 19, is shown in Figure \ref{fig:tp_ripples}. This example shows that this approach is promising -- the baseline in the corrected spectrum is significantly improved. However, \updates{we note that we only performed this test on a subset of the fields, and we found that it performed poorly for the HNCO and \hcop SPWs due to their narrow bandwidths and broad line emission.} More detailed testing is \updates{therefore} required on all fields and spectral windows to ensure that \updates{any final} procedure is robust \updates{across the full dataset} and does not incorrectly subtract or change the profile of real emission (see the middle row of Figure \ref{fig:tp_ripples}). We therefore do not apply this technique to the initial ACES data release.

\subsubsection{Caveats and the problem with HCO+}
\label{sec:tp_hcop}

\begin{figure}
    \centering
    \includegraphics[width=\columnwidth]{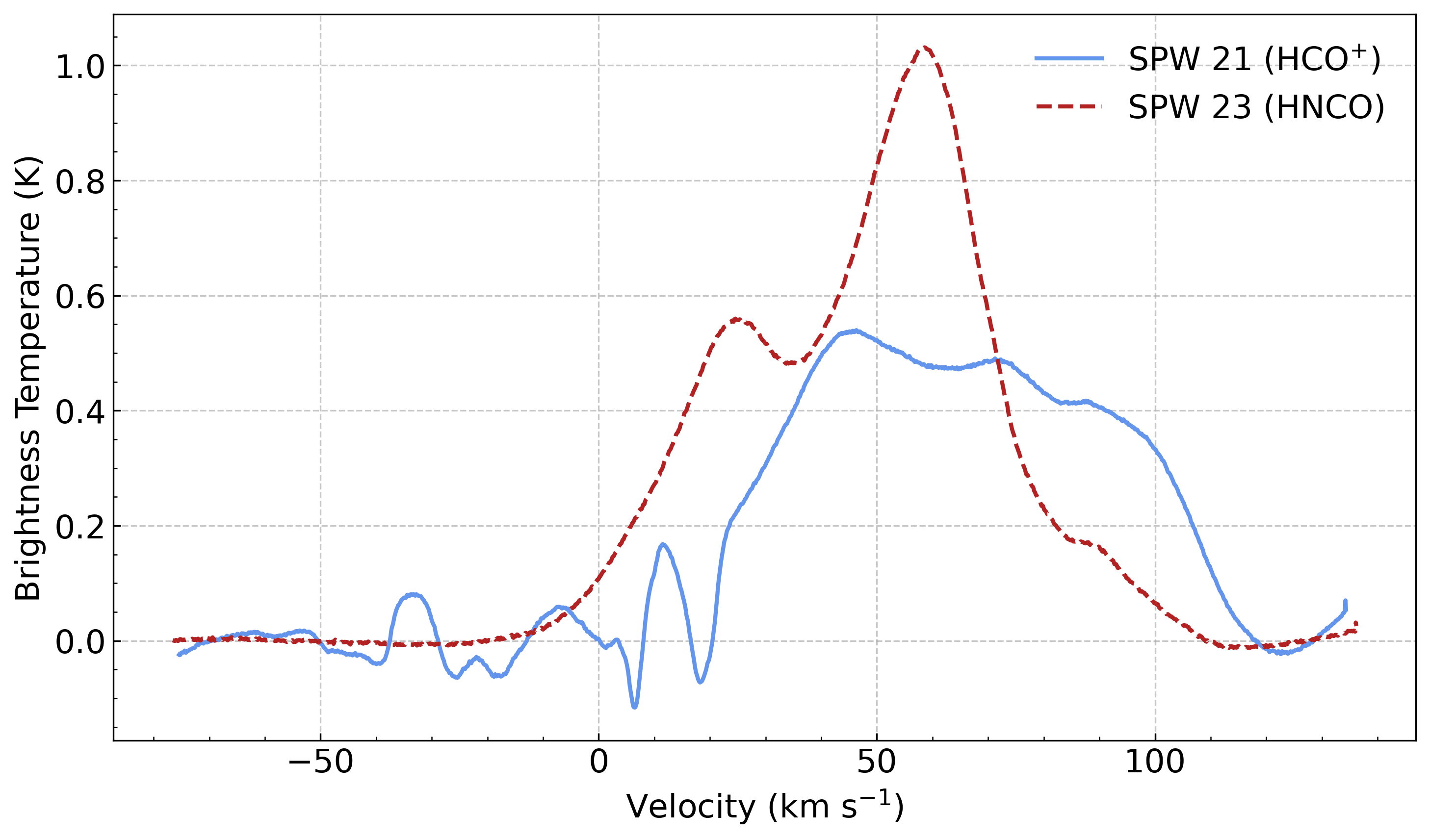}
    \caption{Comparison for field \texttt{j} between the HNCO and \hcop Total Power data. Shown are the mean spectra for each, where the spectra have been spatially averaged across the full field.}
    \label{fig:tp_meanspec_bls}
\end{figure}

The problems relating to the TP baseline subtraction (Section \ref{sec:tp_line_id}) predominantly affect the narrow SPWs containing the HNCO and \hcop emission presented in this paper. The \hcop window (TP SPW 21) is particularly problematic because the line profiles are typically broader than those of HNCO, and so the channel identification and baseline subtraction are more challenging. Figure \ref{fig:tp_meanspec_bls} shows the mean spectra for the HNCO and \hcop TP data for field \texttt{j}, which is an especially challenging case.  A potential approach to fix/minimise these issues is to manually perform the line channel identification, and to re-run the pipeline with a fit order of 0, similar to the approach taken for the 12m and 7m data (Section \ref{sec:contsub}). 

As the baseline subtraction issues are fairly widespread for the \hcop data, and because it is not trivial to remove the ripples for all SPWs, we did not attempt to resolve these issues for the first data release. We therefore encourage those using ACES products containing TP data to keep these caveats in mind.

However, given the severity of the baseline subtraction issues for the \hcop data, we have produced the following different versions of the \hcop cubes:

\begin{itemize}
    \item 12m7mTP, with the caveat that the data may be affected by poor baseline corrections
    \item Interferometric-only (12m+7m), with the caveat that the zero spacing data are missing, and thus total flux is not recovered
    \item 12m7mMopra, where the Mopra single dish data from the Mopra CMZ survey \citep{Jones2012} are used in place of the TP data, with the caveat that the spectral grid is degraded to \app 2~\kms channel spacing (vs. the \app 0.1~\kms channels for ACES)
\end{itemize}

To create the 12m7mMopra cubes, the ALMA 12m and 7m data are first smoothed to the spectral resolution of the Mopra data and then regridded on to the same spectral grid as the Mopra data, using \texttt{spectral-cube}. We also scale the Mopra data by a main beam efficiency factor of 0.49 to convert from antenna to main beam temperature \citep[e.g.][]{Jones2012, Rathborne2015}, and convert from units of K to Jy/beam prior to combination. \updates{The Mopra data are also affected by baseline ripples \citep{Jones2012}; however, for the \hcop data, the severe baseline subtraction issues were the primary motivating factor for using the Mopra data as a replacement, not the ripples.}

We release these different \hcop cubes alongside the general ACES data release as an acknowledgement of the problems with the TP data, and to ensure that the community has access to the products without these issues. We are exploring potential methods to minimise these problems, which may be reflected in a future data release if successful.

Due to the significant issues affecting the TP \hcop data, all subsequent figures, plots, and statistics in this paper use the 12m7mMopra \hcop data, unless otherwise specified.

\updates{While the bandwidth of the HNCO SPW is also narrow, and hence potentially susceptible to issues with baseline identification, we found that the baseline subtraction performed by the ALMA single dish pipeline was acceptable, whereas the \hcop baselines often rendered the data unusable. This is likely due to the fact that the \hcop line profiles are typically broader and more complex (e.g., Figure \ref{fig:tp_meanspec_bls}).}

\begin{figure*}
    \centering
    \includegraphics[width=\columnwidth]{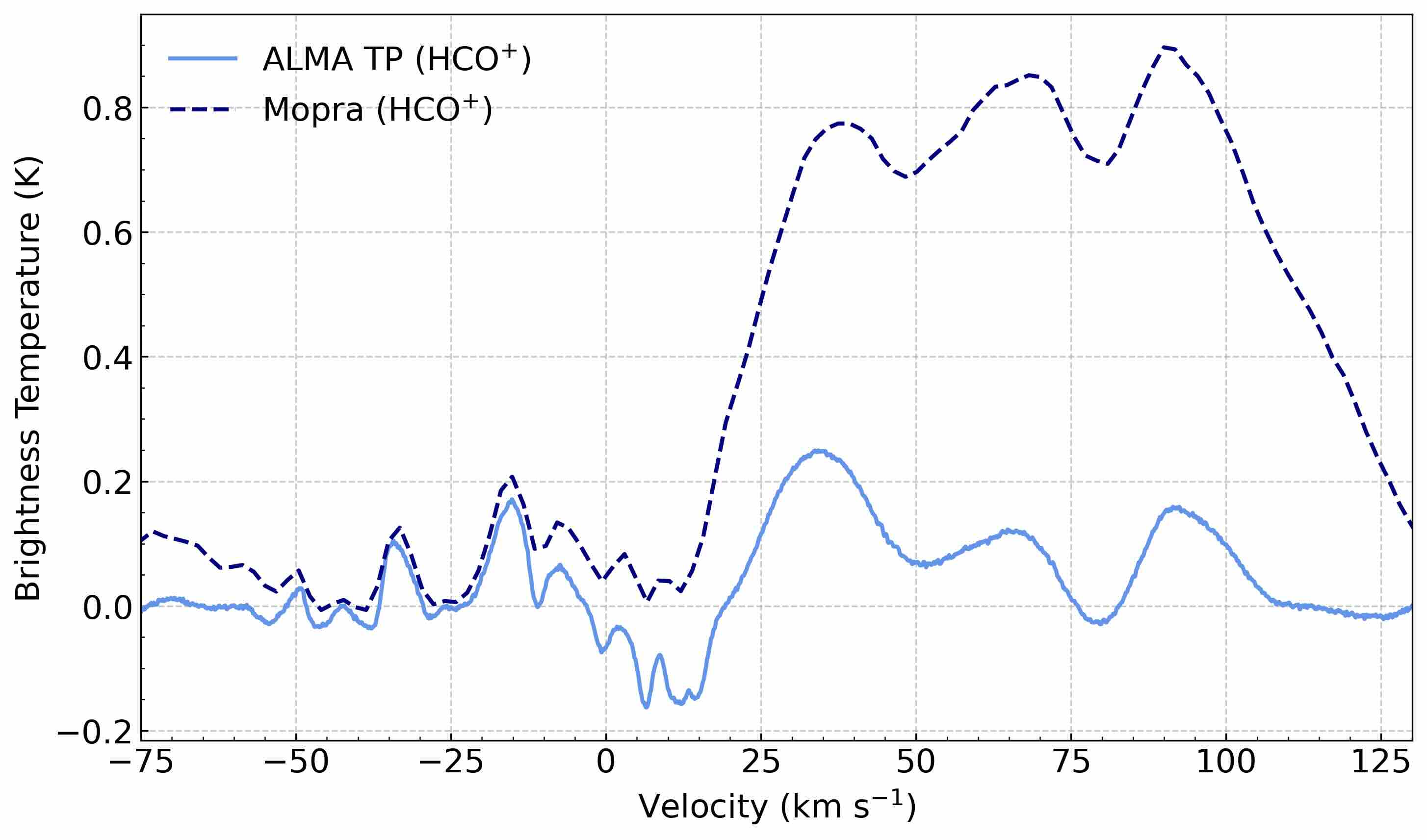}
    \includegraphics[width=\columnwidth]{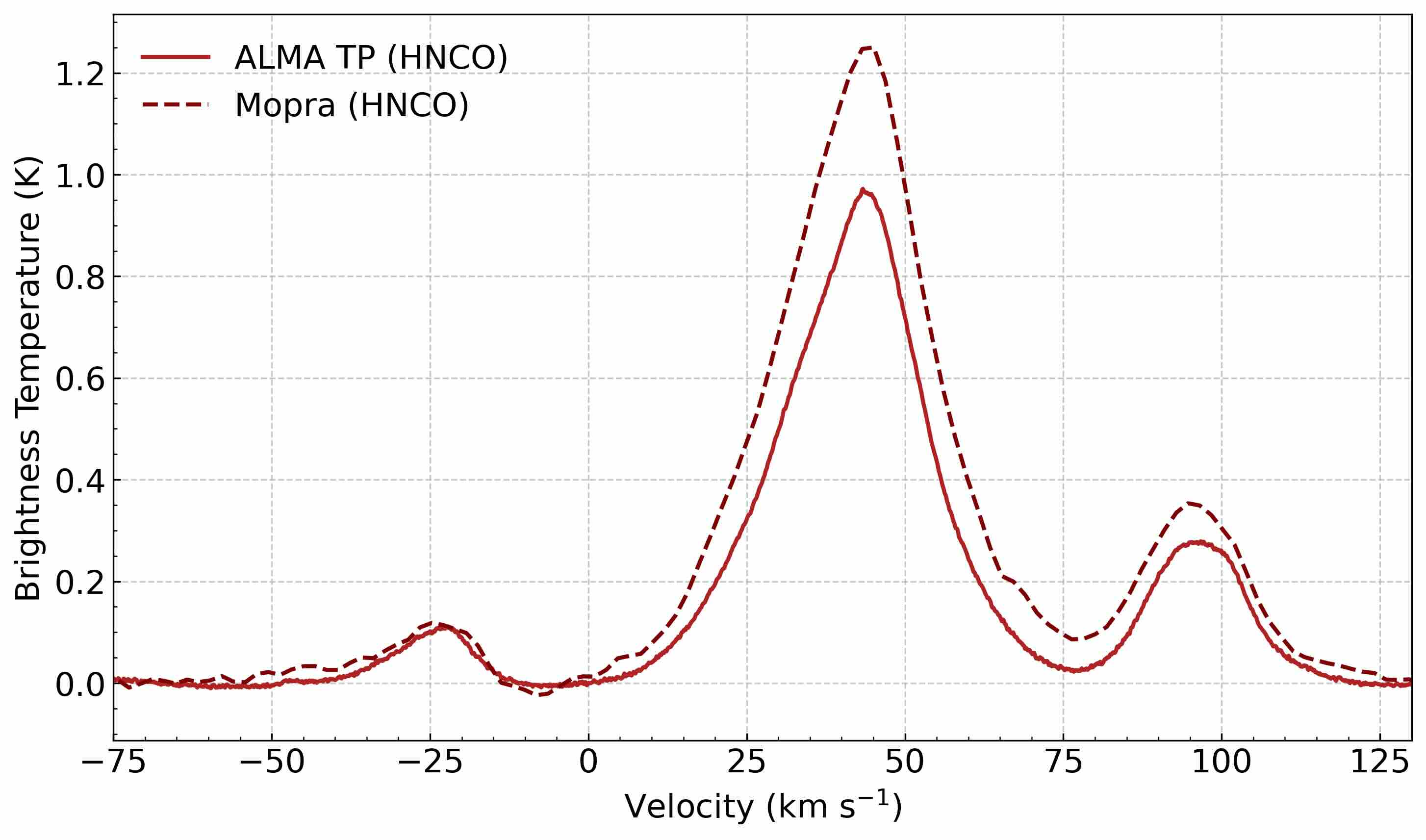}
    \caption{\updates{Comparison for field \texttt{ah} between ALMA Total Power and Mopra spectra. The left panel shows the \hcop spectra, while the right panel shows the HNCO. In all cases, the spectra shown are spatially averaged over the spatial extent of the ALMA TP data. The TP data are converted to units of K and the Mopra data are converted from antenna temperature to main beam temperature using a conversion factor of 0.49 (see Section \ref{sec:tp_hcop}).}}
    \label{fig:tp_mopra_bl_comparison}
\end{figure*}

\updates{Figure \ref{fig:tp_mopra_bl_comparison} shows a comparison between the spatially-averaged ALMA Total Power data and the Mopra data for ACES field \texttt{ah}, for both HNCO and \hcop. This comparison shows that the HNCO spectra are well matched between the two datasets, whereas the \hcop spectra look significantly different -- the line emission has clearly been mistakenly subtracted by the pipeline, which has altered the line shape and amplitude.}

\updates{As HNCO is the primary kinematic tracer for the dense molecular gas, it is crucial that we maintain high spectral resolution to facilitate the science goals, which require the ability to trace the kinematics down to the sub-sonic regime. Given that the TP baseline subtraction is acceptable, we therefore do not use the Mopra data in place of the TP data for the combined HNCO cubes, as this would degrade the spectral resolution by a factor of \app 20, which is too coarse for the requirements of ACES.}

\section{Analysis}
\label{sec:analysis}

\subsection{Cube statistics}
\label{sec:cubestats}

\begin{figure*}
    \centering
    \includegraphics[width=\columnwidth]{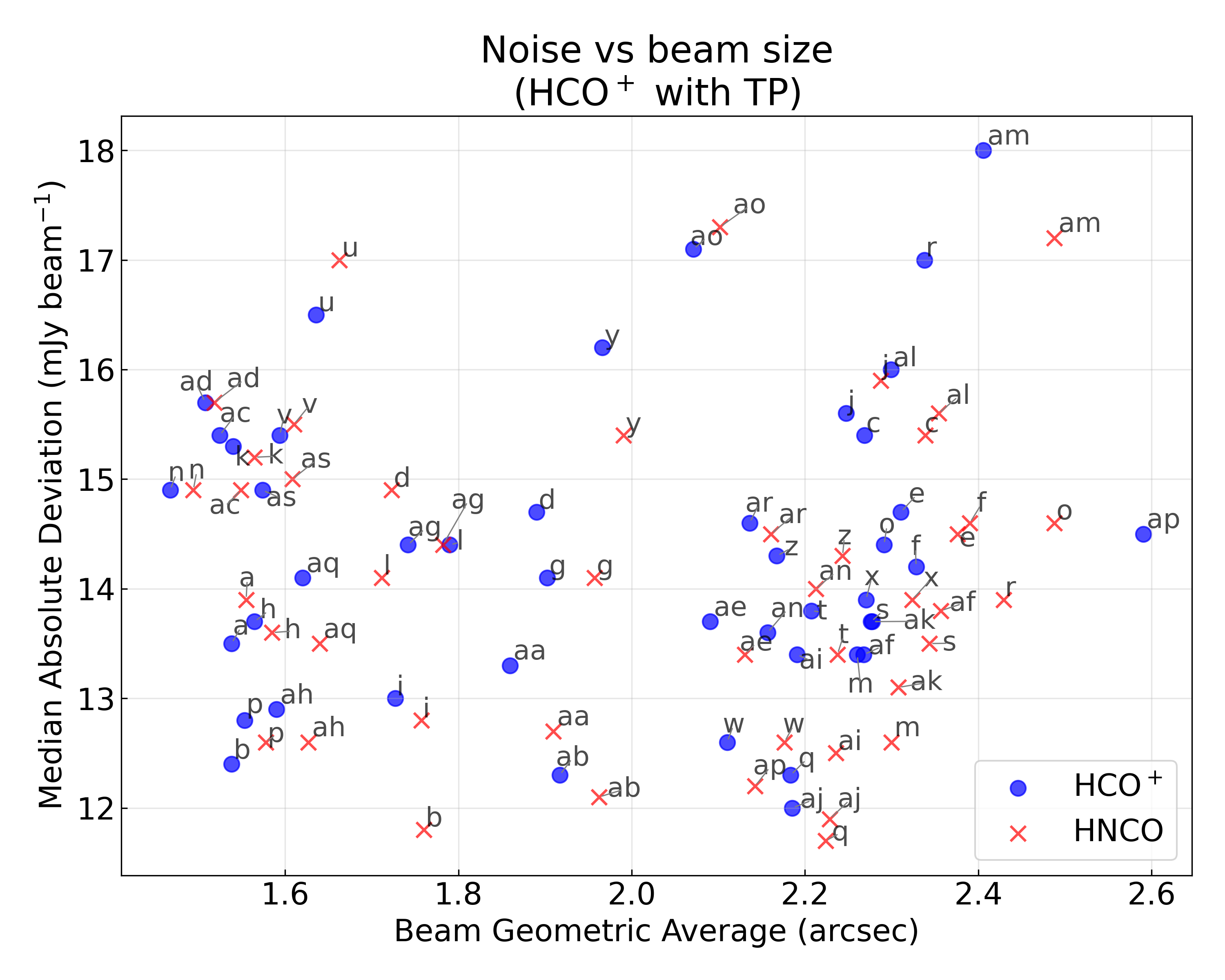}
    \includegraphics[width=\columnwidth]{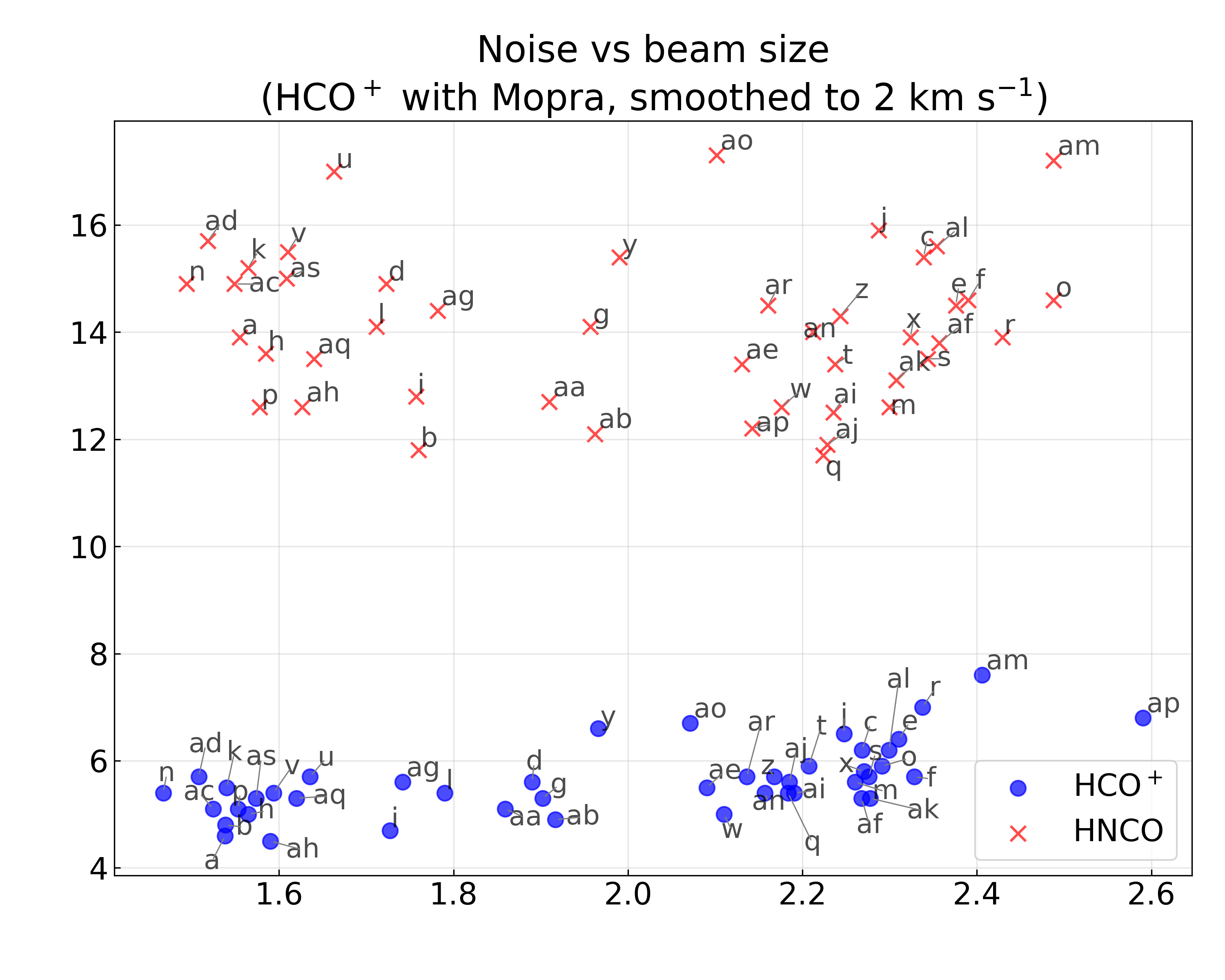}
    \caption{Noise vs. beam size for all 45 ACES fields, for both \hcop and HNCO. Blue circle markers correspond to \hcop and red cross markers to HNCO. The corresponding region name (\texttt{a}--\texttt{as}) is given for each marker, with grey lines connecting labels and markers in crowded locations. The beam size is the geometric average of the major and minor axes. The noise is estimated as the median absolute deviation of the full spectrum. The left figure shows the statistics when measured using the 12m7mTP cubes for both HNCO and \hcop. The right figure shows the same, but instead using the 12m7mMopra cubes for the \hcop data (see Section \ref{sec:tp_hcop}). The noise is significantly lower in this case as the \hcop data are spectrally regridded and smoothed to match the Mopra data, resulting in a channel spacing of \app 2~\kms, vs. 0.1~\kms for the ALMA data. Note that the left and right plots have different y-axis ranges, for improved clarity.}
    \label{fig:cube_stats}
\end{figure*}

We compute global statistics for the HNCO and \hcop cubes for all 45 regions. The statistics are given \updates{for \hcop in Table \ref{tab:hcopcubestats} and for HNCO in \ref{tab:hncocubestats}}. For the \hcop data, separate statistics are shown for the 12m7mTP and 12m7mMopra cubes, where applicable. Figure \ref{fig:cube_stats} shows the noise level and average beam size per field per SPW, where the beam is the geometric average calculated as $\sqrt{\textrm{B$_{\textrm{maj}}$B$_{\textrm{min}}$}}$, and the noise is estimated across the full cube via the median absolute deviation.

Overall there is a fairly large spread in beam size and noise, with median values of 2.1\arcsec and 14.1~mJy~beam$^{-1}$, respectively\footnote{Note that the quoted median noise value is measured across the HNCO and \hcop cubes using the ALMA 12m7mTP data. The noise in the \hcop 12m7mMopra cubes is significantly lower, with a median value of 5.5~mJy~beam$^{-1}$ (see the right panel of Figure \ref{fig:cube_stats}). This is due to the spectral smoothing required to match the ALMA and Mopra data along the frequency axis (see Section \ref{sec:tp_hcop}).}. The requested angular resolution for ACES was 1.5\arcsec, with an upper tolerance of 1.8$\arcsec$ for QA2. Note however that these values correspond to the specified representative frequency, which for ACES was in the upper sideband continuum spectral window at 100.5~GHz. Scaling these values to the frequency of the HNCO window, the upper limit for the requested resolution is 2.06\arcsec. As can be seen in Figure \ref{fig:cube_stats}, this was not achieved for many fields -- only \app 47\% of fields fall below this limit. Though we note that the upper limit resolution of 1.8$\arcsec$ was achieved for all fields when considering the aggregate continuum, which is dominated by the two upper sideband spectral windows (\citetalias{Ginsburg2025}).

\subsection{Moment and P-V maps}
\label{sec:moments}

\begin{figure*}
    \centering
    \includegraphics[width=\textwidth]{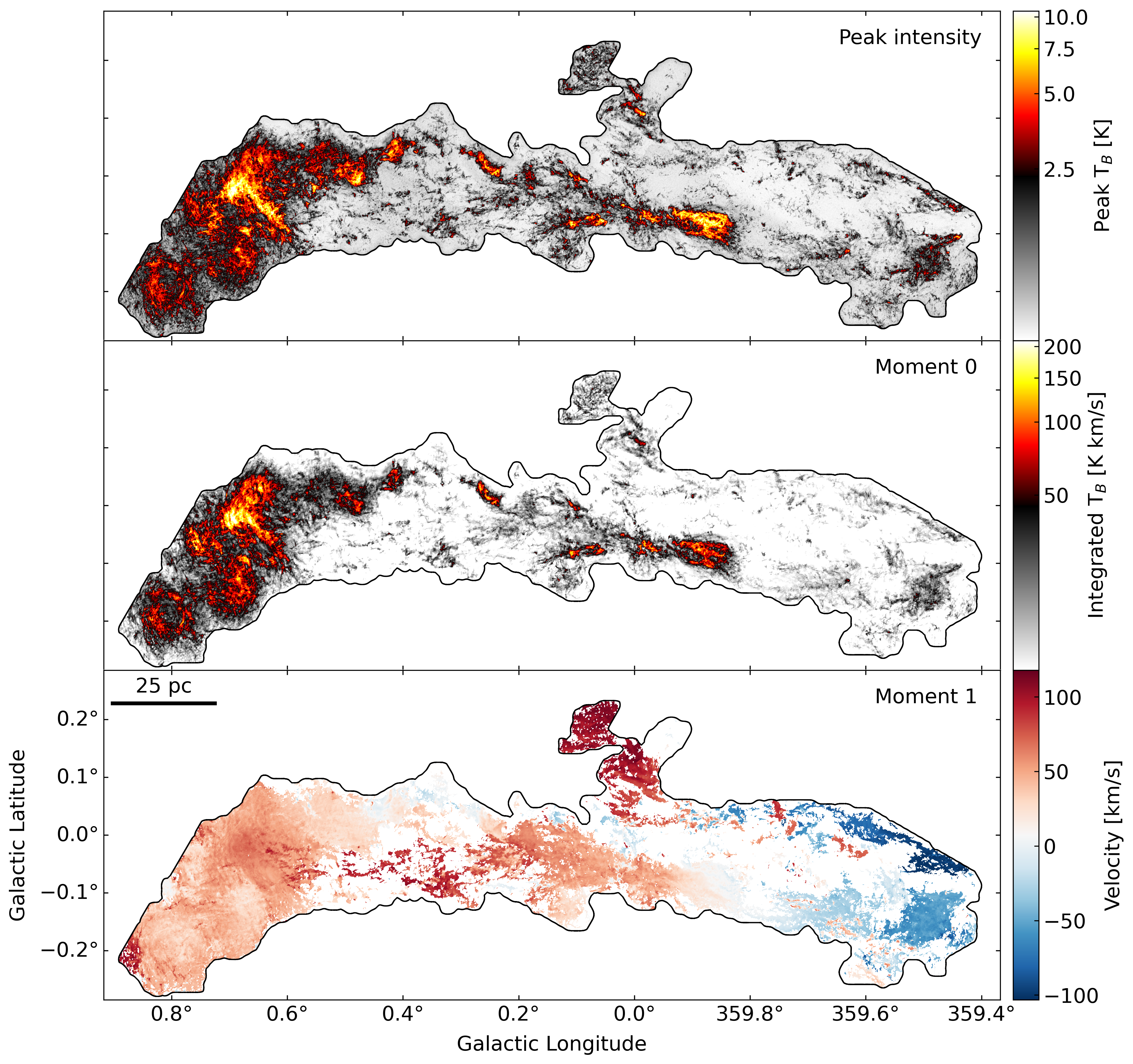}
    \caption{Masked maps of the HNCO (4-3) emission. The panels show the peak intensity (\textit{top}), integrated intensity (\textit{centre}), and the weighted velocity field (\textit{bottom}). The peak intensity and integrated intensity maps are shown on an \textit{asinh} stretch, while the velocity field is linear. The peak intensity map is masked only using edge masking to suppress elevated noise at field edges. The integrated intensity and velocity maps are masked using the binary dilation method described in Section \ref{sec:moments}, combined with the edge masking. The black contour corresponds to the overall footprint of ACES.}
    \label{fig:HNCO_moms}
\end{figure*}

\begin{figure*}
    \centering
    \includegraphics[width=\textwidth]{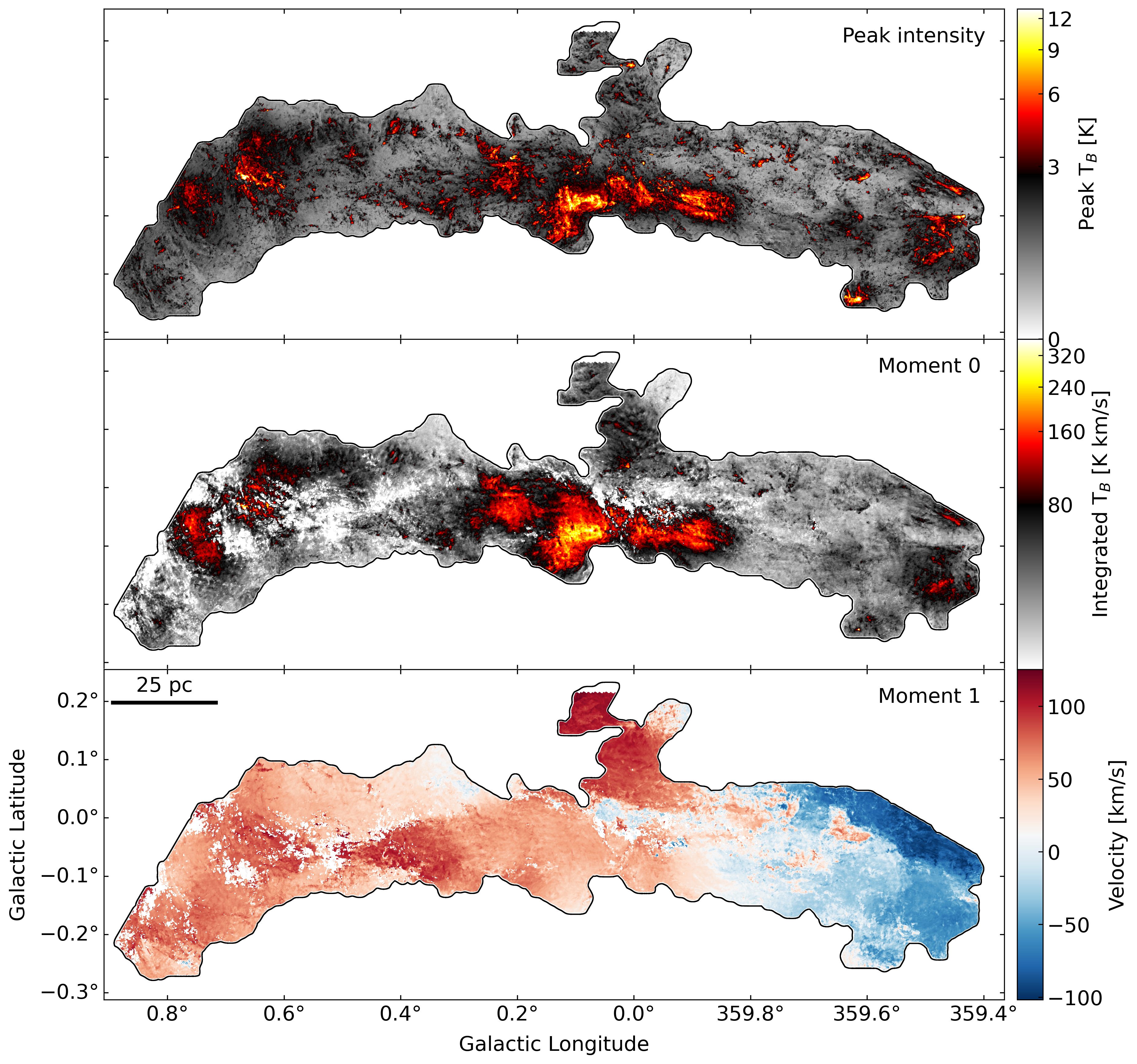}
    \caption{Masked maps of the \hcop (1-0) emission. The panels show the peak intensity (\textit{top}), integrated intensity (\textit{centre}), and the weighted velocity field (\textit{bottom}). The peak intensity and integrated intensity maps are shown on an \textit{asinh} stretch, while the velocity field is linear. The peak intensity map is masked only using edge masking to suppress elevated noise at field edges. The integrated intensity and velocity maps are masked using the binary dilation method described in Section \ref{sec:moments}, combined with the edge masking. The black contour corresponds to the overall footprint of ACES.}
    \label{fig:HCOplus_moments}
\end{figure*}

\begin{figure*}
    \centering
    \includegraphics[width=\textwidth]{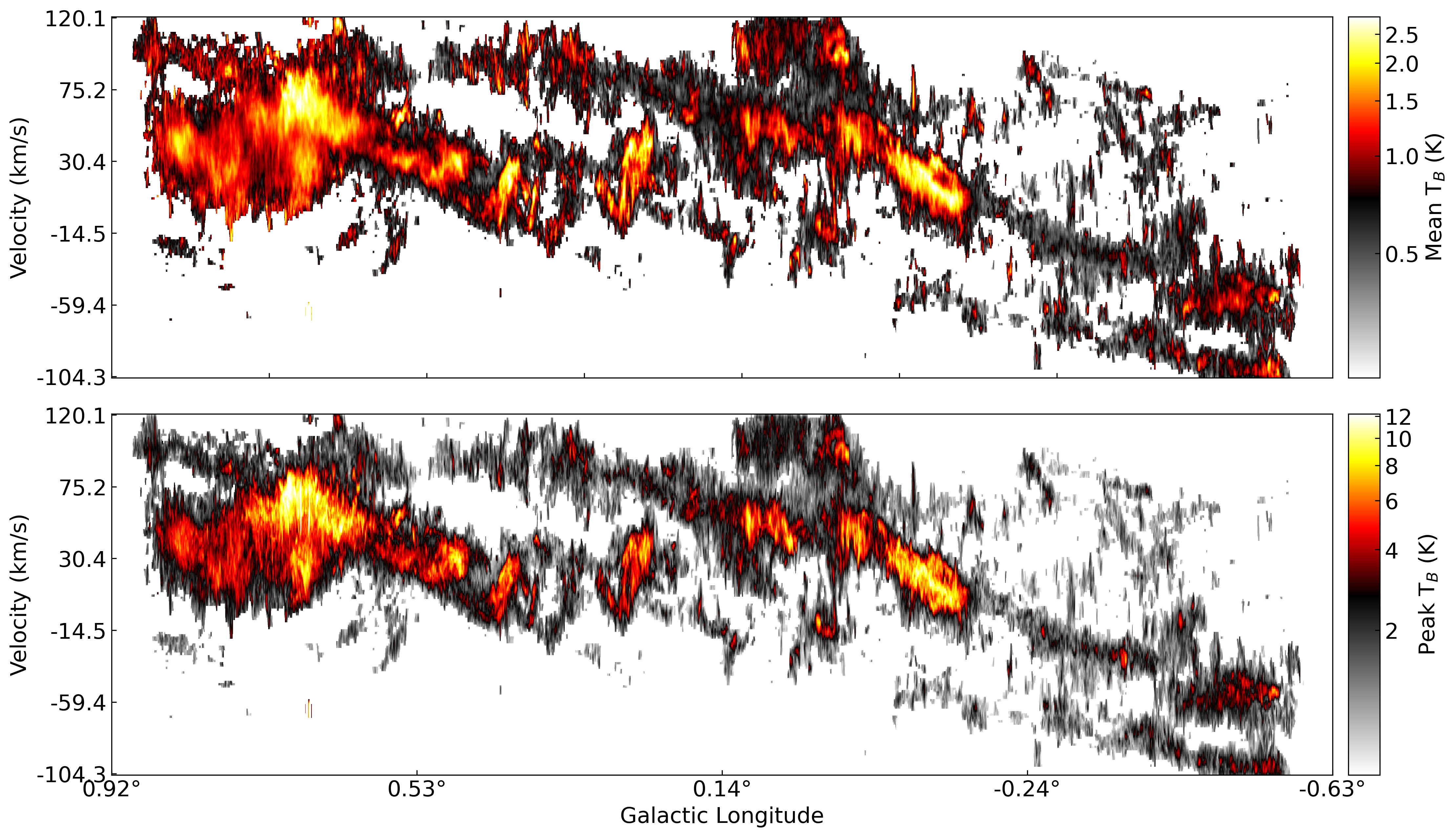}
    \caption{Masked longitude-velocity maps of the HNCO (4-3) emission. The panels show the mean intensity (\textit{top}) and the peak intensity (\textit{bottom}).}
    \label{fig:HNCO_pv}
\end{figure*}

\begin{figure*}
    \centering
    \includegraphics[width=\textwidth]{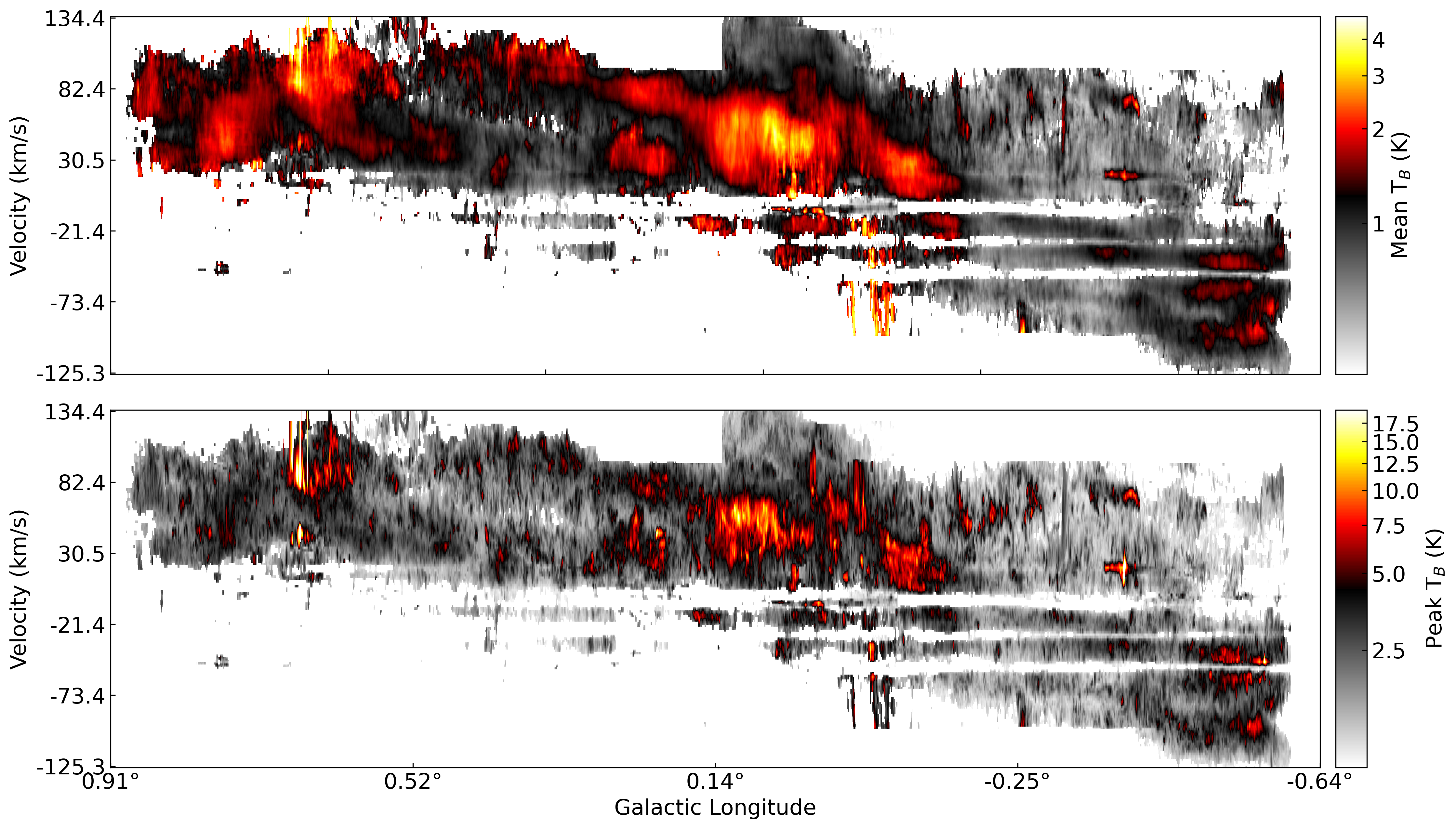}
    \caption{Masked longitude-velocity maps of the \hcop (1-0) emission. The panels show the mean intensity (\textit{top}) and the peak intensity (\textit{bottom}).}
    \label{fig:HCOP_pv}
\end{figure*}

Due to the complexity of the line emission in the CMZ across a large range of spatial scales, it is crucial to mask the emission carefully when making products such as moment maps. To achieve this, we adopt an advanced masking technique similar to those presented in \citet{Dame2001} and \citet{Leroy2021}. We created masked moment maps, peak intensity maps, and position-velocity maps using the full ACES cube mosaics, we do not make these products for individual sub-mosaics. The process is as follows:

\begin{enumerate}
    \item Set a low-significance and a high-significance threshold. We adopted 2.0 and 5.0 $\sigma$, \updates{where $\sigma$ is measured using the median absolute deviation} across all channels.
    \item Create signal masks at each significance threshold.
    \item Prune regions smaller than three times the beam area.
    \item Dilate the high-significance mask (which includes fewer pixels) into the low-significance mask over a number of iterations (i.e. pixels). We used 50 iterations, though in practice the mask does not necessarily expand by this much before it reaches the boundary set by the low-significance mask.
    \item Apply edge masking by eroding the mask inwards to eliminate unreliable/high-noise edge regions (due to primary beam correction effects and internal discontinuities due to differing frequency coverages, see Tables \ref{tab:hcopcubestats} and \ref{tab:hncocubestats}). We used 50 iterations (equivalent to 50 pixels). While this may result in masking out real signal at field edges, we found that too few iterations resulted in prominent linear noise features in the maps, particularly at the boundary between fields with different velocity offsets.
    \item Combine the significance and edge masks and apply this to the cubes to calculate moments and produce position-velocity maps.
\end{enumerate}

While we used the above masking procedure for all other maps, we used a modified version to create the peak intensity maps, whereby we only apply the edge masking to suppress the sub-mosaic edge boundaries, as the other signal masking is not required in the case of identifying the peak intensity at each pixel. 

Masked maps showing the peak intensity, integrated intensity, and velocity field of HNCO and \hcop are shown in Figures \ref{fig:HNCO_moms} and \ref{fig:HCOplus_moments}, respectively. Masked mean and peak longitude-velocity maps of HNCO and \hcop are shown in Figures \ref{fig:HNCO_pv} and \ref{fig:HCOP_pv}, respectively. 

The HNCO emission is very well correlated with the dense gas, where the well-known regions such as Sgr B2, the dust ridge, the 20 and 50~\kms clouds, etc., are very prominent in the peak and integrated intensity maps. The morphology of the \hcop emission is strikingly different, with the Sgr A region dominating the map, particularly in the integrated intensity map. As it is a lower critical density tracer, the \hcop is likely tracing the more diffuse, outer regions of the gas in the molecular clouds, whereas the HNCO is better tracing the denser interiors. 

Both the HNCO and \hcop moment 1 maps clearly show the broad spread in velocities, and the general trend of more negative velocities at more negative longitudes, and vice versa. This is not a smooth gradient, however, and there are many regions where there are spatially overlapping regions of positive and negative velocity components, highlighting the complex kinematics and gas distribution along the line of sight \citep[e.g.][]{Henshaw2016b, Lipman2025, Walker2025}. This is apparent in the PV diagrams in Figures \ref{fig:HNCO_pv} and \ref{fig:HCOP_pv} also, which show that there are often many velocity components at a given longitude.

The \hcop PV maps show at least three prominent blank strips in velocity that span the full longitude range. These are not data artifacts, but absorption features due to foreground material. These absorption features were also noted in \citet{Jones2012} in several tracers, including \hcop (1-0), at velocities of $-$3, $-$28, and $-$52~\kms, which is consistent with what we see in the ACES \hcop data. \citet{Jones2012} attribute this to likely absorption by gas in spiral arms along the line-of-sight at these velocities.

\subsection{\hcop absorption filaments}
\label{sec:hcop_absorption_filaments}

\citet{Bally14} reported the discovery of a network of filaments in G0.253+0.016 (the ‘Brick’) that are seen in \hcop (1-0) absorption, which they categorised as ‘Broad-line absorption’ (BLA) and ‘Narrow-line absorption’ (NLA) filaments, based on the line-widths over which the filaments are observed. The origin of such absorption filaments is unknown. \citet{Bally14} proposed that the BLAs may be associated with the large-scale non-thermal filaments (see the MeerKAT data in Figure \ref{fig:overview}) and may be tracing \hcop molecules gyrating about the magnetic fields, or that they may be tracing shocks in the foreground CMZ gas that we are seeing edge-on. The NLAs are proposed to be tracing optically-thick gas on the foreground surface of the cloud, which they also suggest is expanding.

\begin{figure*}
    \centering
    \includegraphics[width=\textwidth]{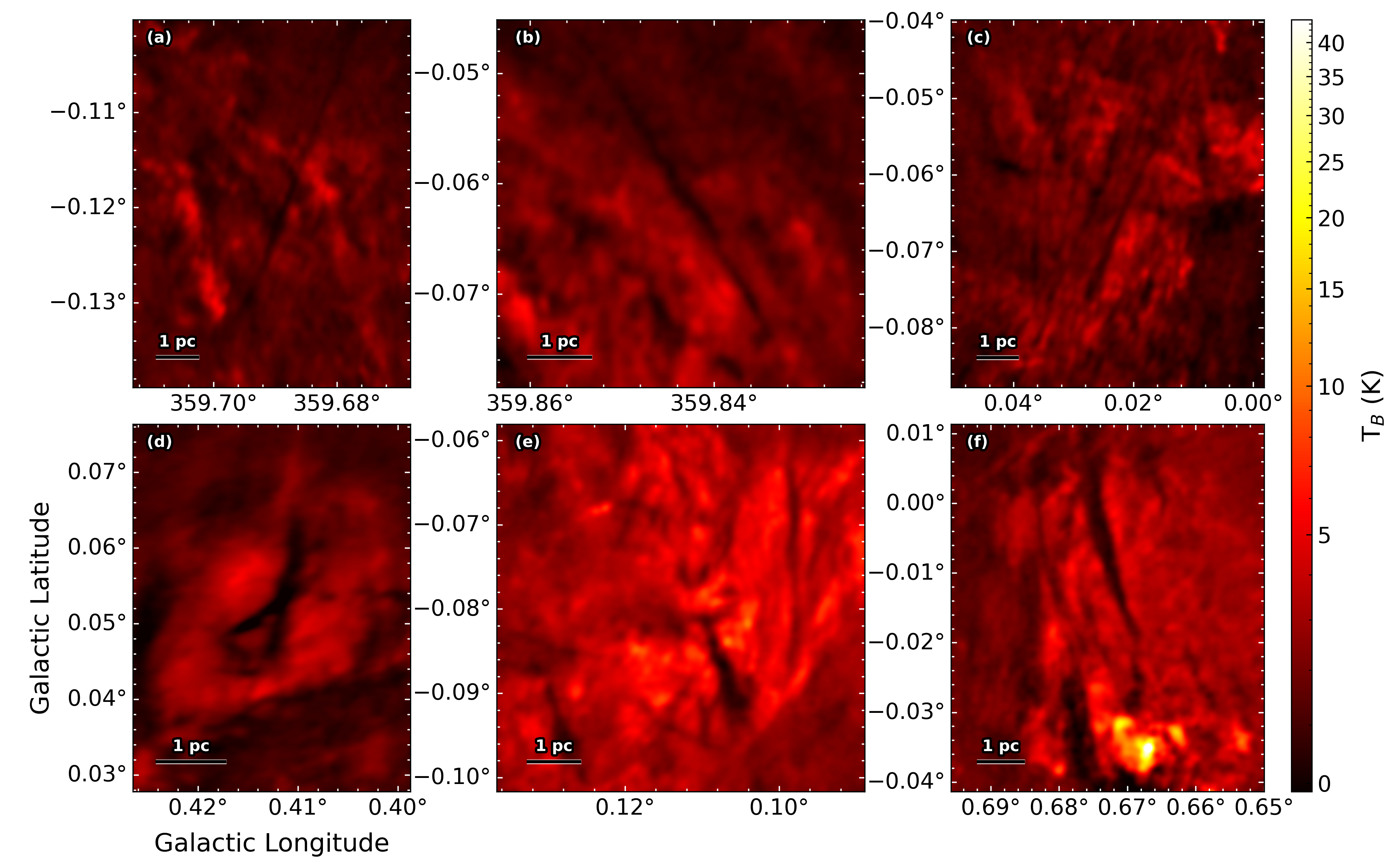}
    \caption{The figure shows examples of the \hcop absorption filaments that are ubiquitous throughout the ACES \hcop data. Each panel shows a single channel from the \hcop cube. a) G359.689-0.121 (not coincident with named molecular cloud), b) 20~\kms cloud, c) Sgr A region, d) dust ridge cloud D, e) the ‘three little pigs’ complex, and f) Sgr B2.}
    \label{fig:hcop_absorption_filaments}
\end{figure*}

Our ACES \hcop (1-0) data reveal that these absorption filaments are abundant throughout the CMZ. Figure \ref{fig:hcop_absorption_filaments} shows some representative examples of such absorption features across a range of Galactic coordinates. These features typically appear to be linear, and span several parsecs in length. They are most prominent in high-density regions, particularly towards known molecular clouds. Some of the largest and most striking absorption features are observed towards Sgr B2, dust ridge cloud D, the ‘three little pigs’ clouds, and the 20~\kms cloud (see Figure \ref{fig:hcop_absorption_filaments}).

We do not attempt to catalogue or characterise these \hcop absorption features here, as that is beyond the scope of this data paper and presents a significant challenge, as there are no existing software packages that could catalogue such features in an automated way. Rather, we simply want to bring attention to these features. ACES reveals that they are ubiquitous throughout the CMZ, and presents an opportunity to better understand their origin. 

\subsection{Comparison to other data}
\label{sec:data_comparison}

We compare the ACES HNCO and \hcop data with other CMZ surveys and targeted observations, both to showcase the quality of the ACES data and to validate it against existing ALMA data.

\subsubsection{Mopra CMZ survey}
\label{sec:mopra_comparison}

The Mopra CMZ survey \citep{Jones2012} observed the CMZ with the 22-m Mopra telescope in 20 different molecular lines at 3~mm, including HNCO and \hcop. The coverage is comparable to ACES, although the coverage in Galactic longitude extends further into positive longitudes, up to $l$ \app 1.8$^{\circ}$.  The survey has an angular resolution of \app 38\arcsec\ and spectral resolution of \app 3.6~\kms.

To compare the ACES and Mopra data on the same intensity scale, we first apply a correction to the Mopra data to convert the brightness temperature from units of antenna temperature to main beam temperature. To do this we divide the data by a main beam efficiency factor of 0.49 \citep{Jones2012, Rathborne2015}.

\begin{figure*}
    \centering
    \includegraphics[width=0.8\textwidth]{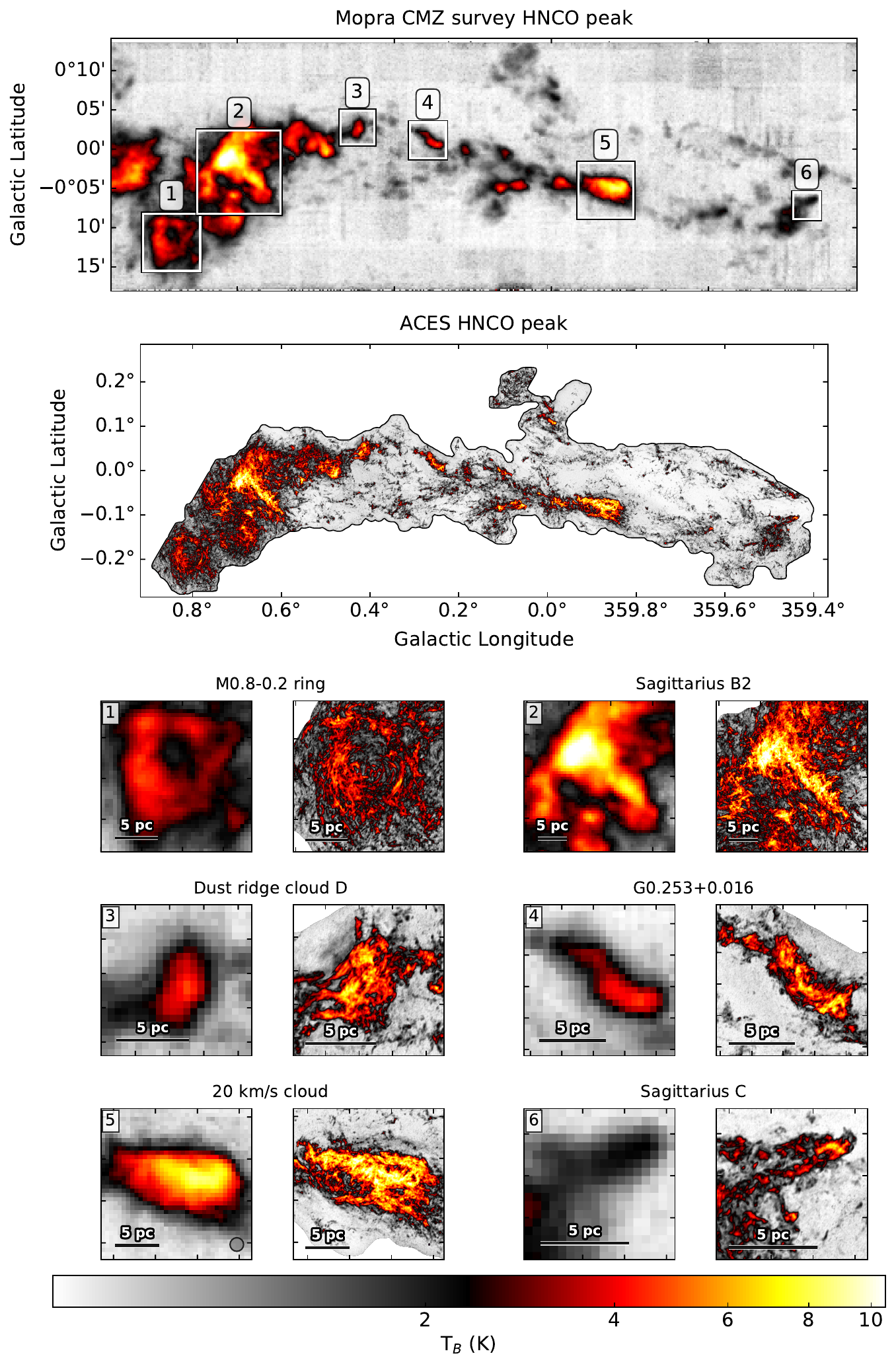}
    \caption{Comparison of the peak HNCO (4-3) emission between the Mopra CMZ survey \citep{Jones2012} and ACES. The top two panels show the peak intensity from Mopra (top) and ALMA (bottom). The subsequent six pairs of panels show zoom-ins of specific regions for comparison, where for each pair the Mopra data is on the left and the ACES data on the right. The numbers for each region are highlighted in the upper left corner of the Mopra zoom-in, where the corresponding number and area are shown in the Mopra overview panel (top). All panels are on a common colour-scale. Reference beams are shown in the lower right of the panels for the 20~\kms cloud.}
    \label{fig:HNCO_mopra_alma}
\end{figure*}

Figure \ref{fig:HNCO_mopra_alma} shows a comparison of the HNCO peak intensity between the Mopra CMZ survey and ACES, along with zoom-ins highlighting specific regions of interest. Overall we see a good correspondence between the general morphology of the HNCO emission and the peak brightness, with the ACES data revealing significant substructure in the gas as a result of the \app 15x increase in angular resolution.

\begin{figure*}
    \centering
    \includegraphics[width=\textwidth]{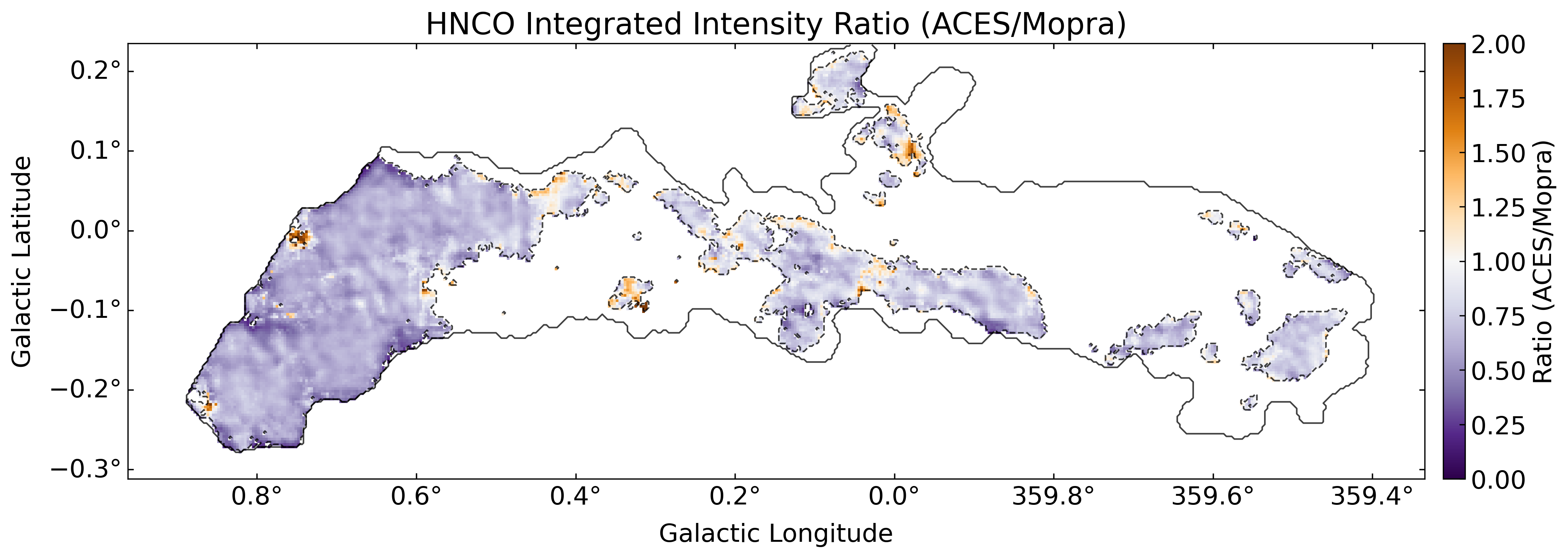}
    \caption{Comparison of the HNCO (4-3) emission between the Mopra CMZ survey \citep{Jones2012} and ACES. The figure shows the ratio map of the ACES HNCO integrated intensity divided by the Mopra HNCO integrated intensity. Both cubes were integrated over the same velocity range, and were both masked following the procedure detailed in Section \ref{sec:moments}. The ACES data were reprojected and smoothed prior to the comparison. The solid contour shows the ACES footprint, and the dashed contour shows the footprint of the masked ratio map.}
    \label{fig:HNCO_mopra_alma_diff}
\end{figure*}

There are some notable differences between the ACES and Mopra maps. In particular the Sagittarius C region (region 6 in Figure \ref{fig:HNCO_mopra_alma}) is significantly fainter in peak HNCO intensity in the Mopra data. This could potentially arise due to the emission in Sgr C being dominated by compact sources, rather than extended emission (see Figure \ref{fig:HNCO_mopra_alma_diff}). However, the sensitivity of Mopra may also be a contributing factor. While several regions have good correspondence (e.g. regions 1, 2 and 5; M0.8-0.2 ring, Sgr B2 and the 20~\kms cloud, respectively), regions 3 (dust ridge cloud D), 4 (the Brick), and 6 (Sgr C) are clearly different both in terms of their peak fluxes and emission morphology.

\updates{To compare the integrated intensities directly, we created a ratio map of the ACES data to the Mopra data (Figure \ref{fig:HNCO_mopra_alma_diff}), after reprojecting and smoothing the ACES data to match the resolution of the Mopra survey}. The two datasets were masked following the procedure described in Section \ref{sec:moments}, and were integrated over the same velocity range ($-$125 to $+$135~\kms). Overall this comparison shows that we are doing a reasonable job of recovering the larger scale structure with ACES. The mean value across the masked region is 0.71, indicating that we are generally under-recovering flux compared to the Mopra data. Though we note that this is sensitive to the choice of the main beam efficiency factor used to convert from antenna to main beam temperature for the Mopra data. We again used a factor of 0.49 for consistency with the literature \citep{Jones2012, Rathborne2015}. There are some regions where ACES sees greater integrated intensities, though we caution that some of the regions near the edges of the map may be skewed by small values in the Mopra data.

\subsubsection{G0.253+0.016 (the ‘Brick’)}
\label{sec:brick_comparison}

G0.253+0.016 (a.k.a. ‘the Brick’) was observed with ALMA band 3 in Cycle 0 and published by \citet{Rathborne2015}. The spectral setup of ACES is very similar to that used in this Cycle 0 observation, covering many of the same key lines, including the HNCO and \hcop lines presented in this work. Despite being from Cycle 0, where there were significantly fewer available antennas (25 vs. 41 and 43 in the two ACES executions blocks of the Brick cloud), the G0.253+0.016 data have better angular resolution (1.7\arcsec\ vs. 2.6\arcsec) and sensitivity due to the longer time-on-source.

While the ACES data are from the combination of the ALMA 12m, 7m and TP arrays (or 12m, 7m, and Mopra in the case of \hcop), the Cycle 0 data are ALMA 12m-only and Mopra, where the Mopra data came from the MALT90 survey \citep{MALT90_1, MALT90_2} and were scaled by the same main beam correction factor of 0.49 as the Mopra CMZ survey (Sections \ref{sec:tp_hcop} and \ref{sec:mopra_comparison}).

\begin{figure*}
    \centering
    \includegraphics[width=\textwidth]{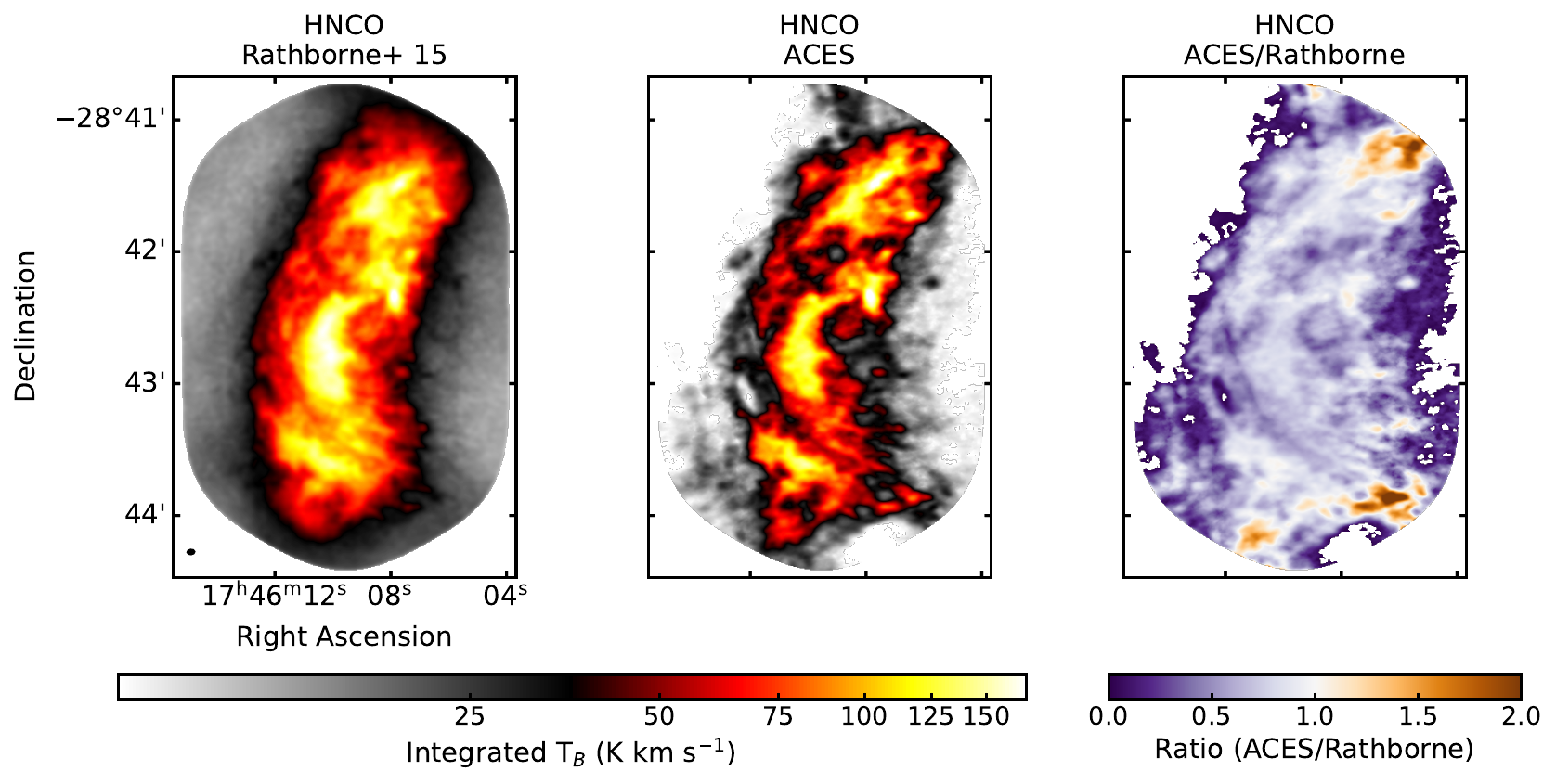}
    \includegraphics[width=\textwidth]{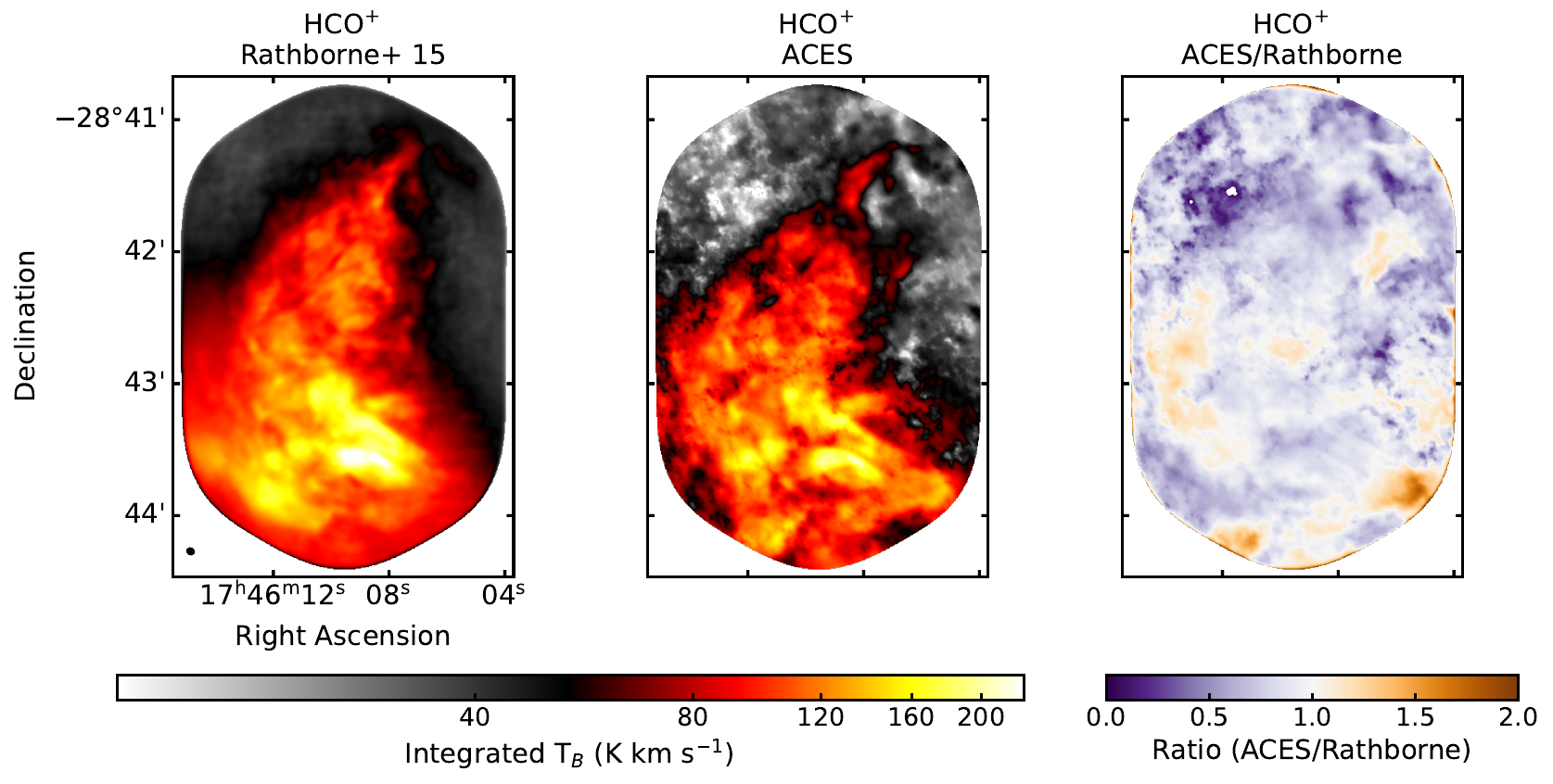}
    \caption{Comparison between \citet{Rathborne2015} and ACES data. The top row shows the HNCO (4-3) integrated emission, while the bottom row shows \hcop (1-0). Shown for each is the integrated intensity of the \citet{Rathborne2015} data (\textit{left}) and the ACES data (\textit{centre}), as well as the ratio image (ACES/Rathborne; \textit{right}). The ACES data have been reprojected onto the same grid as the Rathborne data, and the Rathborne data have been smoothed to match the slightly coarser beam of the ACES data. The beam is shown in the lower left of the first panel. Both datasets were integrated over the same velocity range of $-$125 to $+$135 \kms. NaN values in the ACES and ratio maps are due to the moment masking discussed in Section \ref{sec:moments}. Note that unlike the majority of figures in this paper, these are shown with Right Ascension and Declination to minimise unnecessary white-space.}
    \label{fig:brick_comparison}
\end{figure*}

Figure \ref{fig:brick_comparison} shows a comparison of the HNCO and \hcop integrated intensities for G0.253+0.016 between the ALMA Cycle 0 and ACES data. The Cycle 0 data have been smoothed to match the resolution of the ACES data, and the ACES data have been reprojected to match the Cycle 0 footprint. The ratio of the two (ACES / Cycle 0) are shown in the right column of Figure \ref{fig:brick_comparison}. Note that the \citet{Rathborne2015} data are not primary beam corrected, whereas our data are, which may explain why the ACES data sometimes show slightly higher flux values towards the field edge.

This comparison shows that we are recovering the emission well, especially given that the \citet{Rathborne2015} data are a much deeper observation of a single region. The main difference is that the \citet{Rathborne2015} data better recover the large-scale emission compared to ACES. This is likely due to the fact that we are using ALMA Total Power dishes (12-m diameter), whereas \citet{Rathborne2015} used the much larger Mopra telescope (22-m diameter), which better recovers intermediate scales. The relatively large main beam efficiency correction factor (0.49) that \citet{Rathborne2015} applied to the Mopra data could also contribute to the single dish flux being weighted too heavily, though we note that this would apply to our ALMA+Mopra \hcop data also (see the bottom row of Figure \ref{fig:brick_comparison}, and Section \ref{sec:tp_hcop}).

\section{Conclusions}
\label{sec:conclusions}

\updates{In this paper we have presented the general data reduction procedure for the line data for the ACES Large Program, which observed the CMZ at 3~mm at an angular resolution of \app 2\arcsec. We describe the quality assurance, calibration, and array combination methods used, along with the techniques used for creating the large CMZ-wide mosaics and the associated advanced products (e.g., moment maps, position-velocity maps, etc.).}

\updates{We compare the data with previous small field-of-view ALMA observations to demonstrate that ACES does a good job of recovering the known structure. We also compare ACES with the Mopra 3~mm CMZ survey \citep{Jones2012} to showcase the significant improvements in angular resolution and the richness of substructure that is revealed.}

\updates{We present an overview of the two narrow-bandwidth, high spectral resolution spectral windows that contain the HNCO (4-3) and \hcop (1-0) emission, which are the primary kinematic tracers targeted in ACES. The full suite of products associated with these lines are presented in Appendix \ref{subsec:data_appendix}. These products, along with the other spectral line and continuum products, are released to the community alongside this initial ACES paper series.}

\updates{We highlight the discovery of ubiquitous linear, parsec-scale absorption filaments as traced by \hcop. To date, such \hcop filaments have only been reported in the Brick \citep{Bally14}, and their origin remains a mystery. The ACES data therefore present an opportunity to identify a comprehensive population of these features throughout the CMZ to characterise them and better constrain their nature.}

\updates{The ACES data represent a new standard for mm observations of the CMZ, offering high spectral and spatial resolution across a contiguous coverage, and provide substantial legacy value to the community via an unparalleled insight into the physics, kinematics, and chemistry of the dense gas in the centre of the Milky Way.}

\section*{Acknowledgements}
\updates{
The paper was instigated and led by the ACES data reduction working group, which is coordinated by Adam Ginsburg, Daniel Walker, and Ashley Barnes, and includes (alphabetically) Nazar Budaiev, Laura Colzi, Claire Cooke, Savannah Gramze, Pei-Ying Hsieh, Desmond Jeff, Xing Lu, Jaime Pineda, Marc Pound, Álvaro Sánchez-Monge, and Qizhou Zhang.
ACES is led by Principal Investigator Steven Longmore, together with co-PIs and the management team, John Bally, Ashley Barnes, Cara Battersby, Laura Colzi, Adam Ginsburg, Jonathan Henshaw, Paul Ho, Izaskun Jiménez-Serra, Elisabeth Mills, Maya Petkova, Mattia Sormani, Robin Tress, Daniel Walker, and Jennifer Wallace. 
The ACES proposal was led by Principal Investigator Steven Longmore, together with co-PIs John Bally, Cara Battersby,  Adam Ginsburg, Jonathan Henshaw, Paul Ho, J. M. Diederik Kruijssen,  Izaskun Jiménez-Serra, and Elisabeth Mills. 
The remaining coauthors comprise the ACES team, and its members contributed to the proposal, data analysis, idea development, and/or reading and commenting on the manuscript.}
D.L.W gratefully acknowledges support from the UK ALMA Regional Centre (ARC) Node, which is supported by the Science and Technology Facilities Council [grant numbers ST/Y004108/1 and ST/T001488/1].
R.F. acknowledges support from the grants PID2023-146295NB-I00, and from the Severo Ochoa grant CEX2021-001131-S funded by MCIN/AEI/ 10.13039/501100011033 and by ``European Union NextGenerationEU/PRTR''. 
COOL Research DAO \citep{cool_whitepaper} is a Decentralized Autonomous Organization supporting research in astrophysics aimed at uncovering our cosmic origins.
CB gratefully  acknowledges  funding  from  National  Science  Foundation  under  Award  Nos. 2108938, 2206510, and CAREER 2145689, as well as from the National Aeronautics and Space Administration through the Astrophysics Data Analysis Program under Award ``3-D MC: Mapping Circumnuclear Molecular Clouds from X-ray to Radio,” Grant No. 80NSSC22K1125.
I.J-.S., L.C., and V.M.R. acknowledge support from the grant PID2022-136814NB-I00 by the Spanish Ministry of Science, Innovation and Universities/State Agency of Research MICIU/AEI/10.13039/501100011033 and by ERDF, UE. I.J-.S. also acknowledges the ERC Consolidator grant OPENS (project number 101125858) funded by the European Union. V.M.R. also acknowledges the grant RYC2020-029387-I funded by MICIU/AEI/10.13039/501100011033 and by "ESF, Investing in your future", and from the Consejo Superior de Investigaciones Cient{\'i}ficas (CSIC) and the Centro de Astrobiolog{\'i}a (CAB) through the project 20225AT015 (Proyectos intramurales especiales del CSIC); and from the grant CNS2023-144464 funded by MICIU/AEI/10.13039/501100011033 and by “European Union NextGenerationEU/PRTR.
L.C. also acknowledges support from the "la Caixa" Foundation (ID 100010434, fellowship code LCF/BQ/PR25/12110012).
The authors acknowledge UFIT Research Computing for providing computational resources and support that have contributed to the research results reported in this publication. 
A.G acknowledges support from the NSF under grants CAREER 2142300, AAG 2008101, and particularly AAG 2206511 that supports the ACES large program.
P.-Y. H. acknowledges support from the ADC. Data analysis was in part carried out on the Multi-wavelength Data Analysis System operated by the Astronomy Data Center (ADC), National Astronomical Observatory of Japan.
E.A.C.\ Mills  gratefully  acknowledges  funding  from the National  Science  Foundation  under  Award  Nos. 1813765, 2115428, 2206509, and CAREER 2339670.
M.G.S.-M.\ acknowledges support from the NSF under grant CAREER 2142300. M.G.S.-M.\ also thanks the Spanish MICINN for funding support under grant PID2023-146667NB-I00.
A.S.-M.\ acknowledges support from the RyC2021-032892-I grant funded by MCIN/AEI/10.13039/501100011033 and by the European Union `Next GenerationEU’/PRTR, as well as the program Unidad de Excelencia María de Maeztu CEX2020-001058-M, and support from the PID2023-146675NB-I00 (MCI-AEI-FEDER, UE).
K.M.D acknowledges support from the European Research Council (ERC) Advanced Grant MOPPEX 833460.vii.
X.L.\ acknowledges support from the Strategic Priority Research Program of the Chinese Academy of Sciences (CAS) Grant No.\ XDB0800300, the National Key R\&D Program of China (No.\ 2022YFA1603101), State Key Laboratory of Radio Astronomy and Technology (CAS), the National Natural Science Foundation of China (NSFC) through grant Nos.\ 12273090 and 12322305, the Natural Science Foundation of Shanghai (No.\ 23ZR1482100), and the CAS ``Light of West China'' Program No.\ xbzg-zdsys-202212.
Q. Z. gratefully acknowledges the support from the National Science Foundation under Award No. AST-2206512, and the Smithsonian Institute FY2024 Scholarly Studies Program.
FHL acknowledges support from the ESO Studentship Programme, the Scatcherd European Scholarship of the University of Oxford, and the European Research Council’s starting grant ERC StG-101077573 (`ISM-METALS').
C.~F.~acknowledges funding provided by the Australian Research Council (Discovery Projects DP230102280 and DP250101526), and the Australia-Germany Joint Research Cooperation Scheme (UA-DAAD).
F. M acknowledges financial support from the School of Astronomy at the Institute for Research in Fundamental Sciences-IPM.
J.K. is supported by the Royal Society under grant number RF\textbackslash ERE\textbackslash231132, as part of project URF\textbackslash R1\textbackslash211322.
M.C.\ gratefully acknowledges funding from the DFG through an Emmy Noether Research Group (grant number CH2137/1-1).
MCS acknowledges financial support from the European Research Council under the ERC Starting Grant ``GalFlow'' (grant 101116226) and from Fondazione Cariplo under the grant ERC attrattivit\`{a} n. 2023-3014.
J.Wallace gratefully acknowledges funding from National Science Foundation under Award Nos. 2108938 and 2206510. 
F.N.-L. gratefully acknowledges financial support from grant PID2024-162148NA-I00, funded by MCIN/AEI/10.13039/501100011033 and the European Regional Development Fund (ERDF) “A way of making Europe”, from the Ramón y Cajal programme (RYC2023-044924-I) funded by MCIN/AEI/10.13039/501100011033 and FSE+, and from the Severo Ochoa grant CEX2021-001131-S, funded by MCIN/AEI/10.13039/501100011033.
RSK acknowledges financial support for the ERC Synergy Grant ``ECOGAL'' (855130) and for the Heidelberg Excellence Cluster ``STRUCTURES'' (EXC 2181 - 390900948). In addition RSK is grateful for funding for project ``MAINN'' (BMWK, 50OO2206) and for project ``STARCLUSTERS'' (DFG KL 1358/22-1). 
S.Z. acknowledges support from the NAOJ ALMA Scientific Research Grant Code 2025-29B.
This paper makes use of the following ALMA data: ADS/JAO.ALMA\#2021.1.00172.L. ALMA is a partnership of ESO (representing its member states), NSF (USA) and NINS (Japan), together with NRC (Canada), NSTC and ASIAA (Taiwan), and KASI (Republic of Korea), in cooperation with the Republic of Chile. The Joint ALMA Observatory is operated by ESO, AUI/NRAO and NAOJ.
The authors are grateful to the staff throughout the ALMA organisation, particularly those at the European ALMA Regional Centre, the Joint ALMA Observatory, and the UK ALMA Regional Centre Node, for their extensive support, which was essential to the success of this challenging Large Program.
The authors thank the anonymous referee, whose feedback was critical for improving the clarity and utility of this paper.

\section*{Software}
% This paper made extensive use of the following software packages: \texttt{astropy}, a community-developed core Python package for Astronomy \citep{astropy:2013, astropy:2018, astropy:2022}, \texttt{CARTA} \citep{carta}, \texttt{CASA} \citep{casa, CASATeam2022}, \texttt{radio-beam} (\url{https://radio-beam.readthedocs.io/en/latest/}), \texttt{reproject} (\url{https://reproject.readthedocs.io/}), and \texttt{spectral-cube} (\url{https://spectral-cube.readthedocs.io/en/latest/}).

\updates{The ACES pipeline is based on a number of open-source astronomy software packages, including astropy \citep{astropy:2013, astropy:2018, astropy:2022}, astroquery \citep{Ginsburg2019astroquery}, spectral-cube \citep{Ginsburg2019spectralcube}, radio-beam \citep{Koch2025radiobeam}, pvextractor \citep{Ginsburg2016pvextractor}, reproject \citep{Robitaille2020reproject}, statcont \citep{Sanchez-Monge2018}, CASA \citep{CASATeam2022}, numpy \citep{2020NumPy-Array}, and matplotlib \citep{hunter2007matplotlib}. This project also made use of \texttt{CARTA} for data visualisation and exploration \citep{carta}.}

\section*{Data Availability}
\updates{All data products and associated documentation can be found at \url{https://almascience.org/alma-data/lp/aces}. To maximise usability and provide a single, comprehensive resource for users, we also provide an online, machine-readable table, which contains the full, detailed list of all data products released for the entire survey, including hyperlinks to the full files from the ALMA Science Archive.}

\updates{All code and data processing issues are available at the public GitHub repository here: \url{https://github.com/ACES-CMZ/reduction_ACES}.}

\updates{The ACES data reduction was a monolithic work that was written up in 5 papers.  If you use the ACES data, please cite the appropriate works, which includes \citet{Longmore2025} and the data papers: continuum \citep{Ginsburg2025} and cubes in high-resolution (this work), medium-resolution \citep{Lu2025}, and low-resolution \citep{Hsieh2025} spectral windows. Note that all papers should be cited for use of any data; the continuum imaging relied on the line papers, and vice-versa.}

% All data products and associated documentation can be found at ... [finalised links will be provided upon acceptance. We are currently working with ESO to coordinate the data release to coincide with the publication of the first round of ACES papers.]

% All code and data processing issues are available at the public GitHub repository here: \url{https://github.com/ACES-CMZ/reduction_ACES}.

\bibliographystyle{mnras}
\bibliography{refs}
\section*{Author Affiliations}
\input{affiliations}

\onecolumn

\appendix

\updates{\section{Data release summary}\label{subsec:data_appendix}}
\noindent
The data products from the ACES survey follow a uniform naming convention. The filenames for the data products presented in this paper (HNCO and \hcop) can be constructed using the templates described below, in conjunction with the information provided in Tables \ref{tab:field_info}, \ref{tab:hcopcubestats}, and \ref{tab:hncocubestats}. 

The following subsections are divided according to different product types which occur at the Member ObsUnitSet (MOUS) and Group ObsUnitSet (GOUS) levels\footnote{For a description of ALMA project structures and naming conventions, please refer to Section 2 of the Cycle 9 QA2 guide: \url{https://almascience.eso.org/documents-and-tools/cycle9/alma-qa2-data-products-for-cycle-9}}.
\\
\subsection{Member-level products (12m and 7m)}
We release the data cubes for all SPWs for the 45 ACES regions. These are member-level products, and correspond to the cubes for the 12m and 7m arrays for each field prior to combination. 

We release the separate cubes for the different arrays, as they have changed from those initially delivered by ALMA due to the changes that were made during reprocessing (see Section \ref{sec:processing}). We do not re-release any of the stand-alone TP products, as we used the same cubes as those available in the ALMA Science Archive (ASA).

In this and subsequent subsections, we describe how to construct the filenames of the released data products by providing filename templates that include placeholder variables. These placeholders can be changed as needed to build the filename of the relevant product.

Following the requirements of the ASA, the filename template is:

\texttt{member.uid\_\_\_A001\_\{\textbf{mous-id}\}.lp\_slongmore.\{\textbf{field\_coords}\}.\{\textbf{array}\}.\{\textbf{freq\_range}\}.cube.pbcor.fits}

The placeholder template components should be replaced as follows:
\begin{itemize}
    \item \texttt{\{\textbf{mous\_id}\}}: The 12m or 7m MOUS ID for a given field, as listed in Table \ref{tab:field_info}.
    \item \texttt{\{\textbf{field\_coords}\}}: The central coordinates of the field from Table \ref{tab:field_info} (e.g., \texttt{G000.073+0.184}).
    \item \texttt{\{\textbf{array}\}}: The ALMA array, either \texttt{12m} or \texttt{7m}.
    \item \texttt{\{\textbf{freq\_range}\}}: The frequency range:
    \begin{itemize}
        \item HNCO (4-3): \texttt{87.9-88.0GHz}
        \item \hcop (1-0):  \texttt{89.1-89.2GHz}
    \end{itemize}
\end{itemize}

For example, to find the 12m-only cube for HNCO for field \texttt{ao} (this field contains the well-known molecular cloud G0.253+0.016, aka the Brick), the resulting filename would be:

\texttt{member.uid\_\_\_A001\_X15a0\_X190.lp\_slongmore.G000.300+0.063.12m.87.9-88.0GHz.cube.pbcor.fits}

\subsection{Group-level products (Combined cubes per region)}
We also release array-combined products per SPW per region (see Section \ref{sec:comb} for details on array combination). As these products combine data from multiple MOUSs, they are deemed to be group-level products. 

The filenames of these products can be constructed as follows:

\texttt{group.uid\_\_\_A001\_X1590\_X30a9.lp\_slongmore.\{\textbf{field\_coords}\}.\{\textbf{arrays}\}.\{\textbf{freq\_range}\}.cube.pbcor.fits}

Where the template components are:
\begin{itemize}
    \item \texttt{\{\textbf{field\_coords}\}} and \texttt{\{\textbf{freq\_range}\}} follow the same definition as for the member-level products.
    \item \texttt{\{\textbf{arrays}\}} specifies the combination of arrays used:
    \begin{itemize}
        \item For HNCO: \texttt{12m7mTP}.
        \item For \hcop: \texttt{12m7mTP}, \texttt{12m7mMopra}, or \texttt{12m7m} (interferometric-only data), as discussed in Section \ref{sec:tp_hcop}.
    \end{itemize}
\end{itemize}

As an example, the corresponding \hcop cube including 12m, 7m, and Mopra data for field \texttt{ao} would be:

\texttt{group.uid\_\_\_A001\_X1590\_X30a9.lp\_slongmore.G000.300+0.063.12m7mMopra.89.1-89.2GHz.cube.pbcor.fits}

\subsection{Group-level products (Full CMZ mosaics)}
Finally, we release the full, contiguous mosaics covering the entire ACES footprint (see Section \ref{sec:mosaics}). Again, since these result from the combination of many MOUSs, the full mosaics are at the group level.

In contrast to cubes for the individual fields, we do not release full mosaics for all SPWs, but rather for 14 specific lines that are generally widespread and bright throughout the CMZ. While here we focus only on the HNCO and \hcop, for completeness, the remaining lines released as full mosaics are CS21, CH3CHO, H13CN, H13COplus, H40alpha, HC15N, HC3N, HN13C, NSplus, SiO21, SO21, and SO32 (note that the formatting of these lines is written to match the filenames). We refer the reader to \citetalias{Lu2025} and \citetalias{Hsieh2025} for more details on these lines and the associated data releases.

Most of the SPWs are sufficiently broad that they contain significant spectral ranges with no line emission. We therefore opted to not produce full mosaics containing these blank channels in the interest of reducing the file sizes. In addition to the data cubes for each line, we also release a suite of advanced products, including moment maps, noise maps, etc. (see below).

The filename template for the full mosaic cubes and associated products is:

\texttt{group.uid\_\_\_A001\_X1590\_X30a9.lp\_slongmore.cmz\_mosaic.\{\textbf{arrays}\}.\{\textbf{molecule}\}.\{\textbf{suffix}}\}

Where the template components are:
\begin{itemize}
    \item \texttt{\{\textbf{arrays}\}} and \texttt{\{\textbf{molecule}\}} are paired:
    \begin{samepage}
    \begin{itemize}
        \item HNCO: \texttt{\{\textbf{arrays}\} = 12m7mTP} and \texttt{\{\textbf{molecule}\} = HNCO}.
        \item \hcop (with TP): \texttt{\{\textbf{arrays}\} = 12m7mTP} and \texttt{\{\textbf{molecule}\} = HCOplus}.
        \item \hcop (with Mopra): \texttt{\{\textbf{arrays}\} = 12m7mMopra} and \texttt{\{\textbf{molecule}\} = HCOplus}.
        \item \hcop (interferometric-only): \texttt{\{\textbf{arrays}\} = 12m7m} and \texttt{\{\textbf{molecule}\} = HCOplus\_noTP}.
    \end{itemize}
    \end{samepage}
    \item \texttt{\{\textbf{suffix}\}} defines the specific data product. Products released with this paper are:
    \begin{itemize}
        \item \texttt{cube.pbcor.fits}: primary beam corrected image cube.
        \item \texttt{cube.downsampled\_spatially.pbcor.fits}: as previous, spatially smoothed to a 5 arcsec beam and then spatially rebinned by a factor of $9\times9$.
        \item \texttt{cube.downsampled\_spatially\_and\_spectrally.pbcor.fits}: as previous, additionally smoothed spectrally and rebinned by a factor of 9 along the spectral axis.
        \item \texttt{integrated\_intensity.fits}: masked integrated intensity map of the full-resolution cube, following the procedure described in Section \ref{sec:moments}.
        \item \texttt{mad\_std.fits}: noise map of the full-resolution cube, following the procedure described in Section \ref{sec:moments}.
        \item \texttt{peak\_intensity.fits}: peak intensity map of the full-resolution cube, following the procedure described in Section \ref{sec:moments}.
        \item \texttt{velocity\_at\_peak\_intensity.fits}: velocity map from the masked full-resolution cube; velocity estimated using the peak intensity of the spectrum for each pixel.
        \item \texttt{PV\_l\_max.fits}: $\ell$–$v$ position–velocity map from the full-resolution cube; maximum intensity taken along Galactic latitude ($b$).
        \item \texttt{PV\_l\_mean.fits}: $\ell$–$v$ position–velocity map from the masked full-resolution cube; mean intensity taken along Galactic latitude ($b$)
        \item \texttt{PV\_b\_max.fits}: $b$–$v$ position–velocity map from the full-resolution cube; maximum intensity taken along Galactic longitude ($\ell$).
        \item \texttt{PV\_b\_mean.fits}: $b$–$v$ position–velocity map from the full-resolution cube; mean intensity taken along Galactic longitude ($\ell$).                
    \end{itemize}
\end{itemize}

As an example, the peak $\ell$–$v$ map for the full ACES mosaic of the interferometric-only \hcop data (i.e., without any TP or Mopra data) would be constructed as:

\texttt{group.uid\_\_\_A001\_X1590\_X30a9.lp\_slongmore.cmz\_mosaic.12m7m.HCOplus\_noTP.PV\_l\_max.fits}

\input{Tables/filename_builder_table}
\input{Tables/hcop_stats}
\input{Tables/hnco_stats}

\updates{\section{Total Power flux recovery}\label{subsec:flux_recovery}}

As discussed in Section \ref{sec:comb}, we primarily tested two approaches to data combination: joint-imaging of the 12m and 7m data, followed by feathering the TP data (see Section \ref{sec:joint_decon_feather}), and feather-only, in which all single array products are sequentially feathered together (see Section \ref{sec:feather_only}).

In principle, both of these methods should recover the total flux in the combined data sets as it is inherently anchored to the Total Power data when feathering. However, as a sanity check to ensure that our combination methods are indeed behaving as expected, we present a brief comparison of total integrated fluxes below.
\\

\subsection{Joint imaging and feather vs. feather-only}

In Section \ref{sec:comb_compare} and Figures \ref{fig:HNCO_imaging_comparison} and \ref{fig:HNCO_imaging_comparison+meanspec} we presented a qualitative comparison between the results of our two different data combination approaches for a single region (field \texttt{t}). Here we present a brief quantitative comparison, whereby we calculate the total integrated flux for the 12m, 7m, TP, 12m7mTP (feather-only), and 12m7mTP (joint imaging + feather). This is performed over a common velocity range and common spatial footprint, the latter of which explicitly excludes the field edges to avoid any regions of elevated noise due to the primary beam correction.

Table \ref{tab:flux_stats} shows a comparison of the total integrated HNCO flux for field \texttt{t} for the various array types and combinations listed. As one would expect, we see that both array combination methods recover the Total Power integrated flux well. The joint imaging and feather flux is closer to the TP, but the difference is small (99.7~\% flux recovery for the feather-only vs. 99.9~\% for the joint imaging + feather).

The fluxes of the 12m and 7m only data are significantly lower, demonstrating the need for single dish data to recover the zero spacing information and total flux. We see that the 7m integrated flux is notably lower than the 12m flux. These interferometer-only values indicate the amount of flux included in the clean model, which roughly corresponds to the flux at the size scale probed by each array. However we caution that these values are also sensitive to things such as imaging parameters, sensitivity, uv coverage and associated artifacts, etc. The key takeaway here is that the fluxes in the combined data for both approaches are consistent with the Total Power.

\input{Tables/flux_stats}

\subsection{Feather-only integrated fluxes of all fields}

To assess the flux recovery of our feather-only combination method, we compared the integrated flux of the final combined 12m7mTP HNCO cubes against the input TP cubes for each of the 45 sub-mosaics. The total flux for each sub-mosaic was calculated over an identical spatial footprint and velocity range, both of which were defined by the combined cube. The edges of the mosaics were masked to exclude noisy pixels in the combined map due to the primary beam correction. 

\begin{figure*}
    \centering
    \includegraphics[width=\textwidth]{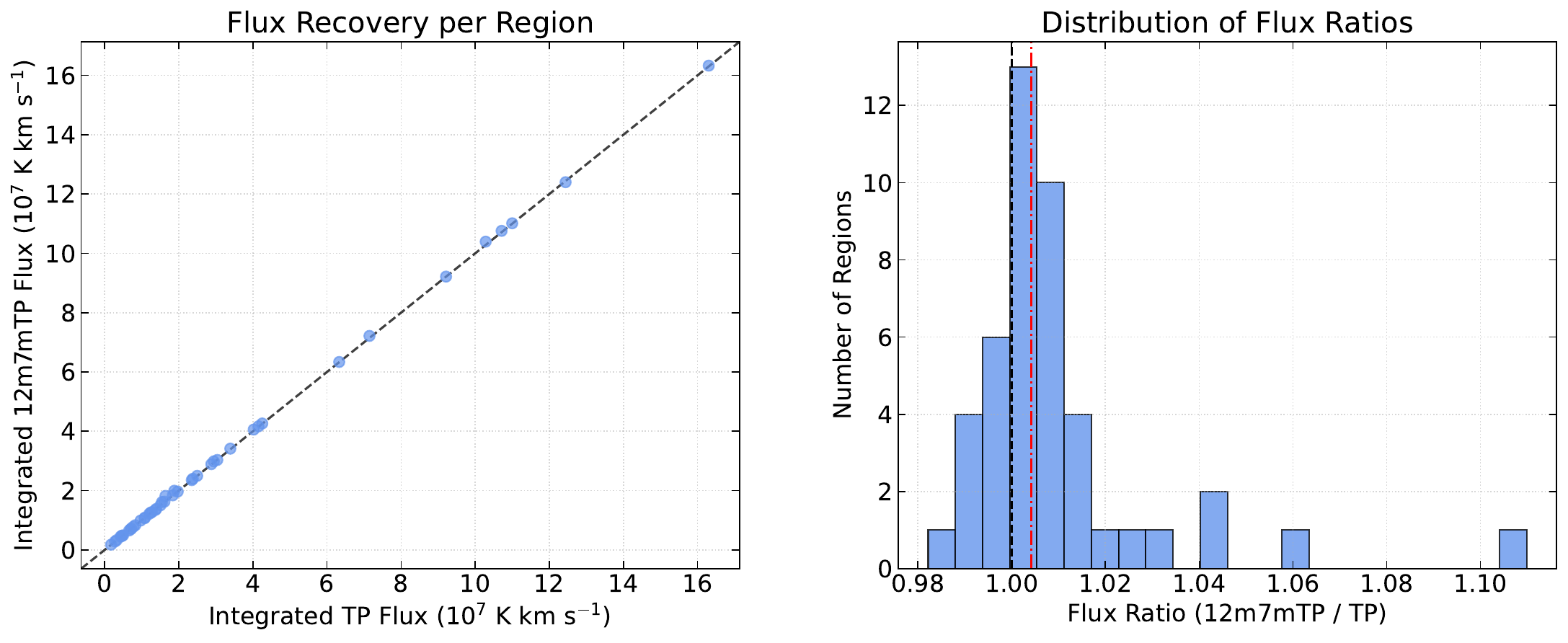}
    \caption{\updates{Comparison of the integrated fluxes between the combined 12m7mTP and the TP HNCO cubes. Integrated fluxes were calculated over a common spatial and spectral extent (Sec. \ref{sec:comb_compare}). The left panel shows a direct comparison of the integrated fluxes, while the right panel shows a histogram of the flux ratios (12m7mTP / TP). In both panels the dashed black line indicates perfect flux recovery, i.e. a 1-to-1 line in the left figure and a flux ratio of 1 in the right. The red dash-dot line in the right panel shows the median ratio of 1.004.}}
    \label{fig:HNCO_flux_recovery}
\end{figure*}

Figure \ref{fig:HNCO_flux_recovery} shows a comparison of these values and demonstrates that we are recovering the flux well in our combined images, with a median 12m7mTP / TP integrated flux ratio of 1.004.

\bsp
\label{lastpage}
\end{document}

%% file: affiliation_macros.tex
\newcounter{affcounter}

\newcommand{\defaffiliationlabel}[1]{%
  \refstepcounter{affcounter}%
  \expandafter\xdef\csname #1\endcsname{\theaffcounter}%
}

\newcommand{\affref}[1]{$^{\csname #1\endcsname}$}

\newcommand{\affrefs}[1]{%
  $^{%
    \@for\@ref:=#1\do{%
      \@ref\@ifnextchar\@nil{}{,}%
    }%
  }$%
}

\newcommand{\affrefTwo}[2]{$^{\csname #1\endcsname,\csname #2\endcsname}$}
\newcommand{\affrefThree}[3]{$^{\csname #1\endcsname,\csname #2\endcsname,\csname #3\endcsname}$}
\newcommand{\affrefFour}[4]{$^{\csname #1\endcsname,\csname #2\endcsname,\csname #3\endcsname,\csname #4\endcsname}$}

\newcommand{\printaffiliation}[2]{%
  $^{\csname #1\endcsname}$#2\\%
}

%% file: affiliation_labels.tex
\defaffiliationlabel{ukarcnode}
\defaffiliationlabel{uflorida}
\defaffiliationlabel{eso}
\defaffiliationlabel{shao}
\defaffiliationlabel{naoc_key}
\defaffiliationlabel{naoj}
\defaffiliationlabel{ice_csic}
\defaffiliationlabel{ieec}
\defaffiliationlabel{umd}
\defaffiliationlabel{mpe}
\defaffiliationlabel{kansas}
\defaffiliationlabel{ljmu}
\defaffiliationlabel{mpia}
\defaffiliationlabel{ari_heidelberg}
\defaffiliationlabel{COOL}
\defaffiliationlabel{eso_chile}
\defaffiliationlabel{jao}
\defaffiliationlabel{nanjing}
\defaffiliationlabel{nanjing_key}
\defaffiliationlabel{cfa}
\defaffiliationlabel{mit}
\defaffiliationlabel{colorado}
\defaffiliationlabel{uconn}
\defaffiliationlabel{cab_csic}
\defaffiliationlabel{iaa_taipei}
\defaffiliationlabel{chalmers}
\defaffiliationlabel{clap}
\defaffiliationlabel{iop_epfl}
\defaffiliationlabel{oaq}
\defaffiliationlabel{inaf_arcetri}
\defaffiliationlabel{jbca}
\defaffiliationlabel{nrao}
\defaffiliationlabel{ita_heidelberg}
\defaffiliationlabel{uva}
\defaffiliationlabel{leiden}
\defaffiliationlabel{anu}
\defaffiliationlabel{iaa_csic}
\defaffiliationlabel{ucn}
\defaffiliationlabel{cassaca}
\defaffiliationlabel{ias}
\defaffiliationlabel{ucl}
\defaffiliationlabel{izw_heidelberg}
\defaffiliationlabel{radcliffe}
\defaffiliationlabel{ipm}
\defaffiliationlabel{ulaserena}
\defaffiliationlabel{iff_csic}
\defaffiliationlabel{gbo}
\defaffiliationlabel{utokyo}
\defaffiliationlabel{umass}
\defaffiliationlabel{aberystwyth}
\defaffiliationlabel{deps}

%% file: authors.tex
\author[D.~L.~Walker \& ACES Team]{Daniel L.~Walker,\affref{ukarcnode}\thanks{E-mail: daniel.walker.astro@gmail.com}\orcidlink{0000-0001-7330-8856}
Adam Ginsburg,\affref{uflorida}\orcidlink{0000-0001-6431-9633}
Ashley~T.~Barnes,\affref{eso}\orcidlink{0000-0003-0410-4504}
Xing Lu,\affrefTwo{shao}{naoc_key}\orcidlink{0000-0003-2619-9305}
Pei-Ying Hsieh,\affref{naoj}\orcidlink{0000-0001-9155-39}
\newauthor
\'Alvaro S\'anchez-Monge,\affrefTwo{ice_csic}{ieec}\orcidlink{0000-0002-3078-9482}
Savannah R. Gramze,\affref{uflorida}\orcidlink{0000-0002-1313-429X}
Nazar Budaiev,\affref{uflorida}\orcidlink{0000-0002-0533-8575}
Marc W.~Pound,\affref{umd}\orcidlink{0000-0002-7269-342X}
\newauthor
Jaime E. Pineda,\affref{mpe}\orcidlink{0000-0002-3972-1978}
Alyssa Bulatek,\affref{uflorida}\orcidlink{0000-0002-4407-885X}
Claire Cook,\affref{kansas}
Jonathan D. Henshaw,\affrefTwo{ljmu}{mpia}\orcidlink{0000-0001-9656-7682}
\newauthor
Katharina Immer,\affref{eso}\orcidlink{0000-0003-4140-5138}
Namitha Issac,\affref{shao}\orcidlink{0000-0002-7881-689X}
Desmond Jeff,\affref{uflorida}\orcidlink{0000-0003-0416-4830}
Fu-Heng Liang,\affrefTwo{ari_heidelberg}{eso}\orcidlink{0000-0003-2496-1247}
Steven N. Longmore,\affrefTwo{ljmu}{COOL}\orcidlink{0000-0001-6353-0170}
\newauthor
Elisabeth A.C. Mills,\affref{kansas}\orcidlink{0000-0001-8782-1992}
Sergio Martín,\affrefTwo{eso_chile}{jao}\orcidlink{0000-0001-9281-2919}
Xing Pan,\affrefThree{nanjing}{nanjing_key}{cfa}\orcidlink{0000-0003-1337-9059}
Thushara G.S. Pillai,\affref{mit}\orcidlink{0000-0003-2133-4862}
\newauthor
Qizhou Zhang,\affref{cfa}\orcidlink{0000-0003-2384-6589}
John Bally,\affref{colorado}\orcidlink{0000-0001-8135-6612}
Cara Battersby,\affref{uconn}\orcidlink{0000-0002-6073-9320}
Laura Colzi,\affref{cab_csic}\orcidlink{0000-0001-8064-6394}
Paul T. P. Ho, \affref{iaa_taipei}\orcidlink{0000-0002-3412-4306}
\newauthor
Izaskun Jim\'enez-Serra,\affref{cab_csic}\orcidlink{0000-0003-4493-8714}
J.~M.~Diederik Kruijssen,\affref{COOL}\orcidlink{0000-0002-8804-0212}
Maya A. Petkova,\affref{chalmers}\orcidlink{0000-0002-6362-8159}
Mattia C. Sormani,\affref{clap}\orcidlink{0000-0001-6113-6241}
\newauthor
Robin G. Tress,\affref{iop_epfl}\orcidlink{0000-0002-9483-7164}
Jennifer Wallace,\affref{uconn}\orcidlink{0009-0002-7459-4174}
J. Armijos-Abenda\~no,\affref{oaq}\orcidlink{0000-0003-3341-6144}
Lucia Armillotta,\affref{inaf_arcetri}\orcidlink{0000-0002-5708-1927}
N. Bijas,\affref{jbca}\orcidlink{0000-0002-6398-7530}
\newauthor
Rojita Buddhacharya,\affref{ljmu}\orcidlink{0009-0004-0685-7678}
Laura A. Busch,\affref{mpe}
Natalie O. Butterfield,\affref{nrao}\orcidlink{0000-0002-4013-6469}
M\'elanie Chevance\affrefTwo{ita_heidelberg}{COOL}\orcidlink{0000-0002-5635-5180}
\newauthor
Samuel Crowe,\affref{uva}\orcidlink{0009-0005-0394-3754}
Ana Karla D\'iaz-Rodr\'iguez,\affref{ukarcnode}\orcidlink{0000-0001-9112-6474}
Katarzyna M. Dutkowska,\affref{leiden}\orcidlink{0000-0003-0980-6871}
Christoph Federrath,\affref{anu}\orcidlink{0000-0002-0706-2306}
\newauthor
Rub\'{e}n Fedriani,\affref{iaa_csic}\orcidlink{0000-0003-4040-4934}
Pablo Garc{\'i}a,\affrefTwo{ucn}{cassaca}\orcidlink{0000-0002-8586-6721}
Simon~C.~O.~Glover,\affref{ita_heidelberg}\orcidlink{0000-0001-6708-1317}
Qi-Lao Gu,\affref{shao}
H Perry Hatchfield\affref{uconn}\orcidlink{0000-0003-0946-4365}
\newauthor
Rebecca J. Houghton,\affref{ljmu}\orcidlink{0000-0002-9723-1088}
Yue Hu,\affref{ias}\orcidlink{0000-0002-8455-0805}
Janik Karoly,\affref{ucl}\orcidlink{0000-0001-5996-3600}
Ralf S.\ Klessen,\affrefFour{ita_heidelberg}{izw_heidelberg}{cfa}{radcliffe}\orcidlink{0000-0002-0560-3172}
Mark R. Krumholz,\affref{anu}\orcidlink{0000-0003-3893-854X}
\newauthor
Farideh Mazoochi,\affref{ipm}\orcidlink{0000-0003-3390-4893}
Francisco Nogueras-Lara,\affrefTwo{iaa_csic}{eso}\orcidlink{0000-0002-6379-7593}
Dylan Par\'e,\affrefTwo{jao}{nrao}\orcidlink{0000-0002-5811-0136}
Denise Riquelme-V\'asquez,\affref{ulaserena}\orcidlink{0000-0001-5389-0535}
\newauthor
V\'ictor M. Rivilla,\affref{cab_csic}\orcidlink{0000-0002-2887-5859}
Miriam G. Santa-Maria,\affrefTwo{uflorida}{iff_csic}\orcidlink{0000-0002-3941-0360}
Anika Schmiedeke,\affref{gbo}\orcidlink{0000-0002-1730-8832}
Yoshiaki Sofue,\affref{utokyo}\orcidlink{0000-0002-4268-6499}
\newauthor
Volker Tolls,\affref{cfa}\orcidlink{0000-0003-1841-2241}
Q. Daniel Wang,\affref{umass}\orcidlink{0000-0002-9279-4041}
Gwenllian M. Williams,\affref{aberystwyth}\orcidlink{0000-0001-5933-2147}
Fengwei Xu,\affref{mpia}\orcidlink{0000-0001-5950-1932},
and Suinan Zhang\affrefTwo{deps}{naoj}\orcidlink{0000-0002-8389-6695}
\\
$^{*}$Author affiliations are listed at the end of the paper
}

%% file: Tables/spectral_setup.tex
\begin{table*}
\caption{ACES spectral configuration. The spectral window (SPW) ID is given for each of the six SPWs, and for each array type, as the indexing differs between the 12m, 7m, and TP data. For each SPW, the lower ($\nu_{\rm L}$) and upper ($\nu_{\rm U}$) frequency bounds, channel widths ($\Delta\nu$, both in terms of frequency and velocity), and bandwidth \updates{(BW, both in terms of frequency and velocity)} are given. A set of representative lines contained in each SPW is also shown. HNCO and \hcop are highlighted as these are the lines presented in this paper. Note that the upper and lower frequency bounds correspond to the bandwidth for the 12m observations.}
\label{tab:spectral_setup}
\begin{tabular}{cccccccccc}
\hline\hline
SPW & SPW & SPW & $\nu_{\rm L}$ & $\nu_{\rm U}$ & $\Delta\nu$ & $\Delta\nu$ & BW & BW & Representative \\
12m & 7m & TP & & & & & & & Lines \\\hline
& & & $\mathrm{GHz}$ & $\mathrm{GHz}$ & $\mathrm{MHz}$ & $\mathrm{km~s^{-1}}$ & $\mathrm{GHz}$ & \kms &  \\\hline
25 & 16 & 17 & 85.9656 & 86.4344 & 0.2442 & 0.849 & 0.46875 & 1617 & HC$^{15}$N 1-0, SO 2$_2$-1$_1$, SiO 2-1 v=1 maser, H$^{13}$CN 1-0\\ 
27 & 18 & 19 & 86.6656 & 87.1344 & 0.2442 & 0.842 & 0.46875 & 1630 & H$^{13}$CO+ 1-0, SiO 2-1, HN$^{13}$C 1-0\\ 
29 & 20 & 21 & 89.1592 & 89.2178 & 0.0305 & 0.103 & 0.05859 & 197 & \textbf{\hcop 1-0}\\ 
31 & 22 & 23 & 87.8959 & 87.9545 & 0.0305 & 0.104 & 0.05859 & 200 & \textbf{HNCO 4-3}\\ 
33 & 24 & 25 & 97.6625 & 99.5375 & 0.4883 & 1.485 & 1.875 & 5701 & CS 2-1, CH$_3$CHO 5(1,4)–4(1,3) A–, H40$\alpha$, SO 2$_3$-1$_2$\\ 
35 & 26 & 27 & 99.5625 & 101.438 & 0.4883 & 1.457 & 1.875 & 5593 & HC$_3$N 11-10\\ 
\hline\hline
\end{tabular}
\end{table*}

%% file: affiliations.tex
\printaffiliation{ukarcnode}{UK ALMA Regional Centre Node, Jodrell Bank Centre for Astrophysics, The University of Manchester, Manchester M13 9PL, UK}
\printaffiliation{uflorida}{Department of Astronomy, University of Florida, P.O. Box 112055, Gainesville, FL 32611, USA}
\printaffiliation{eso}{European Southern Observatory (ESO), Karl-Schwarzschild-Stra{\ss}e 2, 85748 Garching, Germany}
\printaffiliation{shao}{Shanghai Astronomical Observatory, Chinese Academy of Sciences, 80 Nandan Road, Shanghai 200030, P.\ R.\ China}
\printaffiliation{naoc_key}{State Key Laboratory of Radio Astronomy and Technology, A20 Datun Road, Chaoyang District, Beijing, 100101, P. R. China}
\printaffiliation{naoj}{National Astronomical Observatory of Japan, 2-21-1 Osawa, Mitaka, Tokyo 181-8588, Japan}
\printaffiliation{ice_csic}{Institut de Ci\`encies de l'Espai (ICE), CSIC, Campus UAB, Carrer de Can Magrans s/n, E-08193, Bellaterra, Barcelona, Spain}
\printaffiliation{ieec}{Institut d'Estudis Espacials de Catalunya (IEEC), E-08860, Castelldefels, Barcelona, Spain}
\printaffiliation{umd}{University of Maryland, Department of Astronomy, College Park, MD 20742-2421, USA}
\printaffiliation{mpe}{Max-Planck-Institut f\"ur extraterrestrische Physik, Gie\ss enbachstra\ss e 1, 85748 Garching bei M\"unchen, Germany}
\printaffiliation{kansas}{Department of Physics and Astronomy, University of Kansas, 1251 Wescoe Hall Drive, Lawrence, KS 66045, USA}
\printaffiliation{ljmu}{Astrophysics Research Institute, Liverpool John Moores University, 146 Brownlow Hill, Liverpool L3 5RF, The UK}
\printaffiliation{mpia}{{Max Planck Institute for Astronomy, K\"{o}nigstuhl 17, D-69117 Heidelberg, Germany}}
\printaffiliation{ari_heidelberg}{Astronomisches Rechen-Institut, Zentrum f\"{u}r Astronomie der Universit\"{a}t Heidelberg, M\"{o}nchhofstra\ss e 12-14, D-69120 Heidelberg, Germany}
\printaffiliation{COOL}{Cosmic Origins Of Life (COOL) Research DAO, \href{https://coolresearch.io}{https://coolresearch.io}}
\printaffiliation{eso_chile}{European Southern Observatory, Alonso de C\'ordova, 3107, Vitacura, Santiago 763-0355, Chile}
\printaffiliation{jao}{Joint ALMA Observatory, Alonso de C\'ordova, 3107, Vitacura, Santiago 763-0355, Chile}
\printaffiliation{nanjing}{School of Astronomy and Space Science, Nanjing University, 163 Xianlin Avenue, Nanjing 210023, P.R.China}
\printaffiliation{nanjing_key}{Key Laboratory of Modern Astronomy and Astrophysics (Nanjing University), Ministry of Education, Nanjing 210023, P.R.China}
\printaffiliation{cfa}{Center for Astrophysics | Harvard \& Smithsonian, 60 Garden Street, Cambridge, MA 02138, USA}
\printaffiliation{mit}{Haystack Observatory, Massachusetts Institute of Technology, 99 Millstone Road, Westford, MA 01886, USA}
\printaffiliation{colorado}{Center for Astrophysics and Space Astronomy, Department of Astrophysical and Planetary Sciences, University of Colorado, Boulder, CO 80389, USA}
\printaffiliation{uconn}{Department of Physics, University of Connecticut, 196A Auditorium Road, Unit 3046, Storrs, CT 06269, USA}
\printaffiliation{cab_csic}{Centro de Astrobiolog{\'i}a (CAB), CSIC-INTA, Carretera de Ajalvir km 4, 28850 Torrej{\'o}n de Ardoz, Madrid, Spain}
\printaffiliation{iaa_taipei}{Academia Sinica Institute of Astronomy and Astrophysics, Astronomy-Mathematics Building, AS/NTU No.1, Sec. 4, Roosevelt Rd, Taipei 10617, Taiwan}
\printaffiliation{chalmers}{Space, Earth and Environment Department, Chalmers University of Technology, SE-412 96 Gothenburg, Sweden}
\printaffiliation{clap}{{Como Lake centre for AstroPhysics (CLAP), DiSAT, Universit{\`a} dell’Insubria, via Valleggio 11, 22100 Como, Italy}}
\printaffiliation{iop_epfl}{Institute of Physics, Laboratory for Galaxy Evolution and Spectral Modelling, EPFL, Observatoire de Sauverny, Chemin Pegasi 51, 1290 Versoix, Switzerland}
\printaffiliation{oaq}{Observatorio Astron\'omico de Quito, Observatorio Astron\'omico Nacional, Escuela Polit\'ecnica Nacional, 170403, Quito, Ecuador}
\printaffiliation{inaf_arcetri}{INAF Arcetri Astrophysical Observatory, Largo Enrico Fermi 5, Firenze, 50125, Italy}
\printaffiliation{jbca}{Jodrell Bank Centre for Astrophysics, The University of Manchester, Manchester M13 9PL, UK}
\printaffiliation{nrao}{National Radio Astronomy Observatory, 520 Edgemont Road, Charlottesville, VA 22903, USA}
\printaffiliation{ita_heidelberg}{Universit\"{a}t Heidelberg, Zentrum f\"{u}r Astronomie, Institut f\"{u}r Theoretische Astrophysik, Albert-Ueberle-Str 2, D-69120 Heidelberg, Germany}
\printaffiliation{uva}{Dept. of Astronomy, University of Virginia, Charlottesville, Virginia 22904, USA}
\printaffiliation{leiden}{Leiden Observatory, Leiden University, P.O. Box 9513, 2300 RA Leiden, The Netherlands}
\printaffiliation{anu}{Research School of Astronomy and Astrophysics, Australian National University, Canberra, ACT 2611, Australia}
\printaffiliation{iaa_csic}{Instituto de Astrof\'{i}sica de Andaluc\'{i}a, CSIC, Glorieta de la Astronomía s/n, 18008 Granada, Spain}
\printaffiliation{ucn}{Instituto de Astronom\'ia, Universidad Cat\'olica del Norte, Av. Angamos 0610, Antofagasta, Chile}
\printaffiliation{cassaca}{Chinese Academy of Sciences South America Center for Astronomy, National Astronomical Observatories, CAS, Beijing 100101, China}
\printaffiliation{ias}{Institute for Advanced Study, 1 Einstein Drive, Princeton, NJ 08540, USA}
\printaffiliation{ucl}{Department of Physics and Astronomy, University College London, Gower Street, London WC1E 6BT, UK}
\printaffiliation{izw_heidelberg}{Universit\"{a}t Heidelberg, Interdisziplin\"{a}res Zentrum f\"{u}r Wissenschaftliches Rechnen, Im Neuenheimer Feld 225, 69120 Heidelberg, Germany}
\printaffiliation{radcliffe}{Elizabeth S. and Richard M. Cashin Fellow at the Radcliffe Institute for Advanced Studies at Harvard University, 10 Garden Street, Cambridge, MA 02138, U.S.A.}
\printaffiliation{ipm}{Institute for Research in Fundamental Sciences (IPM), School of Astronomy, Tehran, Iran}
% \printaffiliation{villanova}{Department of Physics, Villanova University, 800 E. Lancaster Ave., Villanova, PA 19085, USA}
\printaffiliation{ulaserena}{Departamento de Astronom\'ia, Universidad de La Serena, Ra\'ul Bitr\'an 1305, La Serena, Chile}
\printaffiliation{iff_csic}{Instituto de Física Fundamental (CSIC), Calle Serrano 121-123, 28006, Madrid, Spain}
\printaffiliation{gbo}{Green Bank Observatory, P.O. Box 2, Green Bank, WV 24944, USA}
\printaffiliation{utokyo}{Institute of Astronomy, The University of Tokyo, Mitaka, Tokyo 181-0015, Japan}
\printaffiliation{umass}{Department of Astronomy, University of Massachusetts, Amherst, MA 01003, USA}
\printaffiliation{aberystwyth}{Department of Physics, Aberystwyth University, Ceredigion, Cymru, SY23 3BZ, UK}
% \printaffiliation{kiaa_pku}{Kavli Institute for Astronomy and Astrophysics, Peking University, Beijing 100871, People's Republic of China}
% \printaffiliation{pku_astro}{Department of Astronomy, School of Physics, Peking University, Beijing, 100871, People's Republic of China}
\printaffiliation{deps}{Department of Earth and Planetary Sciences, Institute of Science Tokyo, Meguro, Tokyo, 152-8551, Japan}

%% file: Tables/filename_builder_table.tex
\begin{table}
\centering
\caption{\updates{Field information with central coordinates and MOUS IDs. Note that we include the TP MOUS ID here for reference, but it is not used in any of the filename construction described in Appendix \ref{subsec:data_appendix}. The table is ordered according to the central coordinates of each field by increasing Galactic longitude.}}
\label{tab:field_info}
\begin{tabular}{ccccc}
\hline
Central Coordinates & Field ID & 12m MOUS ID & 7m MOUS ID & TP MOUS ID \\
\hline
G359.448-0.183 & ad & \texttt{X15a0\_X14e} & \texttt{X15a0\_X150} & \texttt{X15a0\_X152} \\
G359.463-0.090 & aj & \texttt{X15a0\_X172} & \texttt{X15b4\_X47} & \texttt{X15b4\_X49} \\
G359.511-0.166 & ag & \texttt{X15a0\_X160} & \texttt{X15a0\_X162} & \texttt{X15a0\_X164} \\
G359.543-0.041 & e & \texttt{X15a0\_Xb8} & \texttt{X15b4\_X39} & \texttt{X15a0\_Xbc} \\
G359.590-0.121 & ae & \texttt{X15a0\_X154} & \texttt{X15a0\_X156} & \texttt{X15a0\_X158} \\
G359.602-0.207 & l & \texttt{X15a0\_Xe2} & \texttt{X15a0\_Xe4} & \texttt{X15a0\_Xe6} \\
G359.622+0.003 & t & \texttt{X15a0\_X112} & \texttt{X15a0\_X114} & \texttt{X15a0\_X116} \\
G359.668-0.073 & ap & \texttt{X15a0\_X196} & \texttt{X15a0\_X198} & \texttt{X3577\_X6c5} \\
G359.704+0.031 & y & \texttt{X15a0\_X130} & \texttt{X15a0\_X132} & \texttt{X15a0\_X134} \\
G359.717-0.135 & ab & \texttt{X15a0\_X142} & \texttt{X15a0\_X144} & \texttt{X15a0\_X146} \\
G359.748-0.025 & q & \texttt{X15a0\_X100} & \texttt{X15a0\_X102} & \texttt{X15a0\_X104} \\
G359.799-0.102 & h & \texttt{X15a0\_Xca} & \texttt{X15a0\_Xcc} & \texttt{X15a0\_Xce} \\
G359.822+0.016 & ai & \texttt{X15a0\_X16c} & \texttt{X15a0\_X16e} & \texttt{X15a0\_X170} \\
G359.874-0.055 & p & \texttt{X15a0\_Xfa} & \texttt{X15a0\_Xfc} & \texttt{X15a0\_Xfe} \\
G359.915-0.108 & x & \texttt{X15a0\_X12a} & \texttt{X15a0\_X12c} & \texttt{X15a0\_X12e} \\
G359.945+0.165 & v & \texttt{X15a0\_X11e} & \texttt{X15a0\_X120} & \texttt{X15a0\_X122} \\
G359.952-0.007 & aq & \texttt{X15a0\_X19c} & \texttt{X15a0\_X19e} & \texttt{X15a0\_X1a0} \\
G359.972+0.089 & am & \texttt{X15a0\_X184} & \texttt{X15a0\_X186} & \texttt{X15a0\_X188} \\
G359.999-0.077 & aa & \texttt{X15a0\_X13c} & \texttt{X15a0\_X13e} & \texttt{X15a0\_X140} \\
G000.029+0.041 & r & \texttt{X15a0\_X106} & \texttt{X15a0\_X108} & \texttt{X15a0\_X10a} \\
G000.073+0.184 & a & \texttt{X1590\_X30aa} & \texttt{X1590\_X30ac} & \texttt{X1590\_X30ae} \\
G000.082-0.037 & o & \texttt{X15a0\_Xf4} & \texttt{X15a0\_Xf6} & \texttt{X15a0\_Xf8} \\
G000.123-0.115 & af & \texttt{X15a0\_X15a} & \texttt{X15a0\_X15c} & \texttt{X15a0\_X15e} \\
G000.162+0.010 & f & \texttt{X15a0\_Xbe} & \texttt{X15a0\_Xc0} & \texttt{X15a0\_Xc2} \\
G000.208-0.063 & s & \texttt{X15a0\_X10c} & \texttt{X15a0\_X10e} & \texttt{X15a0\_X110} \\
G000.286-0.019 & c & \texttt{X15a0\_Xac} & \texttt{X15a0\_Xae} & \texttt{X15a0\_Xb0} \\
G000.300+0.063 & ao & \texttt{X15a0\_X190} & \texttt{X15a0\_X192} & \texttt{X15a0\_X194} \\
G000.323-0.089 & d & \texttt{X15a0\_Xb2} & \texttt{X15a0\_Xb4} & \texttt{X15a0\_Xb6} \\
G000.367+0.029 & as & \texttt{X15a0\_X1a8} & \texttt{X15a0\_X1aa} & \texttt{X15a0\_X1ac} \\
G000.412-0.050 & k & \texttt{X15a0\_Xdc} & \texttt{X15a0\_Xde} & \texttt{X15a0\_Xe0} \\
G000.453+0.060 & al & \texttt{X15a0\_X17e} & \texttt{X15a0\_X180} & \texttt{X15a0\_X182} \\
G000.464-0.104 & ac & \texttt{X15a0\_X148} & \texttt{X15a0\_X14a} & \texttt{X15a0\_X14c} \\
G000.491-0.002 & ar & \texttt{X15a0\_X1a2} & \texttt{X15b4\_X4f} & \texttt{X15a0\_X1a6} \\
G000.538-0.079 & an & \texttt{X15a0\_X18a} & \texttt{X15a0\_X18c} & \texttt{X15a0\_X18e} \\
G000.569+0.044 & z & \texttt{X15a0\_X136} & \texttt{X15a9\_X12d3} & \texttt{X15a0\_X13a} \\
G000.601-0.131 & u & \texttt{X15a0\_X118} & \texttt{X15a0\_X11a} & \texttt{X15a0\_X11c} \\
G000.616-0.033 & b & \texttt{X15a0\_Xa6} & \texttt{X15b4\_X37} & \texttt{X15a0\_Xaa} \\
G000.637+0.068 & ak & \texttt{X15b4\_X4b} & \texttt{X15a9\_X12d5} & \texttt{X15b4\_X4d} \\
G000.665-0.111 & i & \texttt{X15a0\_Xd0} & \texttt{X15b4\_Xc5} & \texttt{X15a0\_Xd4} \\
G000.692+0.005 & m & \texttt{X15b4\_X41} & \texttt{X15b4\_X43} & \texttt{X15b4\_X45} \\
G000.715-0.180 & w & \texttt{X15a0\_X124} & \texttt{X15a0\_X126} & \texttt{X15a0\_X128} \\
G000.741-0.064 & j & \texttt{X15b4\_X3d} & \texttt{X15a0\_Xd8} & \texttt{X15b4\_X3f} \\
G000.774-0.244 & n & \texttt{X15a0\_Xee} & \texttt{X15a0\_Xf0} & \texttt{X15a0\_Xf2} \\
G000.792-0.142 & g & \texttt{X15a0\_Xc4} & \texttt{X15a0\_Xc6} & \texttt{X15a0\_Xc8} \\
G000.835-0.216 & ah & \texttt{X15a0\_X166} & \texttt{X15a0\_X168} & \texttt{X15a0\_X16a} \\
\hline
\end{tabular}
\end{table}

%% file: Tables/hcop_stats.tex
\begin{table*}
\centering
\caption{Global statistics for \hcop array-combined cubes for all ACES regions. Shown for each field are the velocity offsets of the spectral setup (see Section \ref{sec:obs}), the beam major and minor axes, the beam position angle, the image size in pixels (x, y), the number of spectral channels, the peak intensity in the cube, \updates{and the noise, which is estimated from channels in the lowest quartile of the mean spectrum (assumed emission-free) using the scaled median absolute deviation (${\rm MAD}$; where $\sigma_{\rm MAD} = 1.4826\,{\rm MAD}$)}. The statistics for both the 12m7mTP and 12m7mMopra data are shown (see Section \ref{sec:tp_hcop}). Where applicable, values for the 12m7mTP are shown first, and the those for the 12m7mMopra data are second (separated by a /).}
\label{tab:hcopcubestats}
\begin{tabular}{cccccccccccc}
\hline \hline
Central & V$_{\textrm{off}}$ & B$_{\textrm{maj}}$ & B$_{\textrm{min}}$ & BPA & Image & No. & Pixel & S$_{peak}$ & $\sigma_{MAD}$ & S$_{peak}$ & $\sigma_{MAD}$ \\
Coordinates & & & & & size & channels & size & & & & \\
& (\kms) & (\arcsec) & (\arcsec) & (deg) & (pix) & & (\arcsec) & (mJy~beam$^{-1}$) & (mJy~beam$^{-1}$) & K & K \\
\hline
G359.448-0.183 &  0  &  1.96  &  1.16  &  -75  &  (1172, 2554)  &  1920 / 110  &  0.20  &  309.2 / 214.2  &  15.7 / 5.7  &  20.9 / 14.5  &  1.1 / 0.4 \\
G359.463-0.090 &  -30  &  2.50  &  1.91  &  85  &  (1433, 1280)  &  1913 / 109  &  0.31  &  927.4 / 803.4  &  12.0 / 5.6  &  29.8 / 25.8  &  0.4 / 0.2 \\
G359.511-0.166 &  0  &  1.97  &  1.54  &  -80  &  (1721, 1713)  &  1910 / 109  &  0.24  &  391.3 / 286.0  &  14.4 / 5.6  &  19.8 / 14.5  &  0.7 / 0.3 \\
G359.543-0.041 &  -30  &  2.81  &  1.90  &  88  &  (1434, 1464)  &  1915 / 110  &  0.28  &  530.6 / 338.4  &  14.7 / 6.4  &  15.3 / 9.7  &  0.4 / 0.2 \\
G359.590-0.121 &  0  &  2.30  &  1.90  &  83  &  (1335, 1359)  &  1913 / 109  &  0.30  &  383.9 / 229.5  &  13.7 / 5.5  &  13.5 / 8.1  &  0.5 / 0.2 \\
G359.602-0.207 &  0  &  1.79  &  1.79  &  -79  &  (2408, 1517)  &  1915 / 110  &  0.26  &  1291.8 / 1072.8  &  14.4 / 5.4  &  61.9 / 51.4  &  0.7 / 0.3 \\
G359.622+0.003 &  0  &  2.40  &  2.03  &  74  &  (1288, 1287)  &  1913 / 109  &  0.31  &  486.7 / 391.1  &  13.8 / 5.9  &  15.3 / 12.3  &  0.4 / 0.2 \\
G359.668-0.073 &  0  &  3.05  &  2.20  &  29  &  (1336, 1365)  &  1917 / 110  &  0.30  &  594.4 / 512.3  &  14.5 / 6.8  &  13.6 / 11.7  &  0.3 / 0.2 \\
G359.704+0.031 &  0  &  2.26  &  1.71  &  -86  &  (1454, 1131)  &  1915 / 109  &  0.27  &  427.4 / 231.4  &  16.2 / 6.6  &  17.0 / 9.2  &  0.6 / 0.3 \\
G359.717-0.135 &  0  &  2.03  &  1.81  &  88  &  (1487, 1254)  &  1913 / 109  &  0.28  &  289.2 / 200.6  &  12.3 / 4.9  &  12.1 / 8.4  &  0.5 / 0.2 \\
G359.748-0.025 &  0  &  2.47  &  1.93  &  87  &  (1337, 1360)  &  1913 / 109  &  0.30  &  419.7 / 395.3  &  12.3 / 5.4  &  13.5 / 12.7  &  0.4 / 0.2 \\
G359.799-0.102 &  0  &  1.80  &  1.36  &  -83  &  (1981, 2105)  &  1912 / 108  &  0.21  &  480.7 / 228.7  &  13.7 / 5.0  &  30.2 / 14.4  &  0.9 / 0.3 \\
G359.822+0.016 &  0  &  2.58  &  1.86  &  78  &  (1369, 1614)  &  1913 / 110  &  0.29  &  486.3 / 407.4  &  13.4 / 5.4  &  15.6 / 13.0  &  0.4 / 0.2 \\
G359.874-0.055 &  0  &  1.80  &  1.34  &  85  &  (1879, 1861)  &  1915 / 109  &  0.22  &  345.2 / 231.6  &  12.8 / 5.1  &  22.0 / 14.8  &  0.8 / 0.3 \\
G359.915-0.108 &  0  &  2.63  &  1.96  &  -87  &  (1313, 786)  &  1916 / 110  &  0.30  &  580.1 / 415.7  &  13.9 / 5.8  &  17.3 / 12.4  &  0.4 / 0.2 \\
G359.945+0.165 &  30  &  2.00  &  1.27  &  -82  &  (1396, 1905)  &  1914 / 110  &  0.19  &  397.3 / 279.9  &  15.4 / 5.4  &  24.0 / 16.9  &  0.9 / 0.3 \\
G359.952-0.007 &  0  &  1.81  &  1.45  &  -83  &  (1817, 1806)  &  1914 / 110  &  0.22  &  1199.2 / 1190.6  &  14.1 / 5.3  &  70.2 / 69.7  &  0.8 / 0.3 \\
G359.972+0.089 &  30  &  3.03  &  1.91  &  -85  &  (2602, 1133)  &  1920 / 110  &  0.30  &  929.3 / 699.3  &  18.0 / 7.6  &  24.7 / 18.6  &  0.5 / 0.2 \\
G359.999-0.077 &  0  &  2.01  &  1.72  &  -77  &  (1615, 1473)  &  1917 / 110  &  0.26  &  516.8 / 409.6  &  13.3 / 5.1  &  23.0 / 18.2  &  0.6 / 0.2 \\
G000.029+0.041 &  30  &  2.97  &  1.84  &  -84  &  (1549, 1411)  &  1917 / 110  &  0.29  &  528.3 / 398.7  &  17.0 / 7.0  &  14.9 / 11.2  &  0.5 / 0.2 \\
G000.073+0.184 &  40  &  1.82  &  1.30  &  -78  &  (2083, 2491)  &  1912 / 109  &  0.20  &  447.3 / 219.3  &  13.5 / 4.6  &  29.0 / 14.2  &  0.9 / 0.3 \\
G000.082-0.037 &  0  &  2.72  &  1.93  &  89  &  (1302, 1246)  &  1915 / 110  &  0.32  &  835.7 / 701.2  &  14.4 / 5.9  &  24.5 / 20.5  &  0.4 / 0.2 \\
G000.123-0.115 &  0  &  2.78  &  1.85  &  89  &  (1437, 1170)  &  1915 / 110  &  0.31  &  763.5 / 732.6  &  13.4 / 5.3  &  22.8 / 21.9  &  0.4 / 0.2 \\
G000.162+0.010 &  0  &  2.78  &  1.95  &  86  &  (1773, 1461)  &  1920 / 110  &  0.28  &  996.1 / 813.2  &  14.2 / 5.7  &  28.2 / 23.0  &  0.4 / 0.2 \\
G000.208-0.063 &  0  &  2.63  &  1.97  &  86  &  (1331, 1362)  &  1917 / 110  &  0.30  &  902.6 / 746.1  &  13.7 / 5.7  &  26.8 / 22.1  &  0.4 / 0.2 \\
G000.286-0.019 &  0  &  2.68  &  1.92  &  86  &  (1427, 1454)  &  1914 / 110  &  0.28  &  711.8 / 468.5  &  15.4 / 6.2  &  21.3 / 14.0  &  0.5 / 0.2 \\
G000.300+0.063 &  0  &  2.48  &  1.73  &  -86  &  (2504, 915)  &  1914 / 109  &  0.28  &  520.9 / 346.4  &  17.1 / 6.7  &  18.7 / 12.4  &  0.6 / 0.2 \\
G000.323-0.089 &  30  &  1.89  &  1.89  &  -63  &  (2050, 1578)  &  1917 / 110  &  0.23  &  441.2 / 211.6  &  14.7 / 5.6  &  19.0 / 9.1  &  0.6 / 0.2 \\
G000.367+0.029 &  0  &  1.92  &  1.29  &  -74  &  (1995, 2043)  &  1912 / 109  &  0.20  &  350.0 / 241.6  &  14.9 / 5.3  &  21.7 / 15.0  &  0.9 / 0.3 \\
G000.412-0.050 &  30  &  1.96  &  1.21  &  -79  &  (2213, 2197)  &  1908 / 108  &  0.18  &  402.3 / 163.2  &  15.3 / 5.5  &  26.1 / 10.6  &  1.0 / 0.4 \\
G000.453+0.060 &  0  &  2.67  &  1.98  &  84  &  (2183, 1440)  &  1920 / 110  &  0.28  &  367.9 / 331.1  &  16.0 / 6.2  &  10.7 / 9.6  &  0.5 / 0.2 \\
G000.464-0.104 &  0  &  1.83  &  1.27  &  -79  &  (1672, 1618)  &  1914 / 110  &  0.19  &  329.5 / 176.4  &  15.4 / 5.1  &  21.8 / 11.7  &  1.0 / 0.3 \\
G000.491-0.002 &  40  &  2.44  &  1.87  &  82  &  (1380, 1417)  &  1913 / 109  &  0.29  &  414.4 / 273.4  &  14.6 / 5.7  &  14.0 / 9.2  &  0.5 / 0.2 \\
G000.538-0.079 &  30  &  2.57  &  1.81  &  83  &  (1326, 1351)  &  1913 / 109  &  0.30  &  515.6 / 362.6  &  13.6 / 5.4  &  17.0 / 12.0  &  0.4 / 0.2 \\
G000.569+0.044 &  40  &  2.61  &  1.80  &  82  &  (1373, 1370)  &  1913 / 109  &  0.29  &  308.6 / 187.6  &  14.3 / 5.7  &  10.1 / 6.1  &  0.5 / 0.2 \\
G000.601-0.131 &  30  &  2.09  &  1.28  &  -77  &  (1381, 2078)  &  1914 / 110  &  0.19  &  281.6 / 159.6  &  16.5 / 5.7  &  16.2 / 9.2  &  0.9 / 0.3 \\
G000.616-0.033 &  40  &  1.74  &  1.36  &  86  &  (1904, 1949)  &  1915 / 110  &  0.21  &  3563.7 / 3532.8  &  12.4 / 4.8  &  231.4 / 229.4  &  0.8 / 0.3 \\
G000.637+0.068 &  30  &  2.47  &  2.10  &  83  &  (757, 1002)  &  1916 / 110  &  0.32  &  475.3 / 445.8  &  13.7 / 5.3  &  14.1 / 13.2  &  0.4 / 0.2 \\
G000.665-0.111 &  30  &  1.83  &  1.63  &  -71  &  (1670, 1633)  &  1915 / 110  &  0.25  &  422.1 / 215.8  &  13.0 / 4.7  &  21.7 / 11.1  &  0.7 / 0.2 \\
G000.692+0.005 &  30  &  2.48  &  2.06  &  80  &  (1364, 1373)  &  1917 / 110  &  0.29  &  4613.3 / 4585.8  &  13.4 / 5.6  &  138.7 / 137.9  &  0.4 / 0.2 \\
G000.715-0.180 &  30  &  2.46  &  1.81  &  -80  &  (1498, 1417)  &  1913 / 109  &  0.28  &  297.6 / 231.6  &  12.6 / 5.0  &  10.3 / 8.0  &  0.4 / 0.2 \\
G000.741-0.064 &  30  &  2.87  &  1.76  &  -87  &  (1469, 1503)  &  1914 / 109  &  0.27  &  535.9 / 430.0  &  15.6 / 6.5  &  16.3 / 13.1  &  0.5 / 0.2 \\
G000.774-0.244 &  30  &  1.84  &  1.17  &  -80  &  (1394, 1833)  &  1914 / 110  &  0.19  &  304.3 / 140.3  &  14.9 / 5.4  &  21.7 / 10.0  &  1.1 / 0.4 \\
G000.792-0.142 &  30  &  2.18  &  1.66  &  85  &  (1596, 1638)  &  1915 / 110  &  0.25  &  326.8 / 232.0  &  14.1 / 5.3  &  13.9 / 9.8  &  0.6 / 0.2 \\
G000.835-0.216 &  30  &  1.93  &  1.31  &  -73  &  (1991, 1932)  &  1912 / 109  &  0.20  &  247.9 / 145.5  &  12.9 / 4.5  &  15.1 / 8.8  &  0.8 / 0.3 \\
\hline\hline
\end{tabular}
\end{table*}

%% file: Tables/hnco_stats.tex
\begin{table*}
\centering
\caption{Global statistics for HNCO array-combined cubes for all ACES regions. Shown for each field are the velocity offsets of the spectral setup (see Section \ref{sec:obs}), the beam major and minor axes, the beam position angle, the image size in pixels (x, y), the number of spectral channels, the peak intensity in the cube, \updates{and the noise, which is estimated from channels in the lowest quartile of the mean spectrum (assumed emission-free) using the scaled median absolute deviation (${\rm MAD}$; where $\sigma_{\rm MAD} = 1.4826\,{\rm MAD}$)}.}
\label{tab:hncocubestats}
\begin{tabular}{cccccccccccc}
\hline \hline
Central & V$_{\textrm{off}}$ & B$_{\textrm{maj}}$ & B$_{\textrm{min}}$ & BPA & Image & No. & Pixel & S$_{peak}$ & $\sigma_{MAD}$ & S$_{peak}$ & $\sigma_{MAD}$ \\
Coordinates & & & & & size & channels & size & & & & \\
& & & & & & & & & & & \\
& (\kms) & (\arcsec) & (\arcsec) & (deg) & (pix) & & (\arcsec) & (mJy~beam$^{-1}$) & (mJy~beam$^{-1}$) & K & K \\
\hline
G359.448-0.183 &  0  &  1.97  &  1.17  &  -75  &  (1176, 2558)  &  1920  &  0.20  &  297.7  &  15.7  &  20.4  &  1.1 \\
G359.463-0.090 &  -30  &  2.56  &  1.94  &  85  &  (1434, 1284)  &  1914  &  0.31  &  329.7  &  11.9  &  10.5  &  0.4 \\
G359.511-0.166 &  0  &  2.01  &  1.58  &  -80  &  (1725, 1715)  &  1910  &  0.24  &  420.2  &  14.4  &  20.9  &  0.7 \\
G359.543-0.041 &  -30  &  2.94  &  1.92  &  89  &  (1437, 1469)  &  1915  &  0.28  &  401.9  &  14.5  &  11.3  &  0.4 \\
G359.590-0.121 &  0  &  2.34  &  1.94  &  83  &  (1339, 1361)  &  1913  &  0.30  &  344.2  &  13.4  &  12.0  &  0.5 \\
G359.602-0.207 &  0  &  1.82  &  1.61  &  -81  &  (2412, 1519)  &  1914  &  0.26  &  321.3  &  14.1  &  17.3  &  0.8 \\
G359.622+0.003 &  0  &  2.43  &  2.06  &  74  &  (1291, 1289)  &  1914  &  0.31  &  344.7  &  13.4  &  10.9  &  0.4 \\
G359.668-0.073 &  0  &  2.33  &  1.97  &  79  &  (1340, 1367)  &  1913  &  0.30  &  312.3  &  12.2  &  10.8  &  0.4 \\
G359.704+0.031 &  0  &  2.29  &  1.73  &  -86  &  (1458, 1133)  &  1915  &  0.27  &  293.8  &  15.4  &  11.7  &  0.6 \\
G359.717-0.135 &  0  &  2.07  &  1.86  &  -89  &  (1490, 1256)  &  1915  &  0.28  &  295.4  &  12.1  &  12.1  &  0.5 \\
G359.748-0.025 &  0  &  2.51  &  1.97  &  87  &  (1339, 1363)  &  1914  &  0.30  &  265.1  &  11.7  &  8.5  &  0.4 \\
G359.799-0.102 &  0  &  1.82  &  1.38  &  -85  &  (1986, 2108)  &  1912  &  0.21  &  432.1  &  13.6  &  27.2  &  0.9 \\
G359.822+0.016 &  0  &  2.63  &  1.90  &  78  &  (1372, 1618)  &  1913  &  0.29  &  320.3  &  12.5  &  10.1  &  0.4 \\
G359.874-0.055 &  0  &  1.83  &  1.36  &  84  &  (1882, 1865)  &  1915  &  0.22  &  283.4  &  12.6  &  18.0  &  0.8 \\
G359.915-0.108 &  0  &  2.70  &  2.00  &  -87  &  (1315, 786)  &  1916  &  0.30  &  605.9  &  13.9  &  17.7  &  0.4 \\
G359.945+0.165 &  30  &  2.01  &  1.29  &  -82  &  (1401, 1909)  &  1914  &  0.19  &  309.9  &  15.5  &  18.9  &  0.9 \\
G359.952-0.007 &  0  &  1.83  &  1.47  &  -84  &  (1821, 1809)  &  1914  &  0.22  &  392.1  &  13.5  &  23.0  &  0.8 \\
G359.972+0.089 &  30  &  3.19  &  1.94  &  -84  &  (2606, 1137)  &  1920  &  0.30  &  448.3  &  17.2  &  11.5  &  0.4 \\
G359.999-0.077 &  0  &  2.12  &  1.72  &  -76  &  (1618, 1477)  &  1917  &  0.26  &  552.1  &  12.7  &  23.9  &  0.6 \\
G000.029+0.041 &  30  &  3.19  &  1.85  &  -82  &  (1553, 1415)  &  1917  &  0.29  &  452.8  &  13.9  &  12.1  &  0.4 \\
G000.073+0.184 &  40  &  1.86  &  1.30  &  -78  &  (2088, 2496)  &  1912  &  0.20  &  265.6  &  13.9  &  17.4  &  0.9 \\
G000.082-0.037 &  0  &  2.99  &  2.07  &  -88  &  (1305, 1250)  &  1915  &  0.32  &  596.7  &  14.6  &  15.2  &  0.4 \\
G000.123-0.115 &  0  &  2.97  &  1.87  &  89  &  (1728, 1440)  &  1915  &  0.28  &  448.4  &  13.8  &  12.8  &  0.4 \\
G000.162+0.010 &  0  &  2.96  &  1.93  &  88  &  (1777, 1463)  &  1920  &  0.28  &  412.0  &  14.6  &  11.4  &  0.4 \\
G000.208-0.063 &  0  &  2.76  &  1.99  &  87  &  (1335, 1365)  &  1917  &  0.30  &  489.5  &  13.5  &  14.1  &  0.4 \\
G000.286-0.019 &  0  &  2.82  &  1.94  &  87  &  (1430, 1456)  &  1915  &  0.28  &  538.7  &  15.4  &  15.6  &  0.4 \\
G000.300+0.063 &  0  &  2.51  &  1.76  &  -86  &  (2506, 917)  &  1914  &  0.28  &  430.5  &  17.3  &  15.4  &  0.6 \\
G000.323-0.089 &  30  &  1.94  &  1.53  &  -65  &  (2054, 1580)  &  1917  &  0.23  &  346.3  &  14.9  &  18.4  &  0.8 \\
G000.367+0.029 &  0  &  1.96  &  1.32  &  -75  &  (2000, 2046)  &  1912  &  0.20  &  360.6  &  15.0  &  22.0  &  0.9 \\
G000.412-0.050 &  30  &  1.99  &  1.23  &  -79  &  (2216, 2198)  &  1912  &  0.18  &  346.1  &  15.2  &  22.4  &  1.0 \\
G000.453+0.060 &  0  &  2.73  &  2.03  &  85  &  (2186, 1443)  &  1920  &  0.28  &  433.2  &  15.6  &  12.4  &  0.4 \\
G000.464-0.104 &  0  &  1.86  &  1.29  &  -79  &  (1677, 1622)  &  1914  &  0.19  &  277.5  &  14.9  &  18.3  &  1.0 \\
G000.491-0.002 &  40  &  2.47  &  1.89  &  82  &  (1383, 1420)  &  1914  &  0.29  &  510.4  &  14.5  &  17.3  &  0.5 \\
G000.538-0.079 &  30  &  2.66  &  1.84  &  81  &  (1330, 1355)  &  1913  &  0.30  &  425.5  &  14.0  &  13.7  &  0.5 \\
G000.569+0.044 &  40  &  2.75  &  1.83  &  81  &  (1377, 1380)  &  1917  &  0.29  &  536.4  &  14.3  &  16.8  &  0.4 \\
G000.601-0.131 &  30  &  2.16  &  1.28  &  -76  &  (1387, 2082)  &  1914  &  0.19  &  345.3  &  17.0  &  19.7  &  1.0 \\
G000.616-0.033 &  40  &  1.76  &  1.76  &  88  &  (1909, 1953)  &  1915  &  0.21  &  1377.3  &  11.8  &  70.3  &  0.6 \\
G000.637+0.068 &  30  &  2.50  &  2.13  &  83  &  (759, 1005)  &  1916  &  0.32  &  421.4  &  13.1  &  12.5  &  0.4 \\
G000.665-0.111 &  30  &  1.86  &  1.66  &  -73  &  (1676, 1636)  &  1915  &  0.25  &  631.1  &  12.8  &  32.3  &  0.7 \\
G000.692+0.005 &  30  &  2.53  &  2.09  &  80  &  (1368, 1377)  &  1917  &  0.29  &  2176.9  &  12.6  &  65.1  &  0.4 \\
G000.715-0.180 &  30  &  2.56  &  1.85  &  -82  &  (1502, 1420)  &  1914  &  0.28  &  452.0  &  12.6  &  15.1  &  0.4 \\
G000.741-0.064 &  30  &  2.94  &  1.78  &  -88  &  (1474, 1506)  &  1915  &  0.27  &  637.7  &  15.9  &  19.3  &  0.5 \\
G000.774-0.244 &  30  &  1.86  &  1.20  &  -80  &  (1399, 1837)  &  1914  &  0.19  &  271.4  &  14.9  &  19.2  &  1.1 \\
G000.792-0.142 &  30  &  2.28  &  1.68  &  84  &  (1600, 1642)  &  1915  &  0.25  &  443.2  &  14.1  &  18.3  &  0.6 \\
G000.835-0.216 &  30  &  2.02  &  1.31  &  -73  &  (1996, 1934)  &  1914  &  0.20  &  297.9  &  12.6  &  17.8  &  0.8 \\
\hline\hline
\end{tabular}
\end{table*}

%% file: Tables/flux_stats.tex
\begin{table*}
\caption{\updates{Total integrated fluxes for ACES field \texttt{t} for single array types (7m, 12m, and TP) and for the combined 12m7mTP data using two different approaches: joint imaging and feather (Section \ref{sec:joint_decon_feather}) and feather-only (Section \ref{sec:feather_only}). All integrated flux values are taken across the same spatial and spectral range, using a spatial mask to exclude noisy field edges in the primary beam corrected maps.}}
\label{tab:flux_stats}
\begin{tabular}{cc}
\hline\hline
Array type & Integrated Flux \\
& Jy \kms \\ \hline
7m-only & 173 \\
12m-only & 310 \\
TP-only & 1334 \\
12m7mTP (feather-only) & 1330 \\
12m7m TP (joint + feather) & 1333 \\
\hline\hline
\end{tabular}
\end{table*}